\documentclass[10pt,journal,compsoc]{IEEEtran}

\ifCLASSOPTIONcompsoc
  \usepackage[nocompress]{cite}
\else
  \usepackage{cite}
\fi
\ifCLASSINFOpdf
\else
\fi
\usepackage{lipsum} 
\usepackage{tabularx}
\usepackage{lipsum,amsmath,multicol}
\usepackage{multirow}
\usepackage{array,booktabs}
\usepackage{lipsum} 
\usepackage{tabularx}
\usepackage{adjustbox}
\usepackage{lipsum,amsmath,multicol}
\usepackage{multirow}
\usepackage{array,booktabs}
\usepackage{threeparttable} 
\usepackage{array}
\usepackage{makecell}

\usepackage{pifont}
\usepackage[usestackEOL]{stackengine}
\strutlongstacks{T}
\usepackage{hyperref}
\usepackage{subcaption,siunitx,booktabs}
\usepackage[table]{xcolor}
\definecolor{lightgray}{gray}{0.9}

\usepackage{float}
\usepackage{ragged2e}
\usepackage{epstopdf}
\AppendGraphicsExtensions{.tif}
\newcommand*{\myfont}{\fontfamily{qcr}\selectfont}

\usepackage{ifthen}
\newboolean{printComments}
\setboolean{printComments}{true}
\newcommand{\shagunr}[1] {{\ifthenelse {\boolean{printComments}} {\small{\color{red} [#1]}} {}}}
\newcommand{\faheemb}[1] {{\ifthenelse {\boolean{printComments}} {\small{\color{blue} [#1]}} {}}}

\begin{document}
%
\title{Design and Implementation of Fragmented Clouds for  Evaluation of Distributed Databases}
%
%
%
%

\author{Yaser Mansouri,  Faheem Ullah, Shagun Dhingra,
        and M. Ali Babar
\IEEEcompsocitemizethanks{\IEEEcompsocthanksitem Authors are with Centre for Research on Engineering Software Technology (CREST) Lab. School of Computer Science, The University of Adelaide, Adelaide, Australia.\protect
\IEEEcompsocthanksitem  Authors email: \{yaser.mansouri, faheem.ullah, shagun.dhingra, ali.babar\}@adelaide.edu.au}
}

%
%

\markboth{Journal of \LaTeX\ Class Files,~Vol.~XXX, No.~XXX, September~2022}%
{Shell \MakeLowercase{\textit{et al.}}: Bare Advanced Demo of IEEEtran.cls for IEEE Computer Society Journals}
%



\IEEEtitleabstractindextext{%
\justify
\begin{abstract}
In this paper, we present a Fragmented Hybrid Cloud (FHC) that provides a unified view of multiple geographically distributed private cloud datacenters. FHC leverages a fragmented usage model in which outsourcing is bi-directional across private clouds that can be hosted by static and mobile entities. The mobility aspect of private cloud nodes has important impacts on the FHC performance in terms of latency and network throughput that are reversely proportional to time-varying distances among different nodes. Mobility also results in intermittent interruption among computing nodes and  network links of  FHC infrastructure. To fully consider mobility and its consequences, we  implemented a layered FHC that leverages Linux utilities and bash-shell programming. We also evaluated the impact of the mobility of nodes on the performance of distributed databases as a result of time-varying latency and bandwidth, downsizing and upsizing cluster nodes, and network accessibility. The findings from our extensive experiments provide deep insights into the performance of well-known big data databases, such as Cassandra, MongoDB, Redis, and MySQL, when deployed on a FHC.  
\end{abstract}

}

\maketitle

\IEEEdisplaynontitleabstractindextext

%
\IEEEpeerreviewmaketitle

\ifCLASSOPTIONcompsoc
\IEEEraisesectionheading{\section{Introduction}\label{sec:introduction}}
\else
\section{Introduction}\label{sec:introduction}
\fi

\IEEEPARstart{A}hybrid cloud makes a seamless and secure connection between public and private clouds to serve workload spikes. This type of cloud model exploits popular \textit{on-demand} usage model \cite{Toosi2014}. A hybrid cloud can also come out with two other flavors of usage models based on objective functions and constraints \cite{Mansouri2019}. These are \textit{fragmented} and \textit{collaborative} usage models in which resources are shared between multiple private clouds and a federation of private and public clouds, respectively \cite{Mansouri2019}\cite{ASSIS2016}. The fragmented usage model is becoming more and more popular due to its advantages. These advantages include improvement in cloud performance, lowering network costs, enhancing resilience, and handling data as per the regional rules and privacy laws. 

The fragmented usage model provides a secure resource sharing between multiple private clouds -- so-called Fragmented Hybrid Cloud (FHC) --, where the amount of resource sharing depends on the size of the private clouds. The request for outsourcing between each pair of private clouds can be bi-directional as opposed to the on-demand usage model in which requests are outsourced from private to public cloud \cite{Mansouri2020}.  In contrast to the on-demand model, in the fragmented usage model, a private cloud can be either hosted by a static or a mobile object. The static hosting objects, as its name implies, the location of the private cloud is fixed over time. In contrast, the mobile nature of objects implies that the location of private clouds can change over time. As a side effect, when mobile objects get closer to each other, the amount of resources available to them increases. Similarly, increasing distance leads to fewer resources being available. Thus, due to mobility of private clouds, the fragmented usage model faces critical challenges: \textbf{network services} in terms of time-varying \textit{latency} and \textit{bandwidth}, and \textbf{intermittent resources} in terms of \textit{computing nodes} and \textit{connection links}. 

Due to the mobility of resources, the distance between private clouds may change that needs to be adjusted to latency. Specifically, longer distances are expressed as higher latencies and more degraded in network throughput\cite{Sakagami1991}\cite{Wu2013}. 
Also, the mobility of resources is a plausible cause for the interruption of network connections between one private cloud with other members in FHC over time. As a consequence of this challenge, computing nodes are removed from and added to FHC over time. Given there can be redundant connection links between private clouds, the impact of intermitent interruption can be avoided (or at least minized). However, the redundant links may create the need of re-routing data through proper network links without interruption in  application services in the case of  intermittent unavailability of links. 

To address the above challenges and explore their impact on the performance of data-intensive applications in FHC, we designed and implemented a layered architecture of FHC  supported by Linux-based utilities. The layered architecture includes \textit{hardware}, \textit{network}, and \textit{application} layers. The hardware layer is a combination of private clouds or private and public clouds. The network layer connects computing nodes through a single network link as one connection type. We also leveraged WireGuard \footnote{WireGuard:\url{https://www.wireguard.com/}} to create a network with redundant connection links, where each connection link represents a connection type. Each connection type possesses different values of latency and bandwidth over time, which are controlled by {\myfont{tcconfig}} as a Traffic Control (TC) wrapper\footnote{tcconfig:\url{https://tcconfig.readthedocs.io/en/latest/pages/introduction/index.html}}. We also exploited {\myfont{tinc}}\footnote{Tinc: \url{https://github.com/gsliepen/tinc}} as a Linux-based utility to handle accessibility in the mesh network upon appearing and disappearing a network link. The application layer includes data-intensive applications like distributed databases as studied in this paper.   

Our main criteria to select distributed databases in this study are popularity, usage, and commercialization by well-known cloud providers. We thus investigate the performance of Cassandra\footnote{Cassandra: \url{https://cassandra.apache.org}}, MongoDB\footnote{MongoDB: \url{https://www.mongodb.com}}, Redis\footnote{Redis: \url{https://redis.io/}}, and MySQL\footnote{MySQL: \url{https://www.mysql.com/}} -- deployed on a FHC infrastructure that can be impacted by  time-varying factors primarily latency  and bandwidth. Furthermore, dynamic run-time links disruption and restoration is likely to affect database-related workload performance. This might lead to removing and adding computing nodes from and to FHC. With redundant links between nodes in FHC, re-routing data to the proper link might have  an effect on the performance of databases. In summary, the contributions of this paper are as follows. 

\begin{itemize}
    \item We designed  and implemented a layered fragmented cloud framework in which a set of Linux-based tools enables us to create a mesh network, control latency and traffic, and handle network accessibility. 
    \item We evaluated the impact of time-varying latency and bandwidth on distributed databases performance. 
    \item We measured the impact of removing/adding computing nodes at application levels on the performance of databases.
    \item We assessed the effect of network accessibility on the responsiveness of databases.  
    \vspace{-5mm}
\end{itemize}


\begin{table*}[t]
\begin{threeparttable}
  \centering
  \tiny
  \caption{Comparison of empirical studies on the evaluation of  distributed databases  in edge-cloud paradigms. Here, \textbf{M} stands for Mobility, \textbf{L} for Latency, \textbf{B} for Bandwidth, \textbf{NA} for Network Accessibility, and \textbf{CA} for Computing Accessibility.}    
   \label{tab:relatedwork}%
     \begin{tabular}{p{1cm}p{1.5cm}p{4cm}p{3.5cm}p{0.3cm}p{0.3cm}p{0.3cm}p{0.3cm}p{0.3cm}p{1cm}}
     
    \toprule
          &    & &   & & \multicolumn{2}{c}{\textbf{Impact of}} \\\hline
    \midrule
    \multicolumn{1}{p{1cm}}{\textbf{Paper}}
    &\multicolumn{1}{p{1cm}}{\textbf{Infrastructure}}
    &\multicolumn{1}{p{2cm}}{\textbf{Databases}}
    &\multicolumn{1}{p{2cm}}{\textbf{Performance metrics}}
    &M  &L &B &NA &CA &  \\\hline
    \cite{Rabl2012}   &Private cloud   & \Longunderstack{Cassandra, HBase, Redis,\\ Voldemort, VoltDB, MySQL} &{Throughput, Latency}     &\ding{56} &\ding{56} &\ding{56} &\ding{56} &\ding{56} &-\\\hline
    
    \cite{Kuhlenkamp2014}   &Public cloud   &Cassandra and HBase &Elasticity, Scalability 
    &\ding{56} &\ding{56}  &\ding{56} &\ding{56} &\ding{56} &-\\\hline
	
	\cite{Klein2015}   &Public cloud  &MongoDB, Cassandra, Riak &Throughput  
	&\ding{56} &\ding{56}  &\ding{56} &\ding{56}  &\ding{56} &Consistency\\\hline
	
	\cite{Gandini2014}   &Public cloud  &MongoDB, Cassandra, HBase &Throughput, Scalability  
	&\ding{56} &\ding{56}  &\ding{56} &\ding{56}  &\ding{56} &Replication \\\hline
	
	\cite{Pereira2018}   &Public cloud  &MongoDB, CouchDB, RethinkDB &Throughput  
	&\ding{56} &\ding{56}  &\ding{56} &\ding{56}  &\ding{56} &Replication \\\hline
	
	\cite{Shankar2014}   &Public cloud  &Cassandra &Latency  
	&\ding{56} &\ding{56}   &\ding{51} &\ding{56}  &\ding{56} &Threads \\\hline
	
	\cite{TANG2019}    &MCC   &SQLLite, CouchDB &\Longunderstack{Energy Conspution,\\ Resource utilization } &\ding{51} &\ding{56}  &\ding{56}  &\ding{56}  &\ding{56} &Threads \\\hline

	\cite{Mansouri2020} &Hybrid cloud & \Longunderstack{Cassnadra, MongoDB, Riak,\\ CouchDB, Redis, MySQL} &throughput, Latency &\ding{56}  &\ding{56} &\ding{56} &\ding{56}  &\ding{56}  &-\\\hline
    
    \cite{Mansouri2020distance} &Hybrid cloud &\Longunderstack{Cassnadra, MongoDB, Riak,\\ CouchDB, Redis, MySQL} &Throughput, Scalability &\ding{56} &\ding{51}  &\ding{56} &\ding{56} &\ding{56}  &-\\\hline
    
    \cite{Mansouri2021energy}  &Hybrid cloud, Edge  &MongoDB, Cassandra, Redis, MySQL
    & \Longunderstack{Energy Consumption,\\ Resource utilization}   &\ding{56} &\ding{51}  &\ding{56}  &\ding{56}  &\ding{56} &-\\\hline
    
    \cite{Seybold19}  &Private cloud  &Cassandra, CouchDB,   & throughput  &\ding{56} &\ding{56}  &\ding{56}  &\ding{56}  &\ding{56} &Elasticity\\\hline
    \cite{Seybold20}  &Private cloud  &Cassandra, CouchDB,   & throughput  &\ding{56} &\ding{56}  &\ding{56}  &\ding{56}  &\ding{51} &Availability\\\hline
    
    \textbf{Our work}  &\textbf{Fragmented hybrid cloud}  &\textbf{MongoDB, Cassandra, Redis, MySQL} &Throughput, Resource utilization &\ding{51} &\ding{51}  &\ding{51}  &\ding{51}  &\ding{51} &-\\\hline
    
    \bottomrule
    \end{tabular}%
    
\end{threeparttable}
\vspace{-5mm}
\end{table*}%

\section{Related Work}
\textbf{Hybrid Cloud Implementation:}
Aamzon Web Services (AWS) Outposts\footnote{Outposts: \url{https://aws.amazon.com/outposts}}, Google Anthos\footnote{Anthos: \url{https://cloud.google.com/anthos}}, and Azure Stack hyper-converged infrastructure (HCI)\footnote{HCI: \url{https://azure.microsoft.com/en-au/products/azure-stack/hci/}} are notable examples of hybrid cloud solutions exploit native public clouds services on premises through native commercial Virtual Private Network (VPN) or third party VPN like WireGuard. 
We automated the implementation of a hybrid cloud  through a secure, zero-cost, and resilient WireGuard to evaluate distributed databases \cite{Mansouri2020}. Authors in \cite{Toosi2018} and \cite{TULI2020} implemented a hybrid cloud to evaluate their proposed algorithms for resource provisioning and task scheduling. All these hybrid clouds use an on-demand usage model \cite{Mansouri2020} in which applications outsource to a public cloud in case of workload spikes.     

Outsourcing across multiple private clouds with intermittent connectivity for computing resources leads to the fragmented usage model of hybrid cloud \cite{Mansouri2020}. The widespread need for resource-moderated/constrained mobile devices creates an incentive for academia and industry sectors to explore resource management across them. 
Several studies have investigated this model of resource sharing in which mobility of devices serving by hybrid cloud is a key parameter. Authors in \cite{Sharma2015} proposed hybrid cloud enabled architecture (SHCEI) in which a secure data transmission between a hybrid cloud and mobile/stationary devices are exchanged. Authors in \cite{ALI2018} reviewed different aspects of private cloud interconnection (i.e., fragmented cloud) in terms of availability, accessibility, performance, disaster recovery and data replication. Differently, we implemented FHC and explored the impact of mobility leading to changes in latency, bandwidth, and network accessibility on the performance of distributed databases. 

\textbf{Cloud Federation:} The usage model of our fragmented cloud is a kind of cloud federation in which the network connection is peer-to-peer. Authors in \cite{LIAQAT2017} reviewed a taxonomy of management techniques issues and solutions including discovery, monitoring, allocation, selection, failure and  pricing of resources. Functional properties (e.g., access, monitoring and provisioning), and non-functional properties (e.g., expansion and interaction) of cloud federation have been studied in \cite{ASSIS2016}. In contrast, we investigated the impact of mobility of cloud resources, which in turn, affects the network density (i.e., the average proximity of nodes in the network) on the performance of distributed databases. Likewise, intermittent infrastructure impacts the performance of distributed databases.  

Mobility, the key feature of our FHC, is common with Mobile Cloud Computing (MCC) and Mobile Edge Computing (MEC). Task offloading/delegation in MCC and MEC is categorized into \textit{distant}\cite{mach2017}\cite{ren2019}\cite{MANSOURI2021}, \textit{proximal} \cite{nayyer2018}\cite{yousefpour2019}\cite{yi2015} and \textit{peer-to-peer nodes} \cite{Hong2019}\cite{chang2014}\cite{TANG2019}. 
Our work falls into the last category, where all mobile nodes share resources to perform sub-tasks through a peer-to-peer connection. Our study, however, significantly differs  in design (multiple connections between mobile cloud), application type (distributed databases), and objective functions.  

\textbf{Performance Evaluation of Distributed Databases on Clouds:}
In \cite{Rabl2012}, authors evaluated six NoSQL and relational databases on private computing clusters in terms of throughput, latency, and storage usage. Kuhlenkamp et al. \cite{Kuhlenkamp2014} studied the elasticity and scalability of Cassandra and HBase \footnote{Hbase: \url{https://hbase.apache.org}} on AWS. Authors in \cite{Klein2015} explored the impact of different consistency models on the throughput of MongoDB, Cassandra and Riak deployed on private and public clouds. Gandini et al. \cite{Gandini2014} explored the impact of the replication factor and  computing flavor on the performance of MongoDB, Cassandra, and HBase deployed on public clouds. The impact of threads number on the performance of MongoDB, CouchDB \footnote{CouchDB: \url{https://couchdb.apache.org}} and RethinkDB  \footnote{RethinkDB: \url{https://rethinkdb.com}} in public cloud has been studied in \cite{Pereira2018}. Several studies explored these databases from an energy consumption perspective as the queries are optimized \cite{Mahajan2017}, or different cloud applications are leveraged \cite{Bani2016}. Seybold et al. \cite{Seybold19} evaluated the elasticity of Cassandra and CouchDb for read and write intensive workloads in private clouds. They also presented a comprehensive methodology for evaluating the availability of Cassandra and CouchDB in case of cloud resource failure \cite{Seybold20}. 
We recently evaluated the throughput of six databases on the on-demand usage model of hybrid cloud \cite{Mansouri2020}. Using the same setting, we explored the impact of distance between private and public clouds on the vertical and horizontal scalability and VM packing (i.e., fewer VMs with a larger size vs. more VMs with smaller size) for these databases \cite{Mansouri2020distance}. We then extended the infrastructure to include edge nodes and measured the energy consumption and resource utilization of four databases for (non-)offloading scenarios \cite{Mansouri2021energy}. The unique aspect of our resaerch reported in this paper is that we here implemented FHC to evaluate the impact of mobility-inducing time-varying latency, bandwidth and intermittent computing and network resources.

\section{Fragmented Hybrid Cloud (FHC)}
This section discusses the challenges associated with FHC and then delineates the corresponding implementation. 

\subsection{Challenges}\label{sec:challanges}
FHCs provide secure resource sharing across Geo-distributed private clouds hosted by static or mobile objects. Mobile hosting objects bring several challenges in terms of \textit{discovery}, \textit{management}, and \textit{mobility} of resources. Resource discovery is related to methods that detect new resources appearing in the FHCs environment. Resource management is associated with consumed resources, remaining resources and potential sharing of resources across private clouds. Resource mobility introduces other challenges, which are the focal points of this paper.  

\textbf{Resource mobility:} This is a consequence of private clouds residing in mobile objects, which in turn, leads to the time-varying distance between them. In fact, the private cloud nodes can move away from each other  and  approach each other. To translate the distance between  cloud nodes, we adjust latency to match distance. Specifically, longer distances are expressed as higher latencies. Likewise, as distance increases between private cloud nodes, the bandwidth size decreases, and vice versa \cite{Sakagami1991}\cite{Yao2008}. Thus, \textit{time-varying latency and bandwidth} are main consequences of resource mobility.
Furthermore, due to the mobility of cloud nodes, FHC is faced with intermittent resources as discussed below.     

\textbf{Computing node accessibility:}
Mobility comes with the consequence of impact on the coverage and speed of connectivity between private cloud nodes. This leads to computing node accessibility in FHCs, where the presence of computing nodes changes over time. Thus, leaving and joining nodes to and from FHCs is another side effect of resource mobility. To investigate the effect of intermittent nodes on applications performance, we remove/add nodesfrom/to FHCs at the application layer (\S\ref{sec:designandImple}). 

\textbf{Network accessibility:}
Due to the mobility of cloud nodes in FHCs, network accessibility is another challenge that can be alleviated through the utilization of redundant  links. Each link represents a connection type differentiated by latency and bandwidth size. Re-routing data between redundant links is another hindrance while  link(s) between a pair of nodes are appeared/disappeared over time. 

\begin{figure}[t]
\centering
\includegraphics[width=0.8\columnwidth]{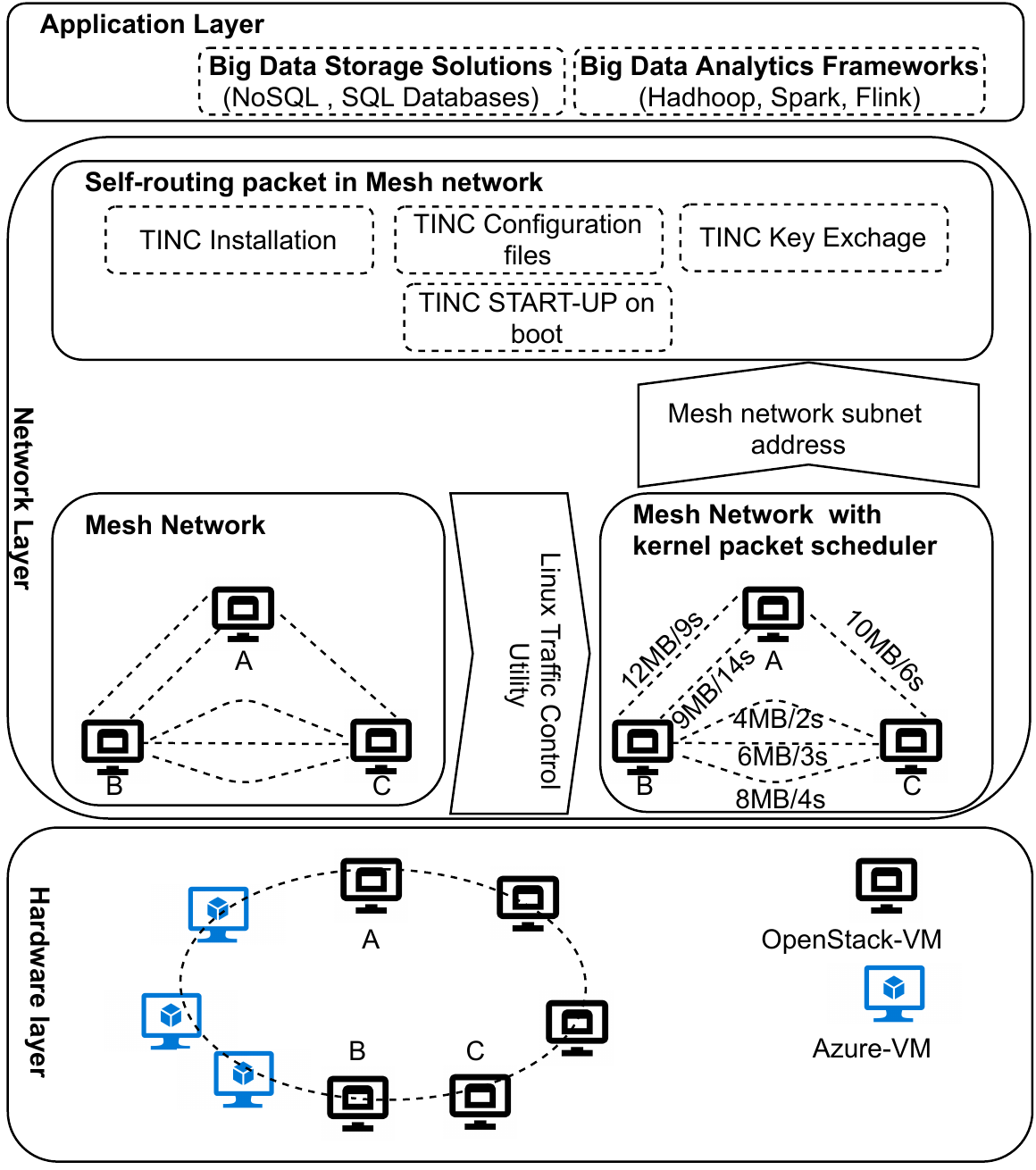}
\vspace{-3mm}
\caption{A layered architecure of  FHC}
\label{fig:frag-arch}
\vspace{-3mm}
\end{figure} 
\subsection{Design and Implementation}\label{sec:designandImple}
From the above-discussed challenges, we designed a layered architecture for FHCs as depicted in Fig. \ref{fig:frag-arch}.  

\textbf{Hardware Layer:} This layer consists of VM nodes spanning between private and public clouds. We leveraged Terraform\footnote{Terraform: \url{https://www.terraform.io}} to deploy infrastructure and exploited WireGuard to build a  performance and cost-efficient network connection between private and public clouds \cite{Mansouri2019}. 

\textbf{Network Layer:} This layer is the core of FHCs and we automated in its implementation through shell script programming to accurately and repeatedly build FHCs. It includes the following components.

\textit{ Mesh network} component provides different connection types between cloud nodes through WireGuard to improve accessibility. In the mesh network, the connection is peer-to-peer and the number between each pair can be different. For example, the number of connections between nodes (A, B) and (B, C) is two and three, respectively. 

\textit{Mesh network with kernel packet scheduler:} This component reflects latency between each pair of nodes and bandwidth size for different connection types. We used tcconfig as a Traffic Control (TC) wrapper to assign different values of latency and bandwidth to connect links between each pair of nodes (Fig. \ref{fig:frag-arch}). For instance, nodes A and B connected through two links with 9 ms and 14 ms in latency, and 12 MB and 9 MB in bandwidth. 

\textit{Self-routing packet in mesh network:} This component exploits {\myfont{tinc}} as  a VPN that leverages tunnelling and encryption to create an overlay network on the mesh network (Fig. \ref{fig:frag-arch}). {\myfont{Tinc}} directly sends data from the source to the destination through the path with minimum network hop between source and destination nodes. Importantly, it handles automatic failover as links in the mesh network are connected/disconnected due to the mobility of nodes.

\begin{figure*}[h]
  \centering
  \subfloat[]{\label{fig:tinc-a}\includegraphics[width=0.25\textwidth]{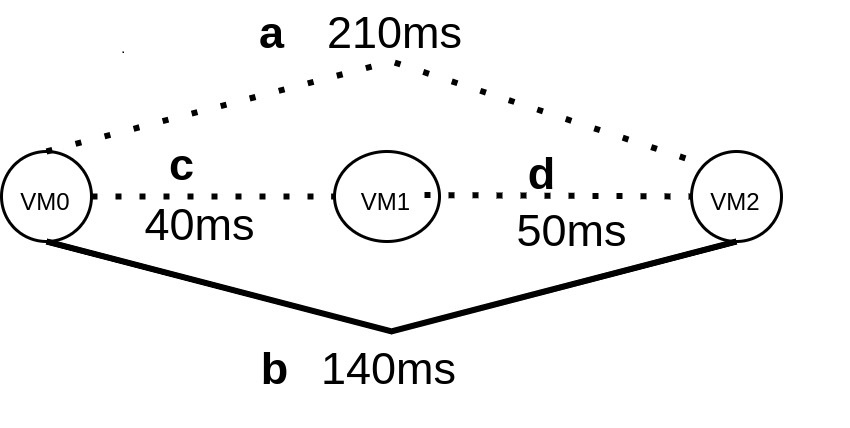}}
  \subfloat[]{\label{fig:tinc-b}\includegraphics[width=0.25\textwidth]{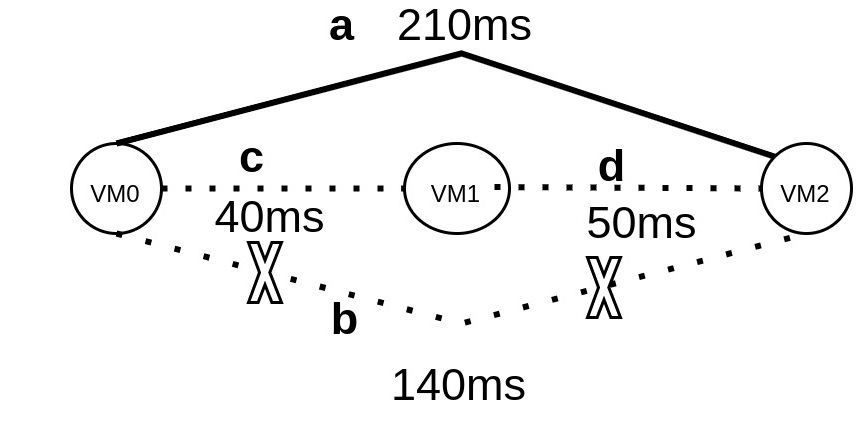}}
  \subfloat[]{\label{fig:tinc-c}\includegraphics[width=0.25\textwidth]{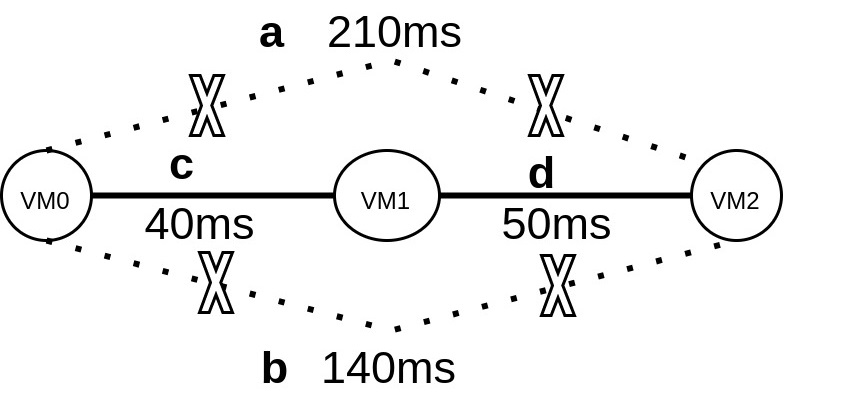}}
  \subfloat[]{\label{fig:tinc-d}\includegraphics[width=0.25\textwidth]{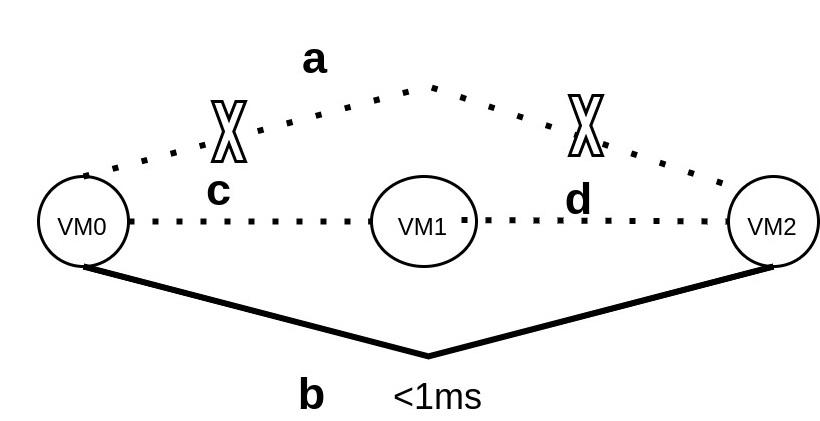}}
  \caption{(a) Tinc sends data via link b, (b) Tinc sends data via link a when link b is disconnected, (c) Tinc sends data via links c and d when both links a and b are disconnected, and (d) Tinc sends data via link b when it gets back.}
\label{fig:tinc-scenarios}
\vspace{-3mm}
\end{figure*}

To clarify the functionality of {\myfont{tinc}}, we created a mesh network by using WireGuard with three nodes (vm0, vm1, and vm2) and four links \textbf{a} with 210 ms, \textbf{b} with 140 ms, \textbf{c} with 40ms, and \textbf{d} with 50 ms (Fig. \ref{fig:tinc-a}). Note that all these latencies are set by {\myfont{tcconfig}}. 
 As depicted in Fig.\ref{fig:tinc-a}, {\myfont{tinc}} initially sends data from node vm0 to vm2 through link b (shown in black line) because the number of hops between these two nodes is one and the latency of link b is less than for link a. When link b is disconnected \ref{fig:tinc-b}, then Tinc sends data through link a due to one hop distance between nodes vm0 and vm1 (Fig. \ref{fig:tinc-b}). 
As shown in Fig. \ref{fig:tinc-c}, {\myfont{tinc}} is resilient to link failure since it sends data through links c and d when both direct links between vm0 and vm2 are disconnected. As direct link b between nodes vm0 and vm2 gets back, {\myfont{tinc}} automatically sends data through the path with one network hop, link b. (Fig. \ref{fig:tinc-d}). Therefore, (i) {\myfont{tinc}} selects the shortest path based on the network hop between source and destination nodes for sending data, and (ii) {\myfont{tinc}} is resilient to the link disconnection so that it requires roughly 30 seconds to send data through an alternative link. 

\textbf{Application Layer:} This layer hosts data-intensive applications including Cassandra, MongoDB, Redis, and MySQL that leverage the underlying resources to serve workloads. Since the availability of cloud nodes in the mobile environment is critical, we investigate these databases architeture to see how to add and remove nodes from and to the cluster.  

\begin{figure*}[t!]
  \centering
  \subfloat[Cassandra]{\label{fig:cass-arc}\includegraphics[height=4cm,width=0.25\textwidth]{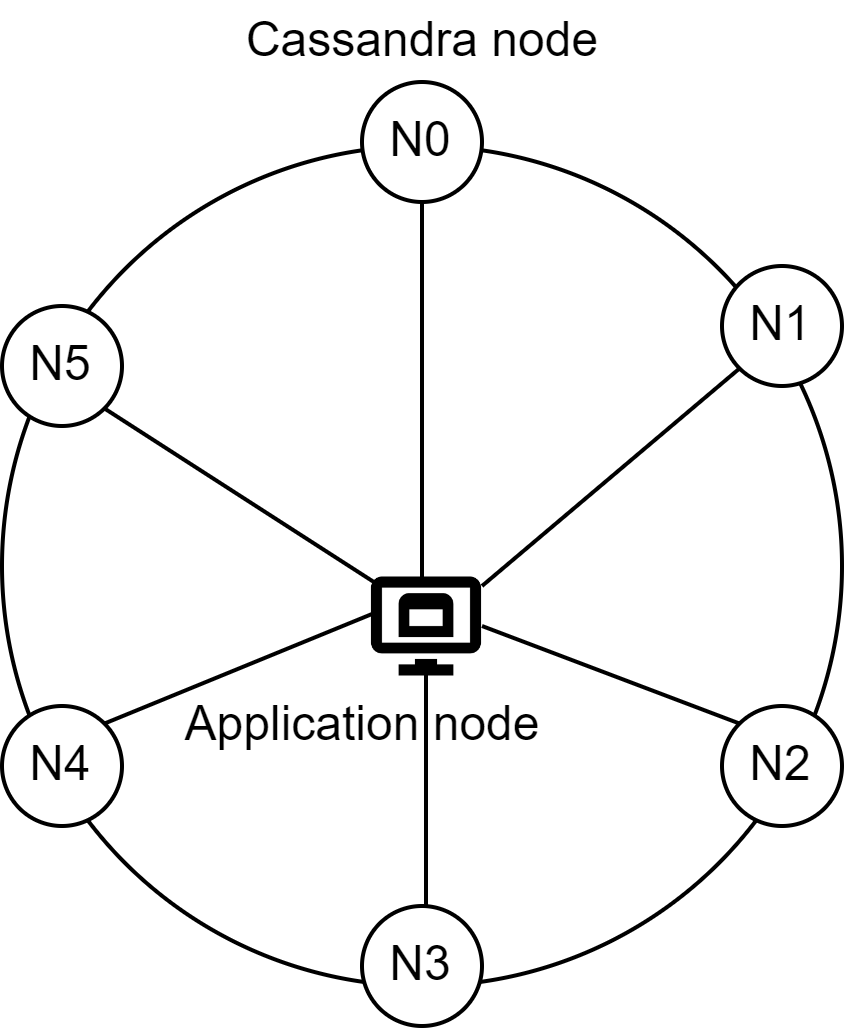}}
  \subfloat[MongoDB]{\label{fig:mongo-arc}\includegraphics[height=4cm,width=0.25\textwidth]{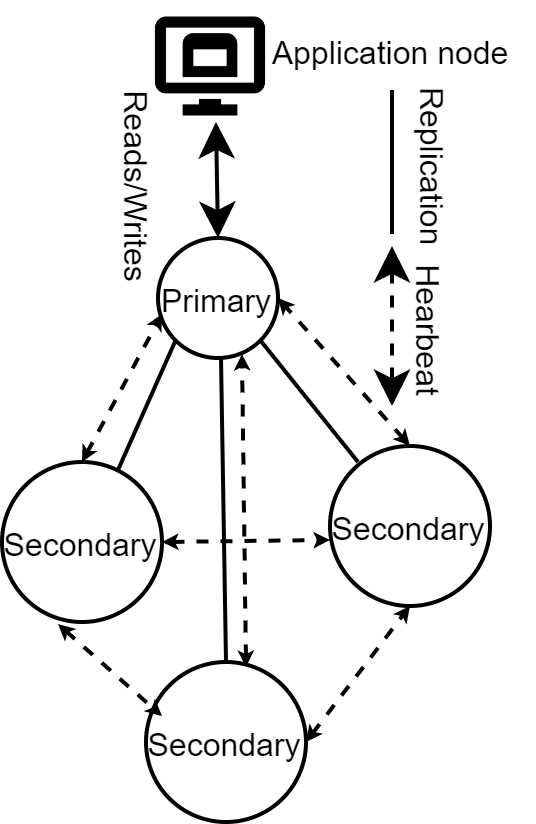}}
  \subfloat[Redis]{\label{fig:redis-arc}\includegraphics[height=4cm,width=0.25\textwidth]{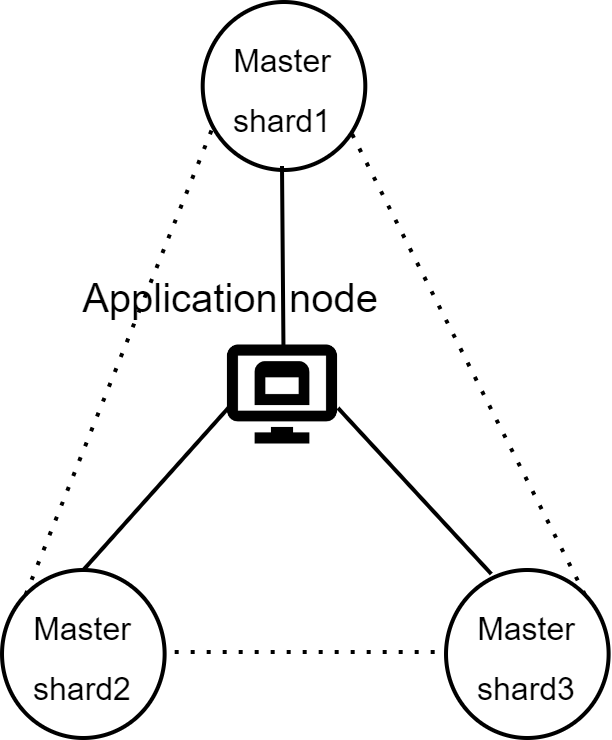}}
  \subfloat[MySQL]{\label{fig:mysql-arc}\includegraphics[height=4cm,width=0.25\textwidth]{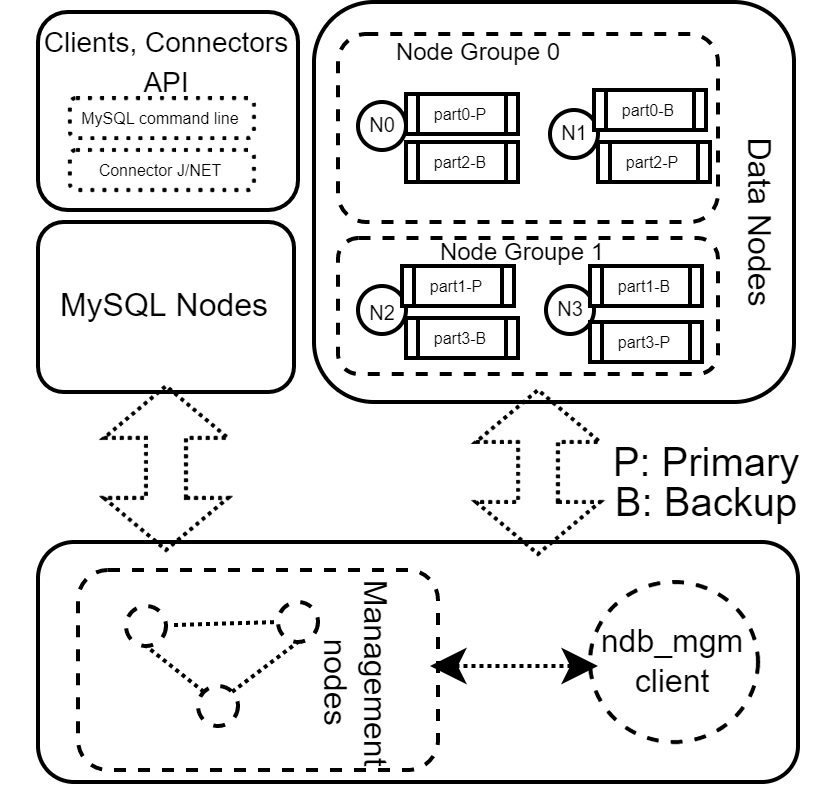}}
  \caption{A schematic architecture of distributed databases  }
\label{fig:database-arch}
\vspace{-3mm}
\end{figure*}

\textit{Cassandra:} 
Cassandra supports a peer-to-peer connection in which each node is connected to all other nodes as depicted in Fig. \ref{fig:cass-arc}. Thus, Cassandra is materless and thus it does not have a single point of failure. Cassandra cluster consists of two components: (i) \textit{seed node} bootstraps a node when it is initially  joined  to a cluster and configures  cluster nodes as a ring and (ii) \textit{goosip protocol} exploits peer-to-peer communication to inform a node about the state of all other nodes through  {\myfont{nodetool status}} command. Cassandra cluster replicates data across all nodes almost equally and all nodes are equal in writes and reads processing. Read and Write Operations can be submitted to any node in the cluster and then the node that received the request must find data on the cluster. Removing and adding nodes to the cluster might have an impact on the performance of Cassandra. A node is removed from the Cassandra cluster through {\myfont{nodetool drain}} command, which stops the node from listening to the client. A node is added to the Cassandra cluster through {\myfont{service Cassandra start}} command. 

\textbf{\begin{table*}[t]
	\caption{Comparison of databases in terms of architecture and features.}\label{tab:database-comp}
	\centering
	\small
	\begin{tabular}{p{1.5cm}p{5.2cm}p{5.2cm}p{2.2cm} p{0.5cm} p{0.5cm}}
		\hline
		        &Remove node API  &Add node API    &Architecture     &Res.$\dagger$  &Rep.     \\\hline\hline
	Cassandra	&nodetool drain \textit{node-ip}   &cassandra service \textit{node-ip}             &Master-less    &YES  &YES \\
	MongoDB	    &rs.remove \textit{node-id}    &rs.add \textit{node-id}           &Master-slave      &NO   &YES \\
	Redis	    &redis-cli --cluster del-node \textit{node-ip,id}    &redis-cli --cluster add-node  \textit{node-ip,id}  &Master-Master    &YES  &NO  \\
	MySQL	    &drop nodegroup \textit{nodegroup-id}$\dagger$    &create nodegroup \textit{nodegroup-id}           &NDBD       &YES  &YES\\\hline
	\end{tabular}
	\begin{tablenotes}
      \tiny
       \item $\dagger$ Res. and Rep. stand for \textit{resharding} and \textit{replication} respectively. 
    \end{tablenotes}
	\vspace{-3mm}
\end{table*}}
\textit{MongoDB:} The cluster of MongoDB offers  high performance, availability and durability for data through \textit{replication} and \textit{sharding} approaches. Replication ensures vertical scalability and sharding inflates horizontal scalability. The core of the replication approach is a replica set, where the primary node and secondary nodes host identical data. The primary node is the only member in the replica set that receives write operations (Fig. \ref{fig:mongo-arc}). Then, the applied write operations are recorded in the primary's \textit{oplog} in order to apply to the secondary nodes. The read operations can be served by both primary and secondary nodes based on the read policies\footnote{Read-preference in MongoDB: \url{https://www.mongodb.com/docs/manual/core/read-preference/}}. It is worth noting that secondary nodes are divided into \textit{active} and  \textit{hidden} nodes. A hidden secondary node neither performs read operations nor goes through election for a new primary node. The architecture of MongoDB simply allows to add and remove secondary nodes, while the primary node cannot be removed. A node is added to and removed from a replica set in MongoDB cluster through {\myfont{rs.add(<node-id>)}} and {\myfont{rs.remove(<node-id>)}} APIs, respectively. Note that these APIs must be submitted to the primary node. 

\textit{Redis:} A Redis cluster is a set of VM nodes that allows us to create any number of Redis databases in a memory pool in the form of key-value pairs. A Redis cluster  consists of a few hundred Redis databases in one of the following types. (i) A simple database (a single master shard), (ii) a pair of a master-slave database, (iii) clustered database containing multiple master shards, each managing a subset of the dataset, and (iv) A highly available clustered database, i.e., multiple pairs of master/slave shards. The last two cluster types shard and replicate data across nodes. We consider the sharded architecture in which the data is split between Redis instances (Fig. \ref{fig:redis-arc}). To split data, Redis allocates a maximum number of 16384 slots to the Redis cluster instances. For example, with three servers, hash slots (0-5500), (5501-11000) and (11001-16383) are allocated to server 1, server 2 and server 3, respectively. To remove or add a node to the Redis cluster, data should be re-sharded and re-balanced between participating nodes in a cluster. Thus, removing/adding a node from/to a Redis cluster is conducted in three steps. (i) Resharding in which data shards are transferred, for example, from the removed node to an arbitrary node in the cluster. This can be conducted through {\myfont{redis-cli --cluster reshard }}. (ii) A  {\myfont{redis-cli --cluster del-node/add-node }} command is submitted the node that should be removed from/added to Redis cluster. (iii) A {\myfont{redis-cli --cluster rebalance}} command is conducted on Redis cluster to re-balance all sharded data across nodes as conducted in all our experiments. 

\textit{MySQL}. This database supports shared-nothing clustering and auto-sharding for data  management. MySQL consists of \textit{data nodes}, \textit{management nodes}, and \textit{SQL nodes} (Fig. \ref{fig:mysql-arc}). Data nodes  are divided into several node groups in which the number of nodes equals the number of replicas for each data shard. For example, with 6 nodes and two replicas for each data shard, the number of node groups is  6/2=3. Management nodes manage  cluster configuration, cluster status, cluster backups and so on through \textit{ndb\_mgm} command. Lastly, SQL nodes support the execution of MySQL queries via command lines, MySQL connector and API. 

MySQL supports removing and adding node groups rather than nodes through  {\myfont{ndb\_mgm create nodegroup }} and {\myfont{ndb\_mgm drop nodegroup}} commands respectively\footnote{Adding NDB Cluster Data Nodes Online: \url{https://dev.mysql.com/doc/refman/5.6/en/mysql-cluster-online-add-node.html}}. For adding a nodegroup, it is required to restart the  management server and conduct a rolling restart of the existing data nodes. Alternatively, without rolling restart, it is possible to reserve nodegroup when the MySQL cluster is initially built. This brings nodes to the MySQL cluster later. For removing a nodegroup, all nodes in the nodegroup must be completely empty of any table data and definition. Since there is currently no way using {\myfont{ndb\_mgm}} or MySQL client to remove all data from a specific data node or nodegroup, this implies that dropping nodegroup performs in a particular condition. In addition to the above time-consuming steps, it is required to manually redistribute data based on relational data model. This makes irrelevent differnces between  MySQL and other databases in terms of removing and adding nodes, which is out of our study scope.

\section{Evaluation}
This section initially defines experimental setup (\S\ref{sec:experisetup}) and then discusses results (\S\ref{sec:results}). 
\subsection{Experiment Setup}\label{sec:experisetup}
\textbf{Infrastructure:}
We selected datacenters in {\myfont{Melbourne}}, {\myfont{Sydney}}, {\myfont{Canberra}}, {\myfont{Pune}}, {\myfont{Singapore}}, {\myfont{Seoul}}, {\myfont{Dubai}} and {\myfont{Virginia}} regions with a range of distance between 725 KM-16671KM. 
The {\myfont{Ping}} and {\myfont{Iperf3}}  have been used to measure latency, download and upload bandwidth between each pair of datacenter for 24 hours, as summarized in Table \ref{tab:latency-bandwidth-measure}. 
Note that the value of upload bandwidth between datacenters is 97-992Mb/s, which is not shown in Table  \ref{tab:latency-bandwidth-measure}, though we considered it in our experiments. We exploited 8 VMs (2 vCPU, 4GB RAM, 40GB Disk) in the private cloud, to emulate the settings across Azure Datacenters. Each VM represents an Azure datacenter. The emulated settings allow us to create large-scale of experimental scenarios and repeatedly conduct experiments with minimal cost. 

\begin{table*}[t]
	\caption{A summary of latency and download bandwidth between 8 datacenters}\label{tab:latency-bandwidth-measure}
	\centering
	\small
	\begin{tabular}{p{2.5cm}p{1.5cm}p{1cm}p{1.3cm}p{1.5cm}p{1.5cm}p{1.5cm}p{1.5cm}p{1cm}}
		\hline
	Datacenter  &Melbourne &Sydney &Canberra &Pune   &Singapore &Seoul &Dubai &Virginia \\\hline\hline
	Melbourne	  &0	&948Mb/s	&990Mb/s	&171Mb/s   &192Mb/s	&164Mb/s	&151Mb/s	&114Mb/s \\
	Sydney	      &189ms	&0	&995Mb/s  &167Mb/s   &206Mb/s	&163Mb/s	&140Mb/s	&121Mb/s\\
    Canberra	  &181ms	&142ms	&0  &173Mb/s   &255Mb/s	&157Mb/s	&138Mb/s	&117Mb/s\\
	Pune          &223ms	&81ms	&64ms	&0	   &489Mb/s   &220Mb/s	&625Mb/s  &119Mb/s\\
	Singapore     &195ms	&35ms	&109ms	&49ms	   &0	&372Mb/s	 &271Mb/s	&111Mb/s\\
	Seoul         &207ms	&172ms	&156ms	&93ms     &153ms	&0	&174Mb/s	&127Mb/s\\
	Dubai    	  &208ms  &168ms	&152ms  &149ms	&147ms   &7ms  &0	&128Mb/s\\
	Virginia      &219ms &165ms &148ms &109ms &142ms &12ms &14ms &0  \\\hline
	  \end{tabular}
	\vspace{-3mm}
\end{table*}

\textbf{Latency and Bandwidth Control:} The latency values in Table \ref{tab:latency-bandwidth-measure} are referenced as \textit{reference values} (RV), and we gradually change these values from 0.2 RV to 1.0 RV with a step of 0.2. Each step of 0.2 RV is called \textit{latencies scaling factor} (LSF) of X, and we thus have LSFs of X, 2X, 3X, 4X, and 5X. Changing these values on-the-fly resembles the mobility of nodes for which as the latency increases the distance between nodes increases and vice versa. All these values are controlled by \textit{tcconfig} in the network layer. Likewise, we exploited \textit{tcconfig} to assign different epochs of bandwidth to the links in  experimental settings.

\begin{figure}[t]
\centering
\includegraphics[width=0.7\columnwidth]{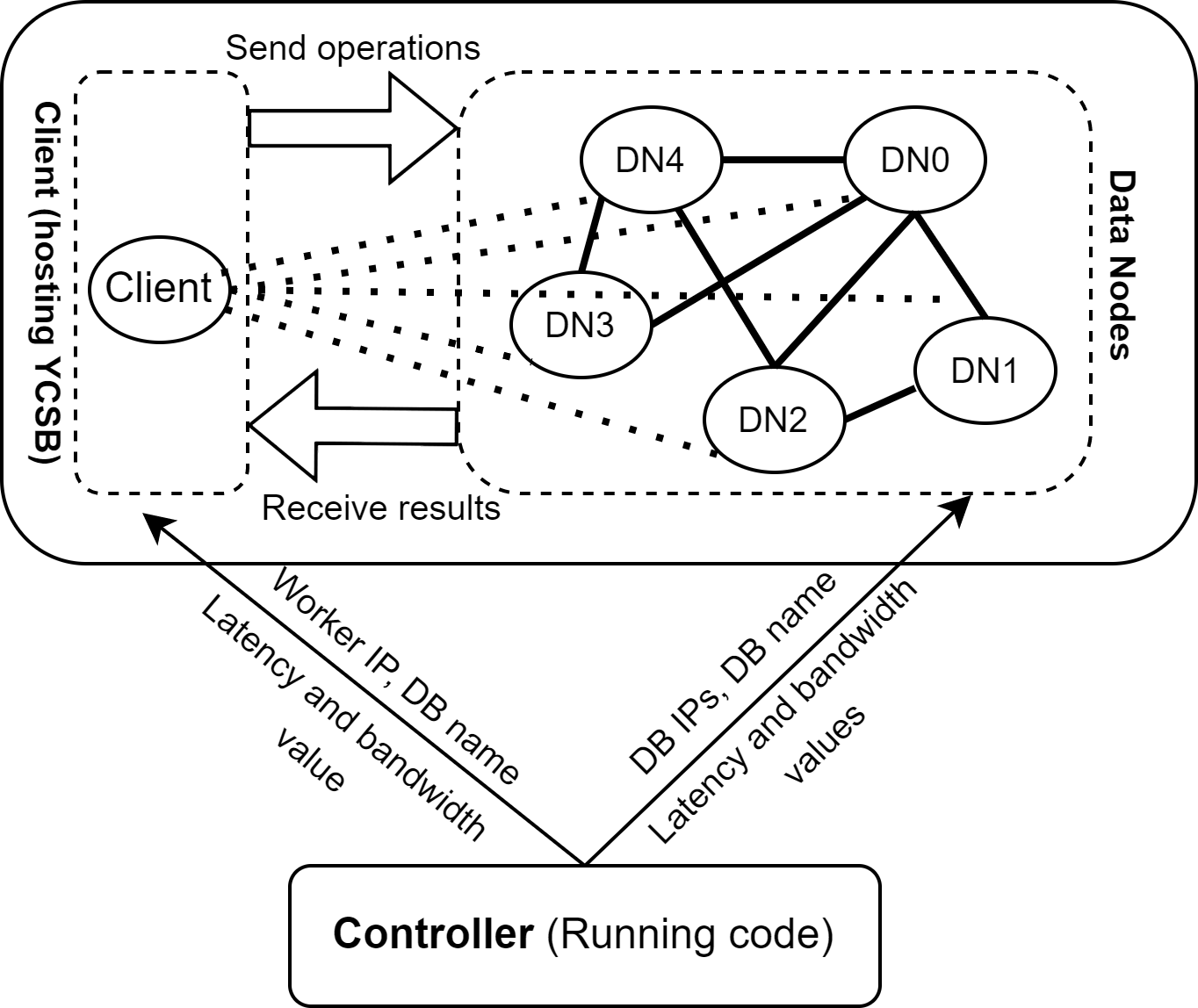}
\vspace{-3mm}
\caption{A schematic of modular components for fragmented hybrid cloud framework}
\label{fig:experiment-framwork}
\vspace{-5mm}
\end{figure}

\textbf{Experimental framework:} We implemented a modular framework (Fig. \ref{fig:experiment-framwork}) to evaluate the performance of databases under different experimental scenarios. The framework helps repetitively and accurately run experiments and reduces interference produced by humans. This framework consists of three components: \textit{controller}, \textit{client} and \textit{data} nodes.
The \textit{controller node} hosts codes to  deploy infrastructure in the private cloud, to install and create a cluster of data nodes for Cassandra ({\myfont{v3.11.13}}), MongoDB ({\myfont{v4.4.15}}), Redis ({\myfont{v6.0.11}}) with Jedis\footnote{Jedis: \url{https://github.com/redis/jedis}} as a Java client ({\myfont{v2.9.0}}), and MySQL ({\myfont{mysql-5.7.22 ndb-7.6,6}}) databases in the \textit{database servers}. 
It also runs codes to apply latency and bandwidth rules to the links between VMs by using \textit{tcconfig} based on the defined epochs.  It also builds a mesh network between nodes by using WireGuard to emulate different connection types between nodes. Lastly, the controller node allows to add or remove a node from the database servers cluster and to emulate network partition by disconnecting links created through WireGuard. 

\textbf{Experimental workloads:} The client node is located in Singapore from which the average distance to all other data center locations is minimum. This means that the client node is in the center of datacenter location. Note that client and data nodes are hosted by different VMs in Singapore. The client node runs YCSB workload ({\myfont{v0.15.0}}) to evaluate databases. YCSB provides  tunable parameter setting and supports 6 pre-defined workloads (Table \ref{tab:workloads}) including  basic operations (read, write, delete, update) and a scan that extracts records for a given key range. We ran YCSB with 10K records, 100K operations (10K operations for 10 times) for workloads A-D and F, and 10K operations (1K operations for 10 times) for workload E. This is a default setting for workloads in our experiments unless indicated otherwise.   

\textbf{Experimental scenarios:}  \textbf{Scenario 1:} We initially measured the impact of time-varying latency between cluster nodes on the throughput of databases (\S\ref{sec:latencyimpact}). We also assessed network traffic distribution between cluster nodes (\S\ref{sec:bandwidthmeasur}). \textbf{Scenario 2:} we measured databases throughput when nodes are added and removed to the cluster on-the-fly (\S\ref{sec:scalingupdown}). \textbf{Scenario 3:} Finally, we compared how much databases are resistant to intermittent networks during running workload (\S\ref{sec:linkremoval}).
\textbf{\begin{table}
	\caption{YCSB Workload Summary}\label{tab:workloads}
	\centering
	\tiny
	\begin{tabular}{p{1.3cm}p{2cm} p{4.5cm}}
		\hline
	Workload 	&Operation         &Description  \\\hline\hline
	A	&Write heavy        &50\% reads, 50\% writes \\
	B	&Read mostly         &95\% reads, 5\% writes \\
	C	&Read only           &100\% reads \\
	D	&Read latest         &Reading newly inserted records \\
	E	&Scan        &Retrieving short ranges of records  \\
	F	&RMW   &Reading, modifying and writing back records   \\\hline
	\end{tabular}
	\vspace{-5mm}
\end{table}}
\subsection{Experiment Results}\label{sec:results}
\subsubsection{The impact of latency on throughput}\label{sec:latencyimpact}
We start by evaluating the impact of latency on the throughput of databases. To compare the throughput of databases, we define \textit{normalized throughput} as $\frac{\mbox{\footnotesize Throughput in LSF}  }{\mbox{\footnotesize Throughput in 5X}} $ , 
where LSF  is X, 2X, 3X, 4X, and 5X. This relation implies that the more normalized throughput is, the better database tolerates latency. 

\begin{figure*}[ht!]
  \centering
  \subfloat[Workload A]{\label{fig:latency-a}\includegraphics[height=3cm,width=0.33\textwidth]{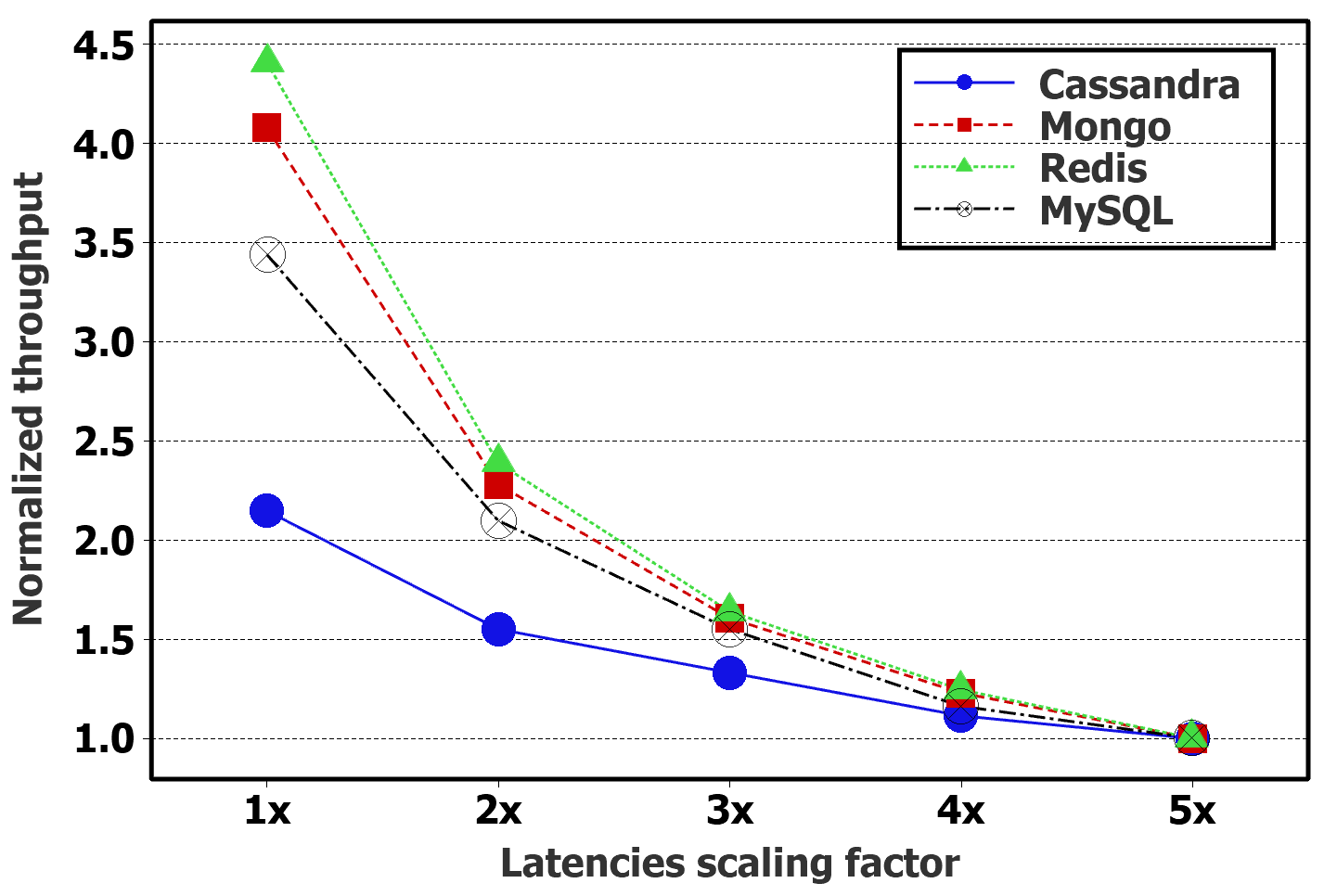}}
  \subfloat[Workload B]{\label{fig:latency-b}\includegraphics[height=3cm,width=0.33\textwidth]{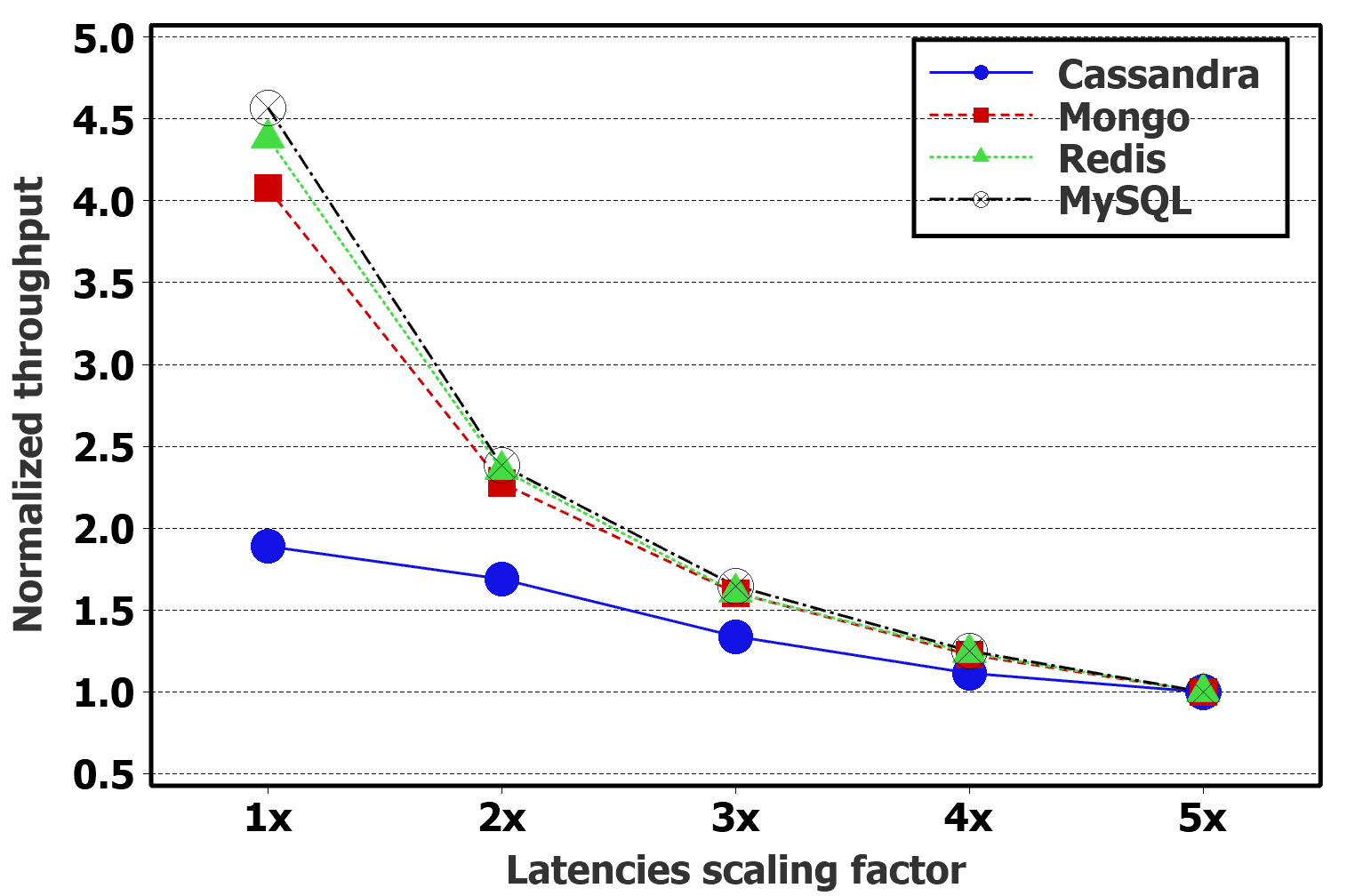}}
  \subfloat[Workload C]{\label{fig:latency-c}\includegraphics[height=3cm,width=0.33\textwidth]{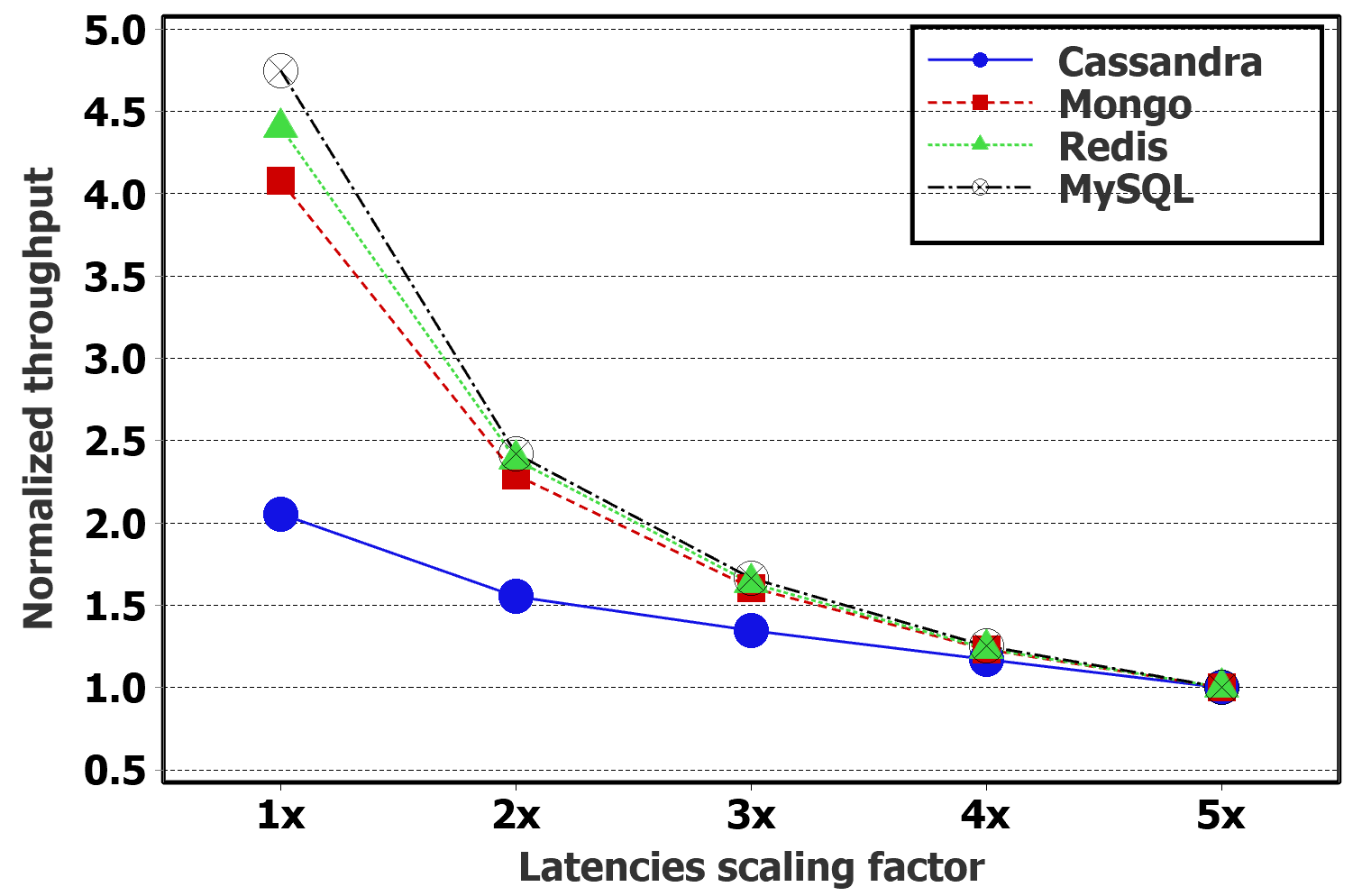}}\\
  \subfloat[Workload D]{\label{fig:latency-d}\includegraphics[height=3cm,width=0.33\textwidth]{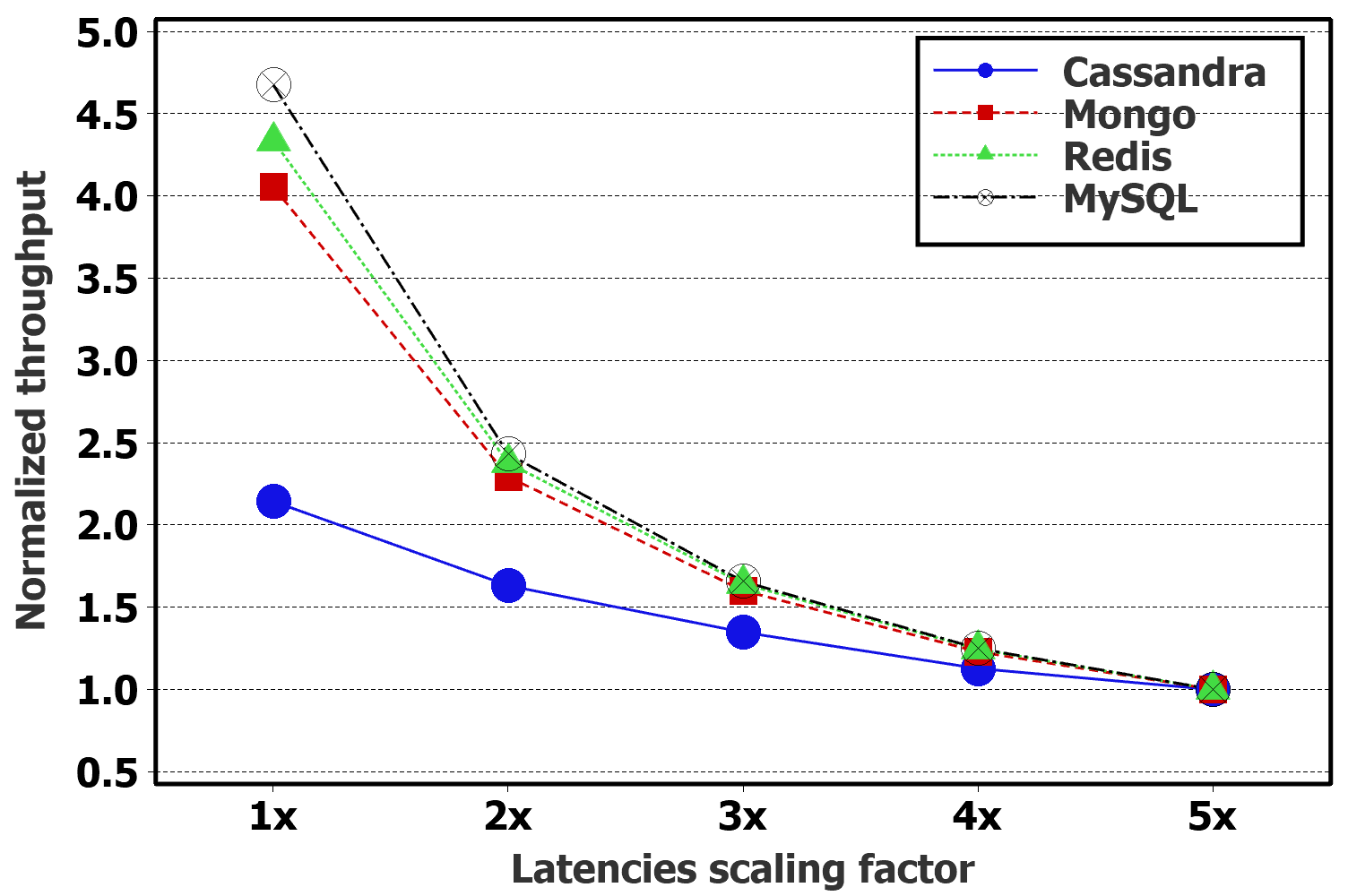}}
  \subfloat[Workload E]{\label{fig:latency-e}\includegraphics[height=3cm,width=0.33\textwidth]{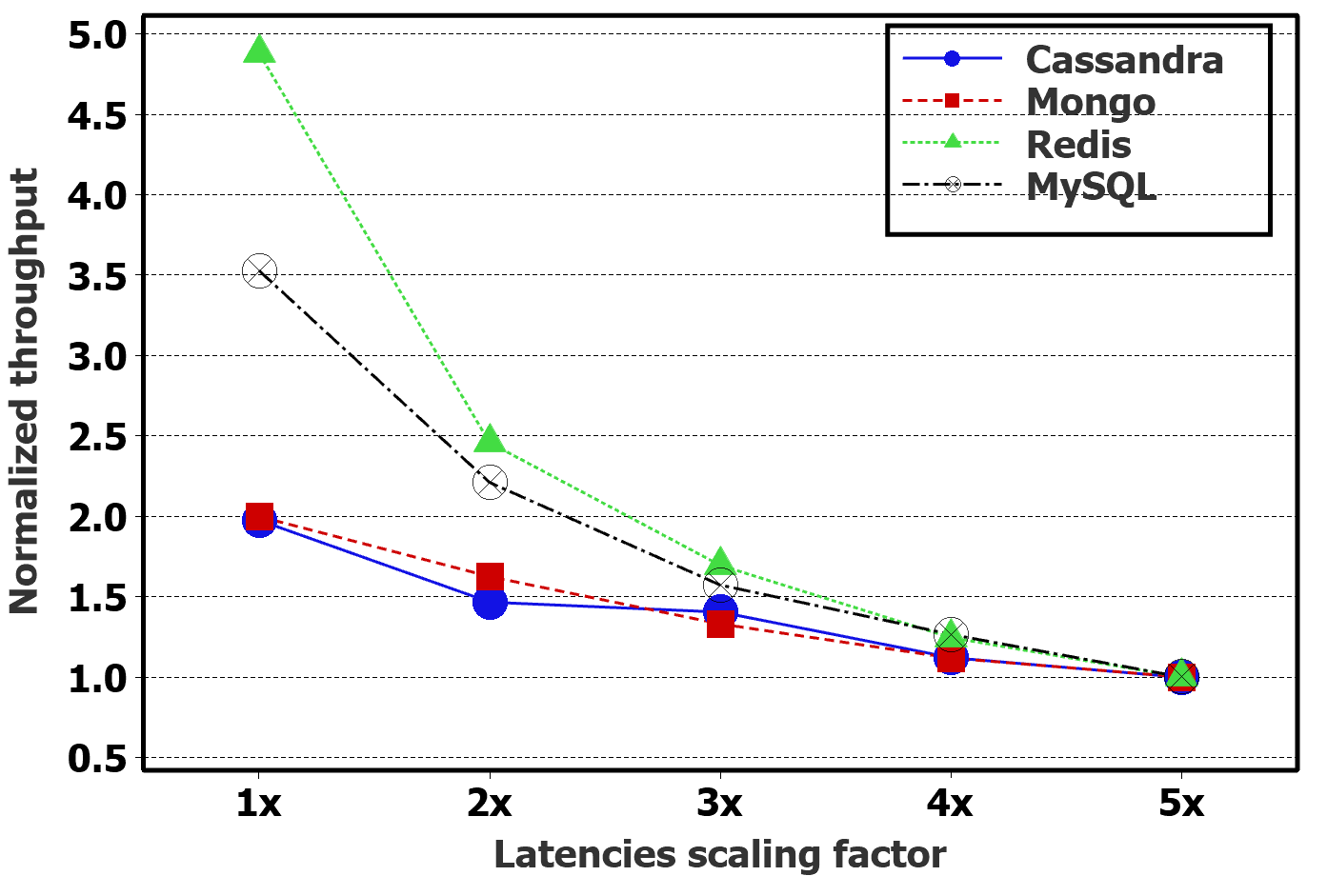}}
  \subfloat[Workload F]{\label{fig:latency-f}\includegraphics[height=3cm,width=0.33\textwidth]{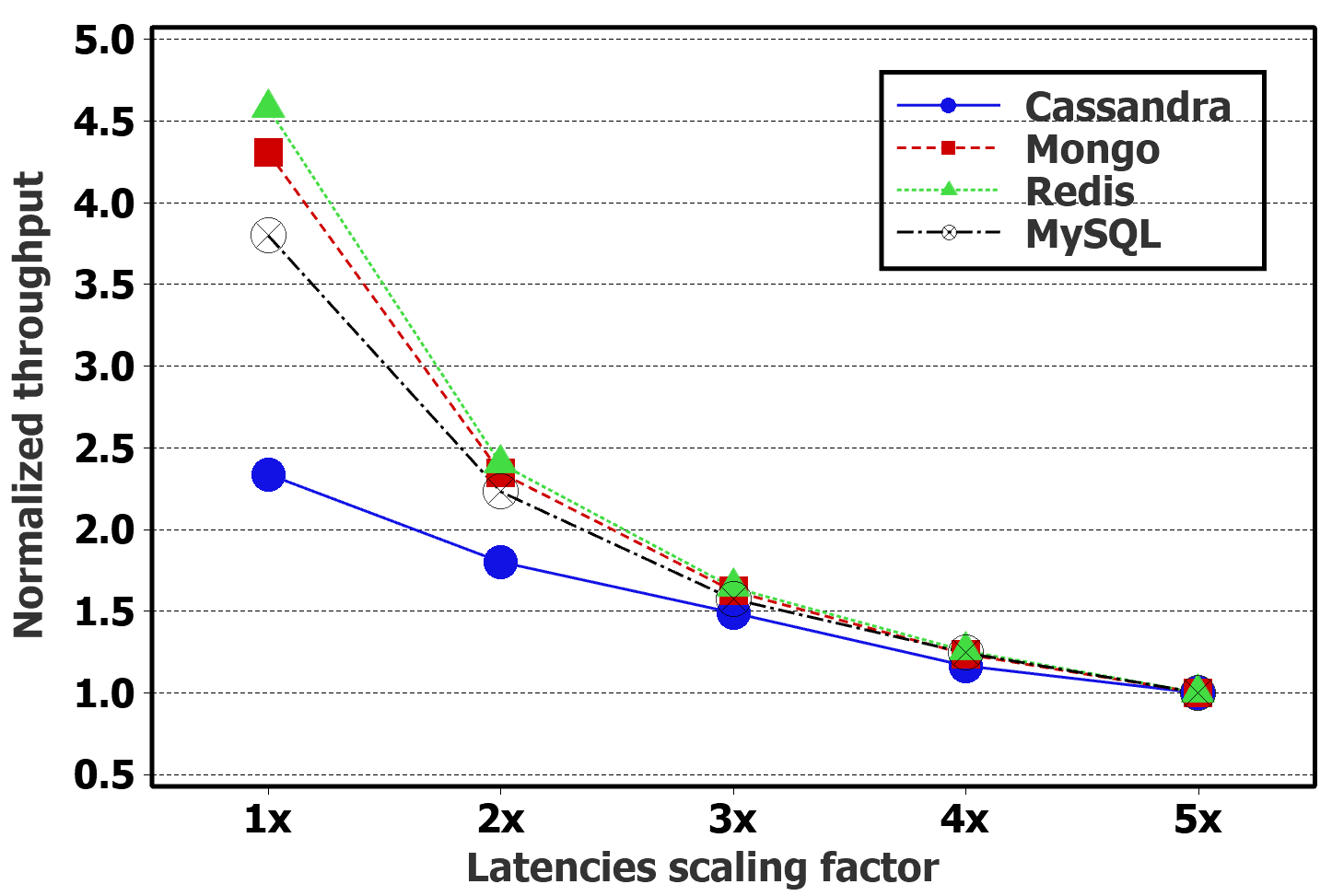}}
  \caption{The impact of LSF varying from X to 5X on the normalized throughput of all databases for workloads A-F.}
\label{fig:latency-impact}
\vspace{-7mm}
\end{figure*}
Fig. \ref{fig:latency-impact} depicts the impact of  latency on the throughput of Cassandra, MongoDB, Redis and MySQL. Cassandra is the most tolerable database against the latency scaling factor in all workloads. For workload A as shown in Fig. \ref{fig:latency-a}, the normalized throughput of Cassandra decreases by 1.38 times (138\%) as LSF increases by two times. MySQL obtains the second rank so that the normalized throughput decreases by 1.6 times as LSF raises from X to 2X. 
By contrast, Redis receives the most impact such that the normalized throughput reduces by 1.84 times for an increment in LSF from X to 2X. MongoDB comes after the Redis as LSF varies between 1X and 3X. For workloads B, C and D (Figs. \ref{fig:latency-b}, \ref{fig:latency-c}, and \ref{fig:latency-d}), network latency dominates the type of workload since these workloads almost have the same values in the normalized throughput.  
For these workloads, Cassandra receives a 50\% reduction in the normalized throughput while other databases face at least four times reduction as LSF changes from 1X to 5X. This is because Cassandra distributes data across all nodes and requests can be submitted to any node. For workload E (Fig. \ref{fig:latency-e}), LSF has the most impact on the normalized throughput of Redis by 4.8 times, while MongoDB and Cassandra perform the best. MySQL holds the second rank for this situation; at most 3.5 times reduction in the normalized throughput for 1X$\leq$LSF$\leq$2X. For workload F with LSF=$1X$, all databases exposed higher normalized throughput as compared to other workloads except workload E (Fig. \ref{fig:latency-f}). This means that workload F has been impacted less than workloads A-D by latency though databases hold the same rank as observed for workload A. This is because workload F is a kind of workload A though workload F includes data modification as well.  
\begin{figure}[t!]
  \centering
  \subfloat[1X]{\label{fig:throughput-1x}\includegraphics[height=3cm,width=0.5\columnwidth]{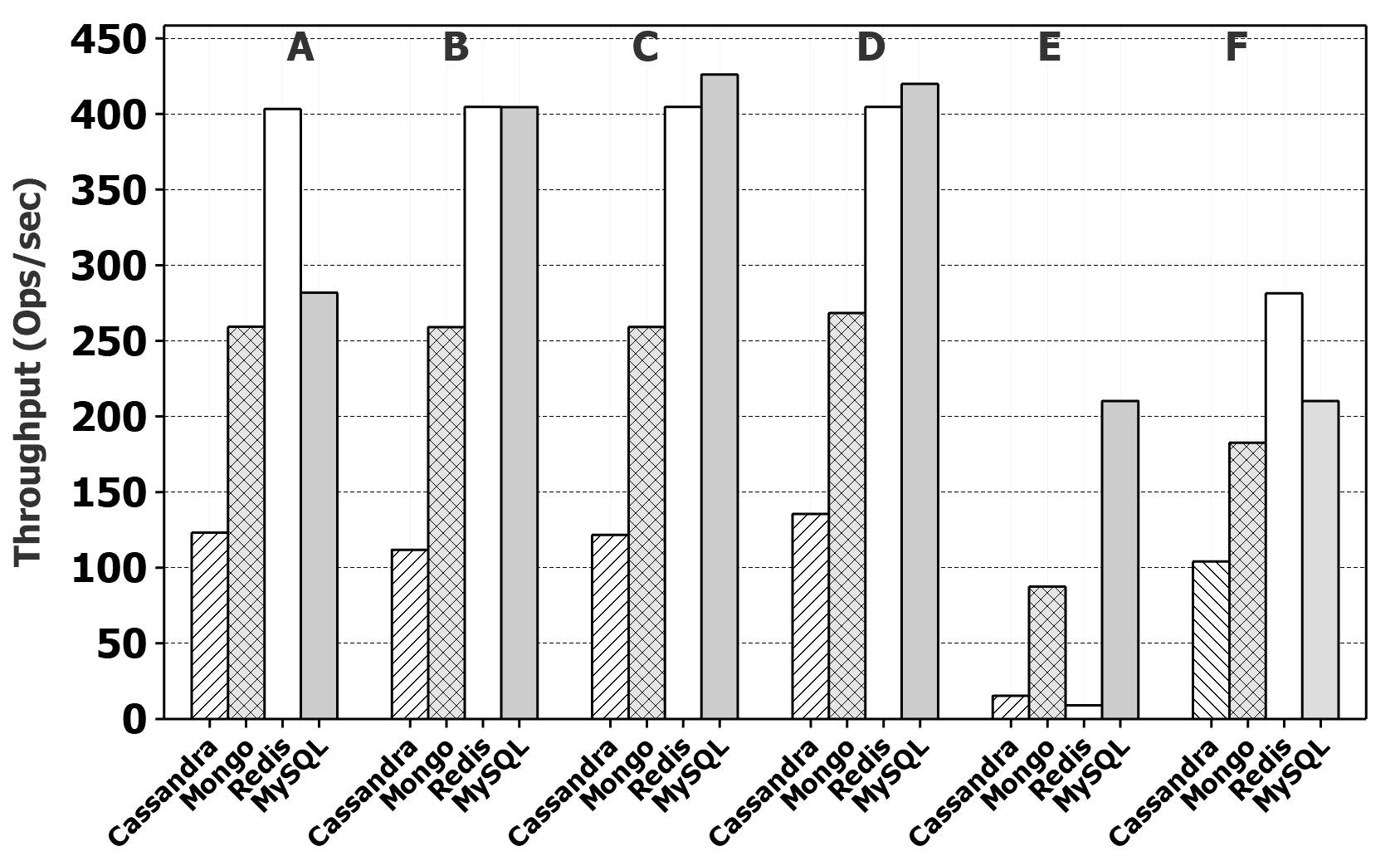}}
  \subfloat[5X]{\label{fig:throughput-5x}\includegraphics[height=3cm,width=0.5\columnwidth]{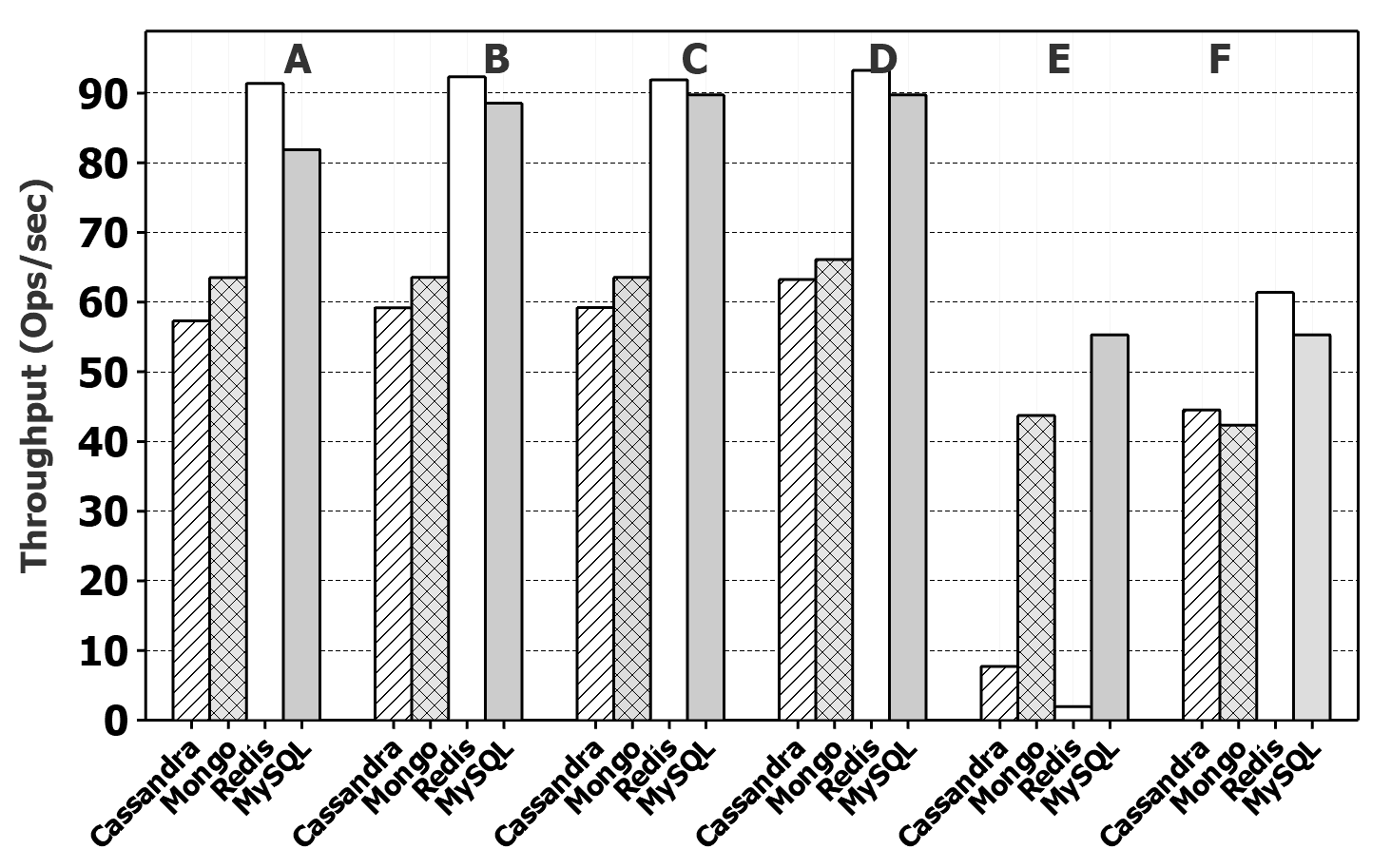}}
 \caption{Throughput of databases for LSF=1X and 5X. } 
\label{fig:throughput-1x5x}
\vspace{-6mm}
\end{figure}

Fig. \ref{fig:throughput-1x5x} presents the absolute values of databases throughput for LSF=1X,5X. As the results show, Redis outperforms all databases in throughput for workloads A and F,  while for workloads B, C, and D Redis and MySQL are competitive. MongoDB holds the third rank among all databases for all workloads, while Cassandra achieves the lowest throughput so that its performance approaches MongoDB throughput as LSF changes from 1X to 5X. These findings validate that the highest tolerance of Cassandra to LSF. All the evaluated databases provide the lowest throughput for workloads E and F with LSF=1X,5X. In contrast, the databases expose the highest throughput for workloads A-D.  

\subsubsection{The amount and distribution of transferred data across cluster nodes}\label{sec:bandwidthmeasur}

\begin{figure*}[ht!]
  \centering
  \subfloat[Workload A]{\label{fig:bandwidth-a}\includegraphics[height=3cm,width=0.33\textwidth]{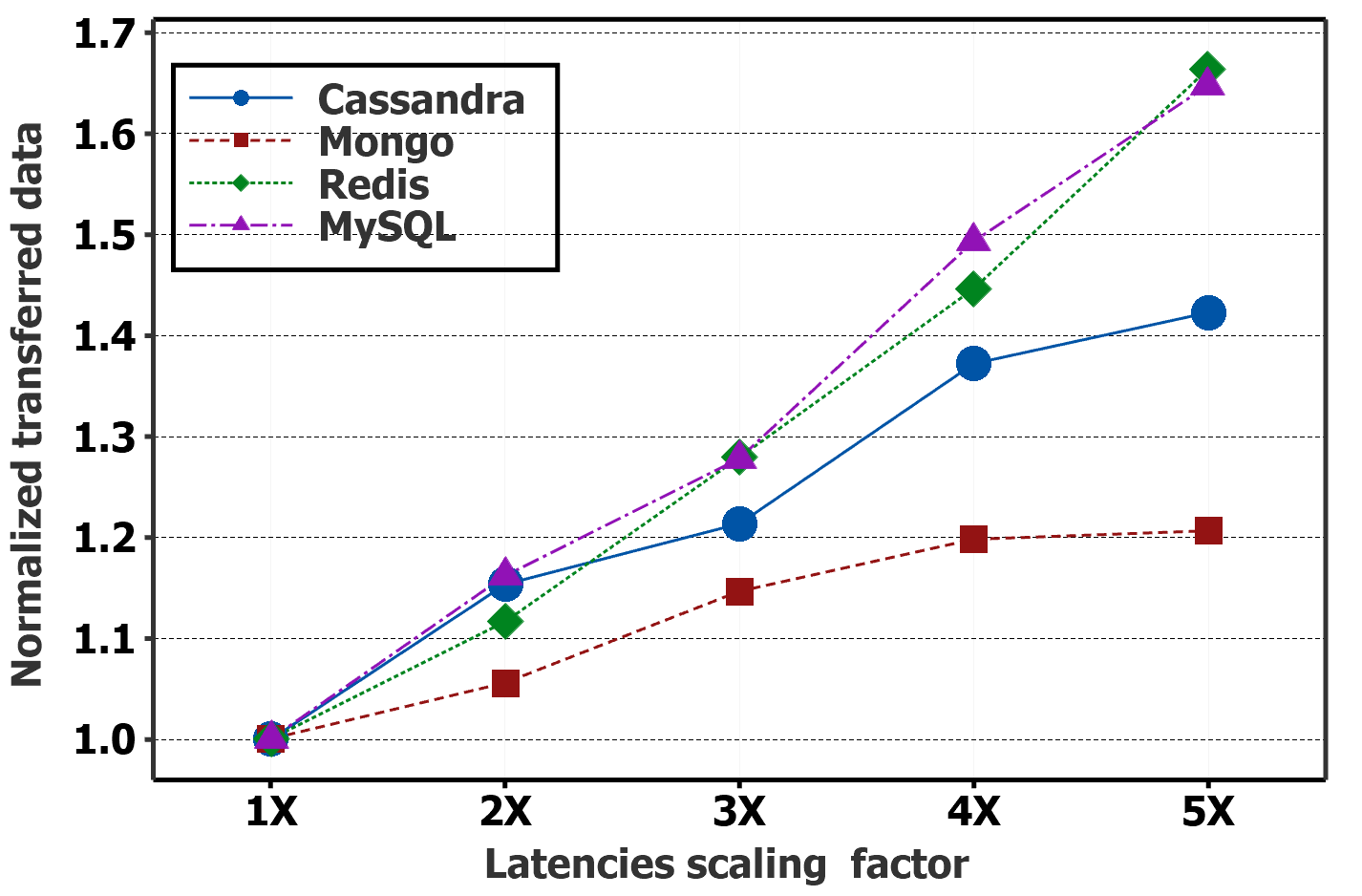}}
  \subfloat[Workload B]{\label{fig:bandwidth-b}\includegraphics[height=3cm,width=0.33\textwidth]{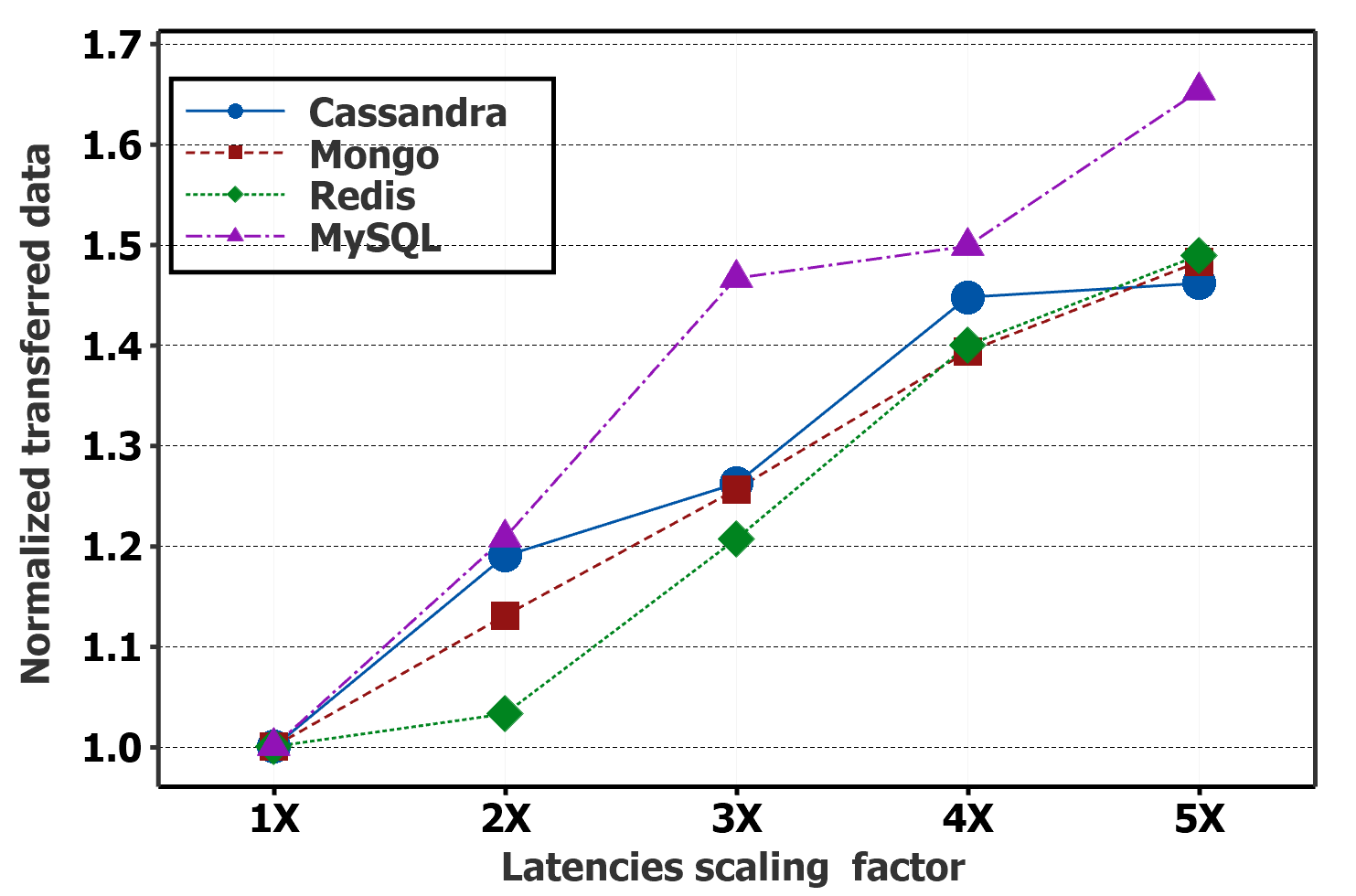}}
  \subfloat[Workload C]{\label{fig:bandwidth-c}\includegraphics[height=3cm,width=0.33\textwidth]{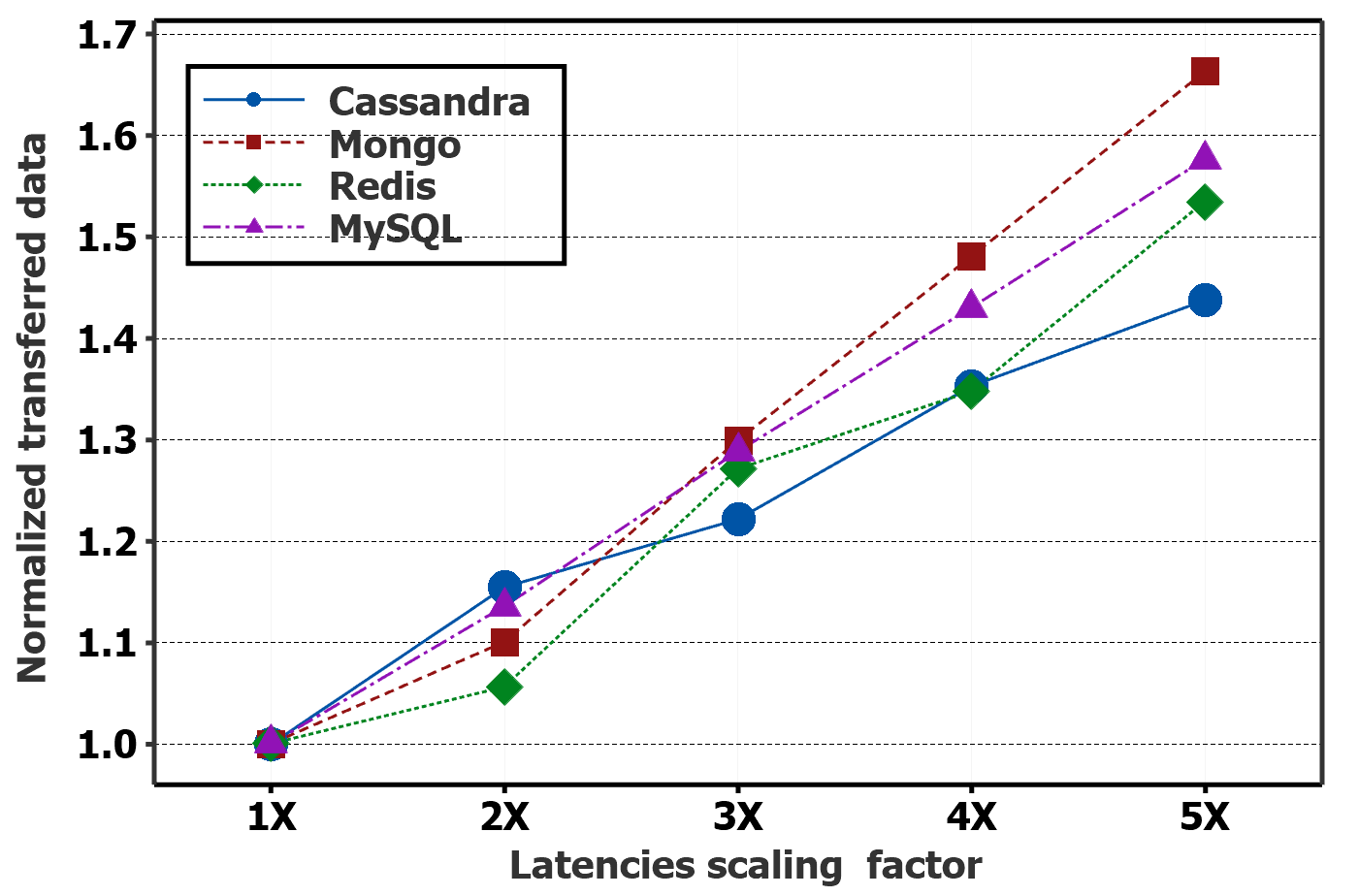}}\\
  \subfloat[Workload D]{\label{fig:bandwidth-d}\includegraphics[height=3cm,width=0.33\textwidth]{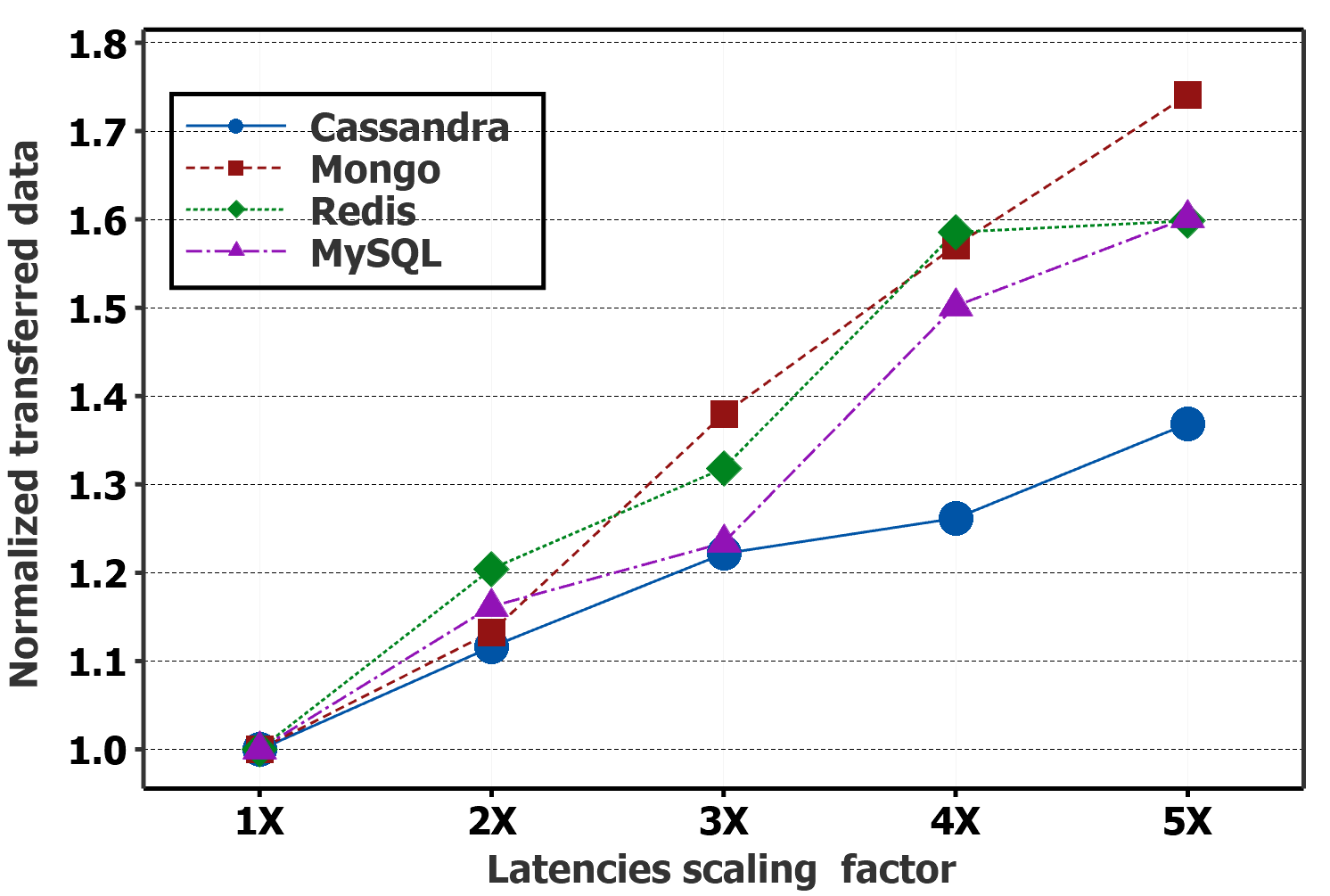}}
  \subfloat[Workload E]{\label{fig:bandwidth-e}\includegraphics[height=3cm,width=0.33\textwidth]{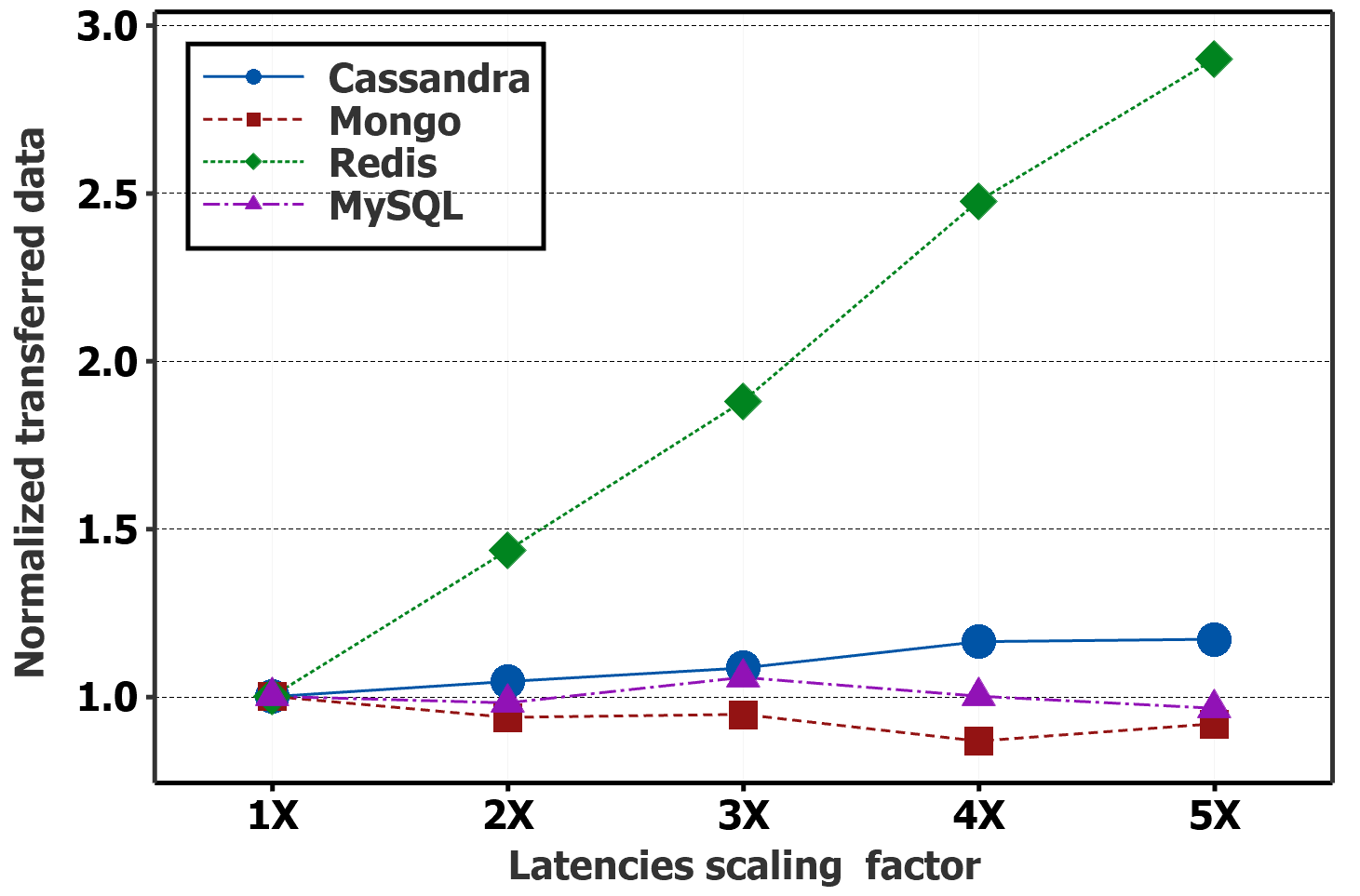}}
  \subfloat[Workload F]{\label{fig:bandwidth-f}\includegraphics[height=3cm,width=0.33\textwidth]{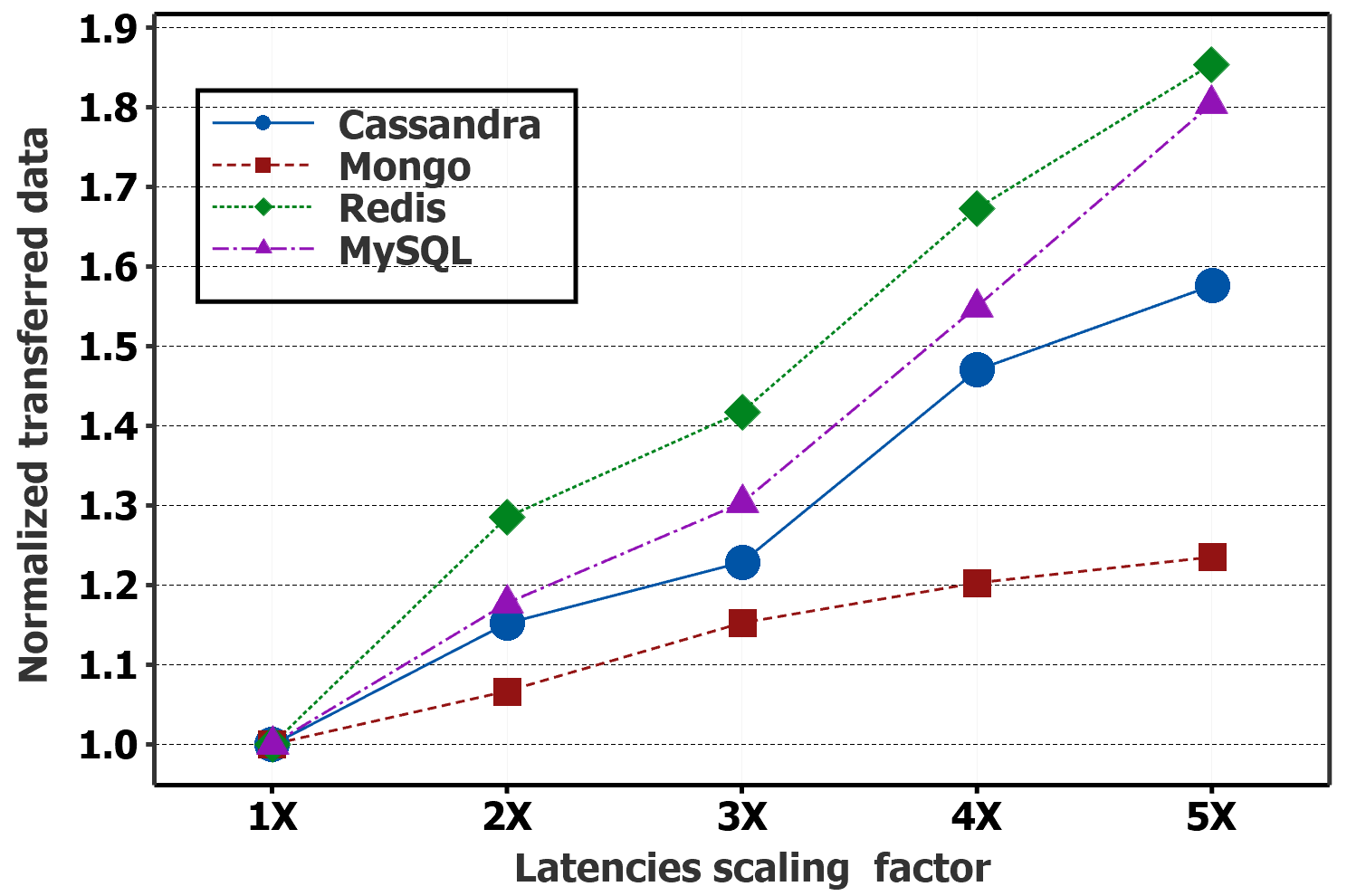}}
  \caption{The impact of LSF varying from X to 5X on the normalized transferred data for all databases for workloads A-F.}
\label{fig:bandwidth-impact}
\vspace{-5mm}
\end{figure*}

Fig. \ref{fig:bandwidth-impact} illustrates the impact of latencies scaling factors changing from 1X to 5X  on the normalized transferred data for databases as workloads A-E are served. \textit{Normalized transferred data} is defined as 
$\frac{\mbox{\footnotesize The amount of transferred data in LSF}  }{\mbox{\footnotesize The amount transferred data in  5X}} $,
where LSF is X, 2X, 3X, 4X, and 5X.

As shown in Fig. \ref{fig:bandwidth-a}, for workload A, MongoDB has the lowest growth rate in the normalized  transferred data when LSF changes from 1X to 5X (20\% at most).  
 Cassandra holds the second rank in the usage of bandwidth as it experienced an increment in the normalized transferred data by 40\% for LSF raises from  1X to  5X. MySQL and Redis performed the worst and exposed up to a 65\% rise in the normalized transferred data. For workload B (Fig. \ref{fig:bandwidth-b}), apart from Redis, all databases preserve the same position in the growth of normalized transferred data as observed for workload A. 
When the databases are running for a longer time (LSF=5X), Redis, MongoDB, and Cassandra converge almost to the same value of normalized transferred data. This result demonstrates that the amount of data to be read is dominated  by the amount of data to be exchanged between cluster nodes for communication. For workload C (Fig. \ref{fig:bandwidth-c}), we observed that Cassandra flips MySQL in the normalized transferred data for a short running (1X$\leq$LSF$\leq$2X) compared to workload B; Whilst for 2X$\leq$LSF$\leq$5X, all databases are competitive in normalized transferred data. For workload D (Fig. \ref{fig:bandwidth-d}), Cassandra and MongoDB have the lowest and highest growth rate in the normalized transferred data. 
For workload E (Fig. \ref{fig:bandwidth-e}), Redis has the highest growth in normalized transferred data, up to 3 times when LSF varies from 1X to 5X. This implies that why Redis is much worse in throughput for workload E. Cassandra, however,  requires about a 25\% increment in normalized transferred data when LSF changes from 1X to 5X. Unlike Redis and Cassandra, MongoDB did not incur an increment in the transferred data since it exploits a full replication strategy. 
For workload F (Fig. \ref{fig:bandwidth-f}), MongoDB, Cassandra, MySQL and Redis have a data transferred growth of 20\%, 60\%, 80\% and 85\% respectively for LSF varying from 1X and 5X.

\begin{figure}[t!]
  \centering
  \subfloat[1X]{\label{fig:tran-data-1x}\includegraphics[height=3cm,width=0.5\columnwidth]{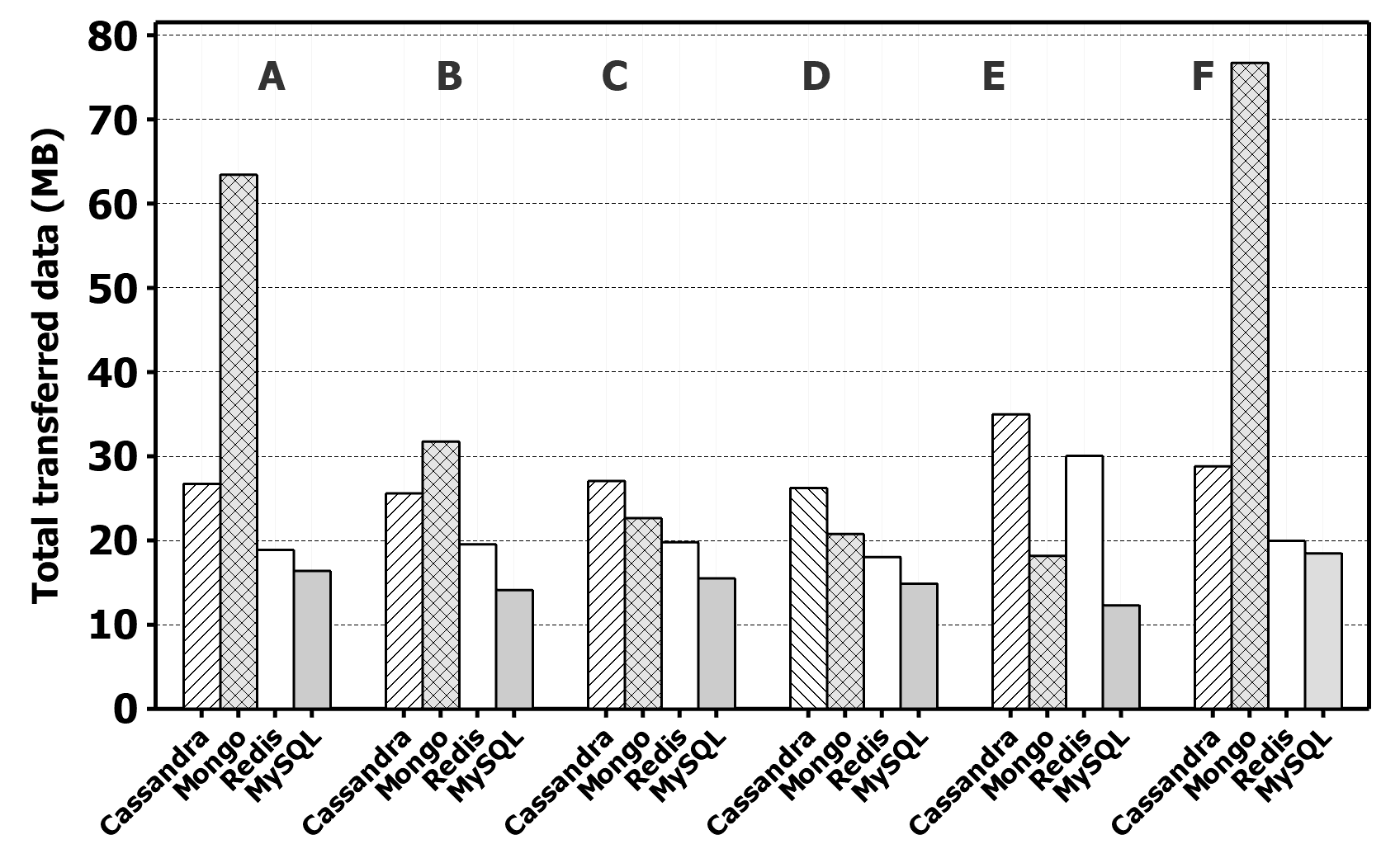}}
  \subfloat[5X]{\label{fig:tran-data-5x}\includegraphics[height=3cm,width=0.5\columnwidth]{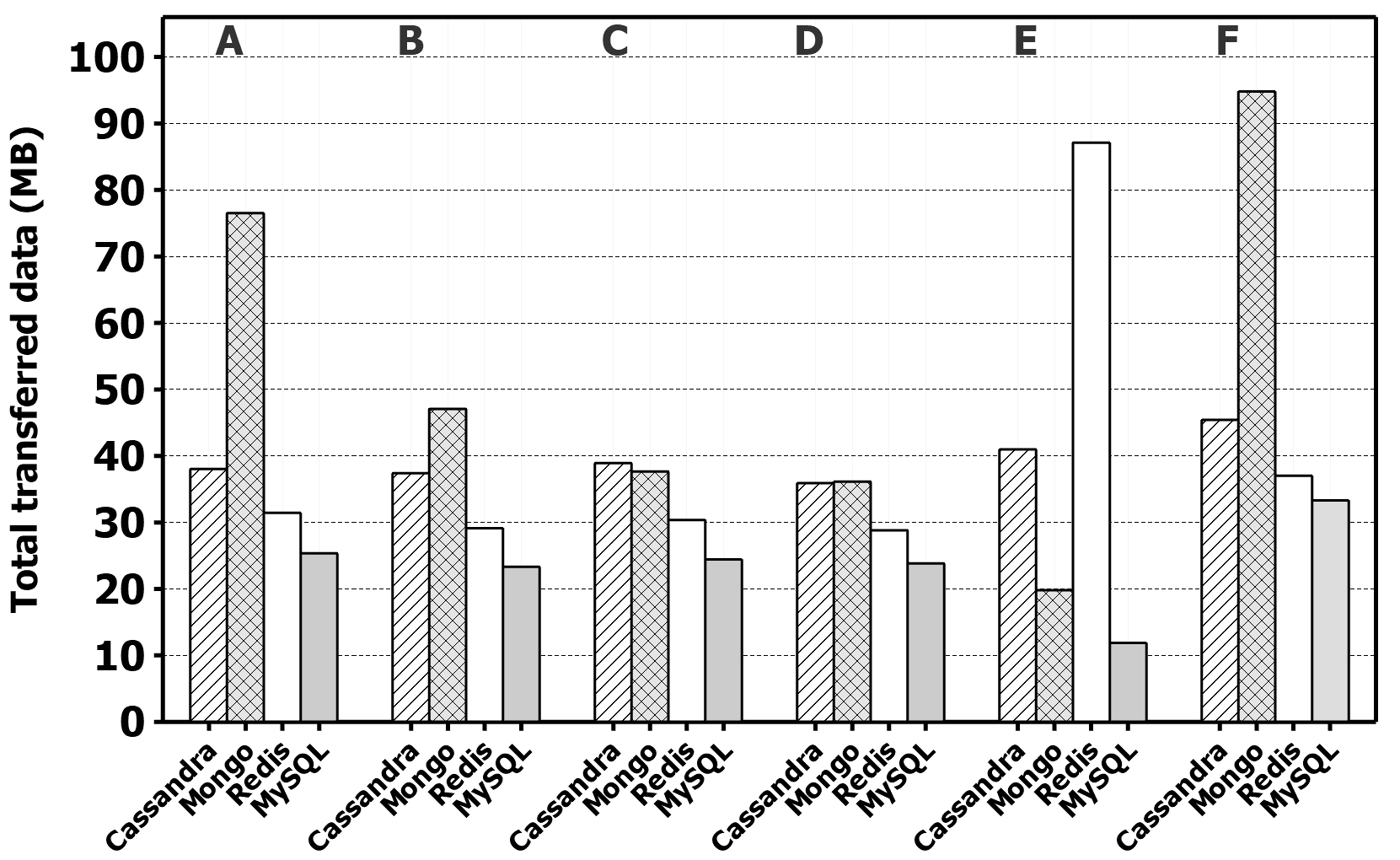}}
 \caption{The total data transfer (MB) for databases deployed on the cluster nodes with LSF=1X and 5X.} 
\label{fig:trans-data-1x5x}
\vspace{-8mm}
\end{figure}

Fig. \ref{fig:trans-data-1x5x} illustrates the total transferred data in MB to serve workload A-F for all databases where LSF=1X,5X. For workloads A-F, MySQL transferred the minimum  amount of data between cluster nodes: 12-13 MB for 1X and 22-36 MB for 5X. This is followed by Redis for all workloads except workload E for which it requires the highest amount of transferred data across cluster nodes. Cassandra and MongoDB come after Redis, which depends on the workload type. For workload A, B, and F, MongoDB transferred data two times more as compared to Cassandra, whilst for workload C and D, Cassandra transferred more data. 

\begin{figure*}[ht!]
  \centering
  \subfloat[Cassandra]{\label{fig:cassand-a-band-dis}\includegraphics[height=5cm,width=0.5\textwidth]{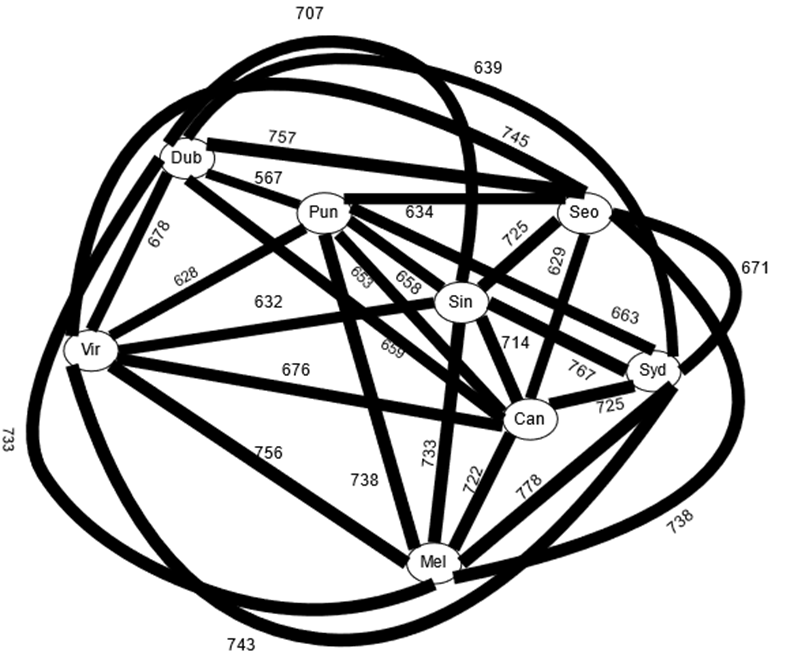}}
  \subfloat[MongoDB]{\label{fig:mongo-a-band-dis}\includegraphics[height=5cm,width=0.5\textwidth]{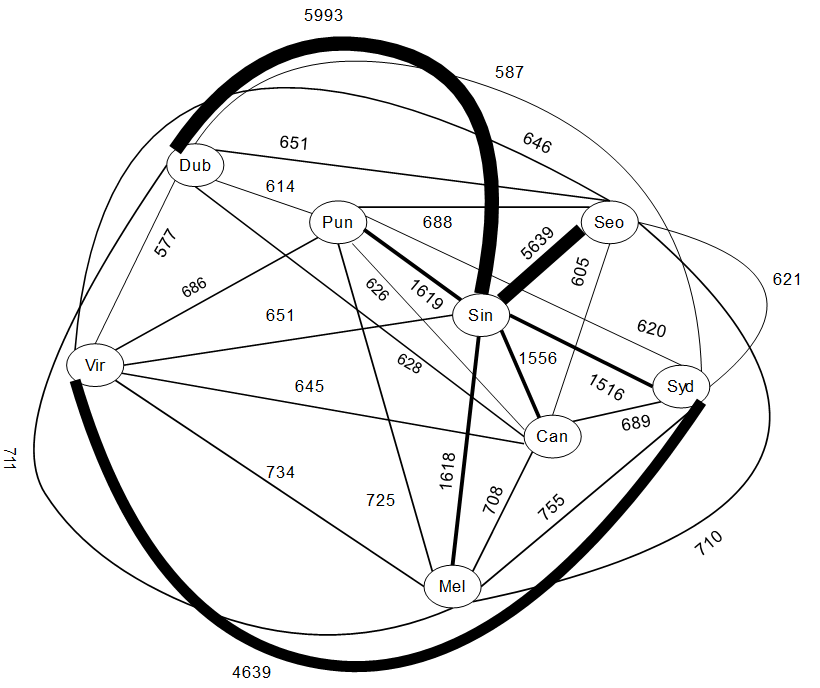}}\\
  \subfloat[Redis]{\label{fig:redis-a-band-dis}\includegraphics[height=5cm,width=0.5\textwidth]{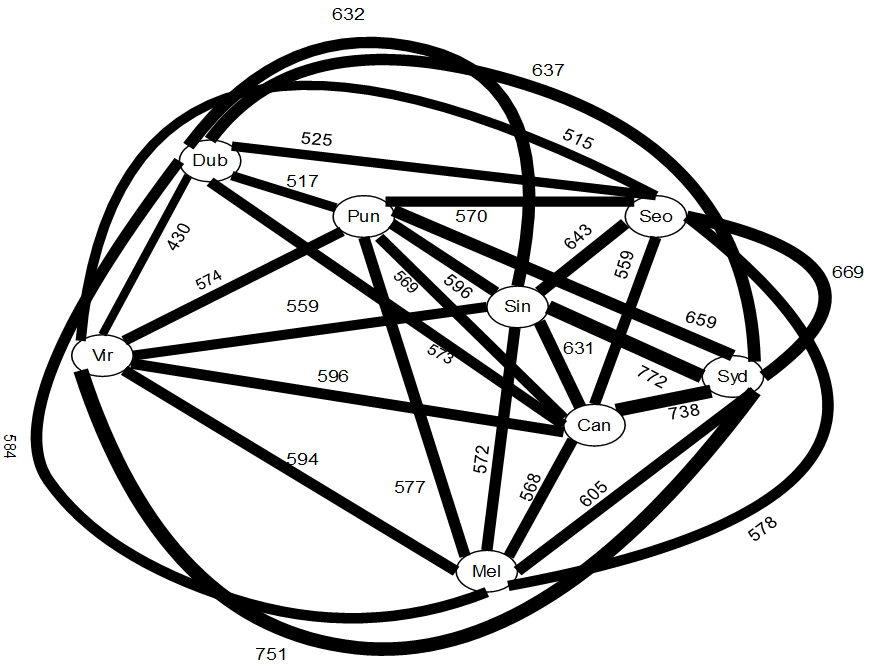}}
  \subfloat[MySQL]{\label{fig:mysql-a-band-dis}\includegraphics[height=5cm,width=0.5\textwidth]{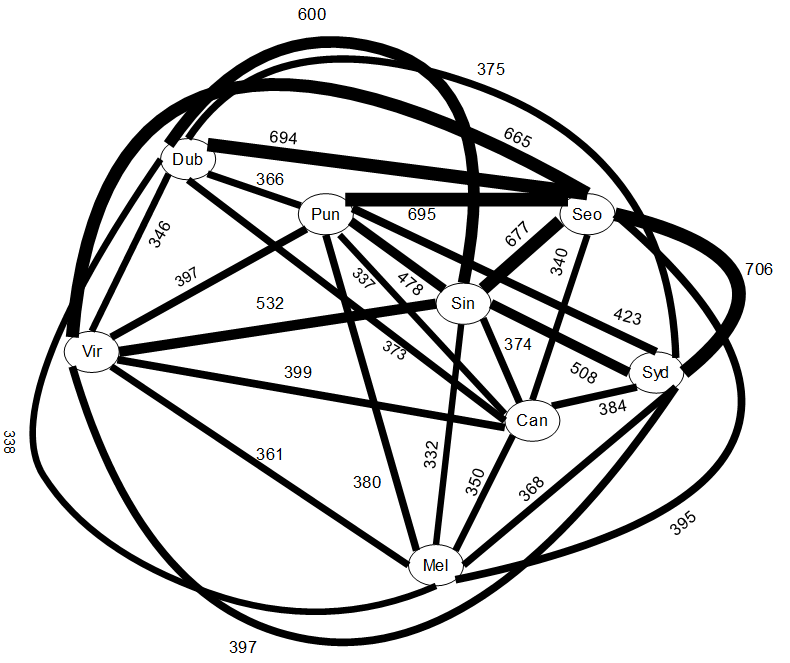}}
  \caption{Network traffic distribution across cluster nodes with latencies scaling factor of 5X for \textbf{workload A}}
\label{fig:band-dis-a}
\vspace{-5mm}
\end{figure*}

\begin{figure*}[ht!]
  \centering
  \subfloat[Cassandra]{\label{fig:cassand-b-band-dis}\includegraphics[height=5cm,width=0.5\textwidth]{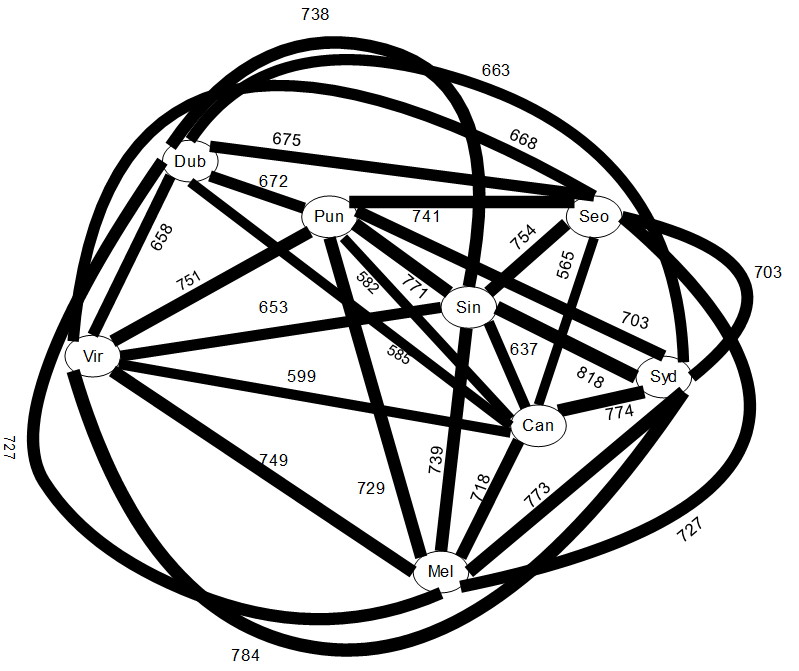}}
  \subfloat[Mongo]{\label{fig:mongo-b-band-dis}\includegraphics[height=5cm,width=0.5\textwidth]{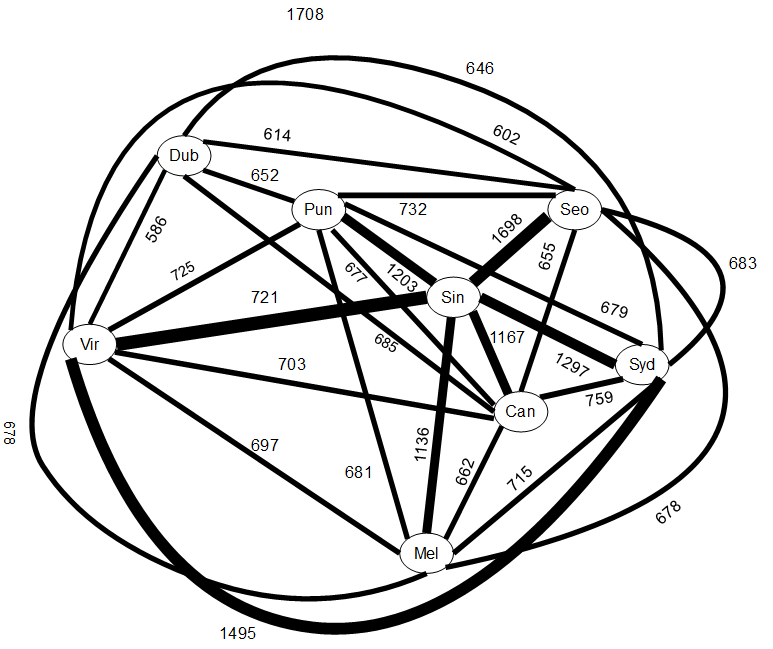}}\\
  \subfloat[Redis]{\label{fig:redis-b-band-dis}\includegraphics[height=5cm,width=0.5\textwidth]{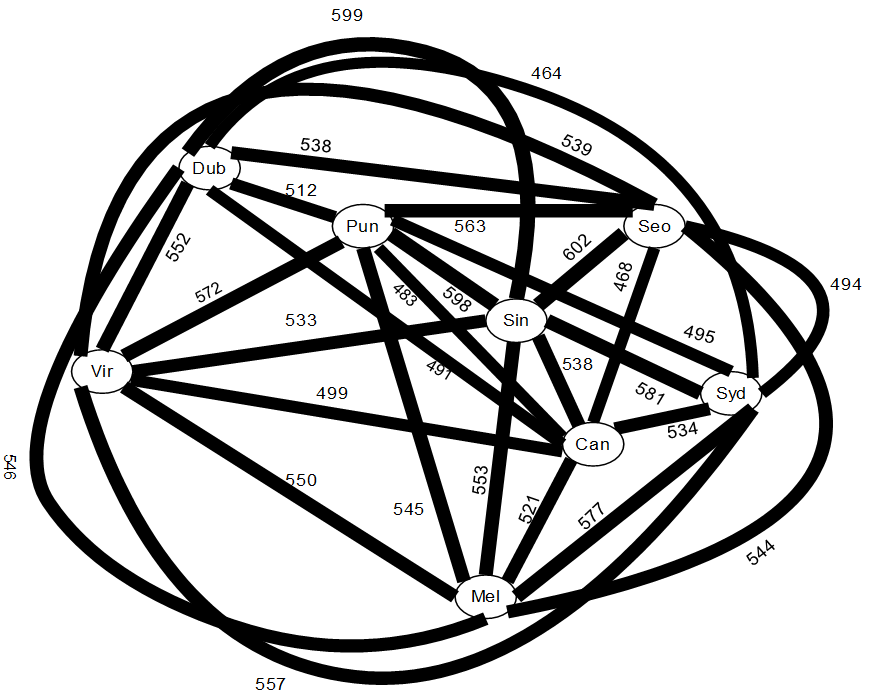}}
  \subfloat[MySQL]{\label{fig:mysql-b-band-dis}\includegraphics[height=5cm,width=0.5\textwidth]{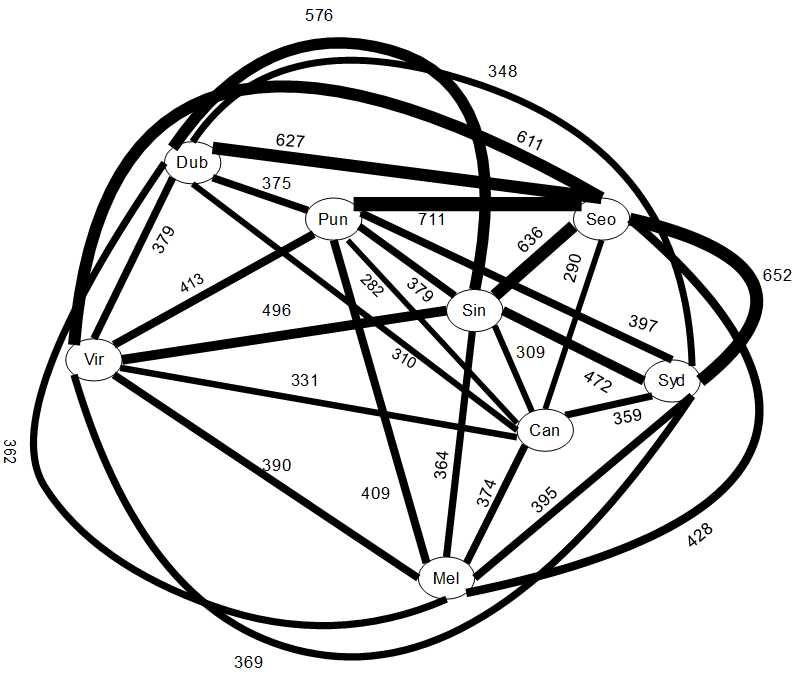}}
  \caption{Network traffic distribution across cluster nodes with latencies scaling factor of 5X for \textbf{workload B}. }
\label{fig:band-dis-b}
\vspace{-5mm}
\end{figure*}

\begin{figure*}[ht!]
  \centering
  \subfloat[Cassandra]{\label{fig:cassand-e-band-dis}\includegraphics[height=5cm,width=0.5\textwidth]{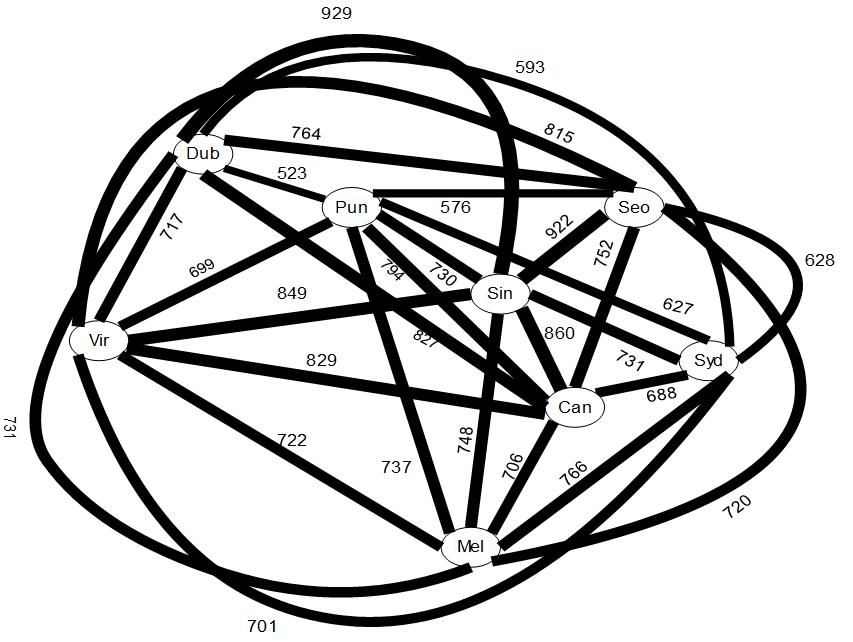}}
  \subfloat[Mongo]{\label{fig:mongo-e-band-dis}\includegraphics[height=5cm,width=0.5\textwidth]{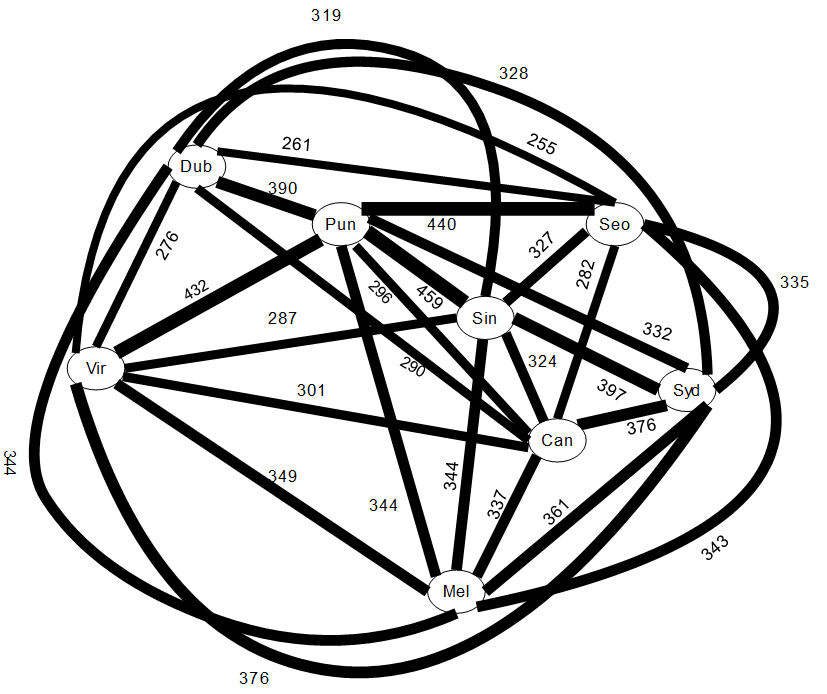}}\\
  \subfloat[Redis]{\label{fig:redis-e-band-dis}\includegraphics[height=5cm,width=0.5\textwidth]{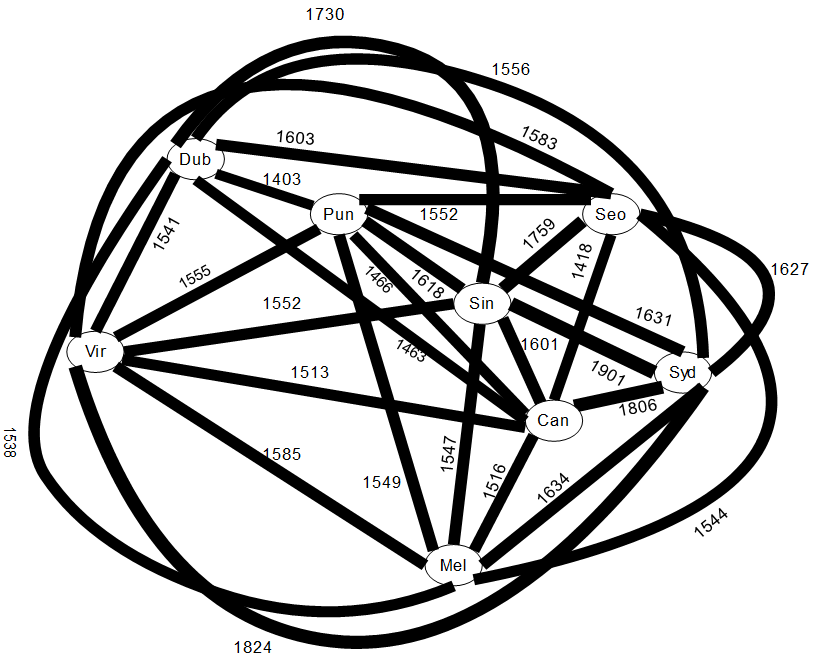}}
  \subfloat[MySQL]{\label{fig:mysql-e-band-dis}\includegraphics[height=5cm,width=0.5\textwidth]{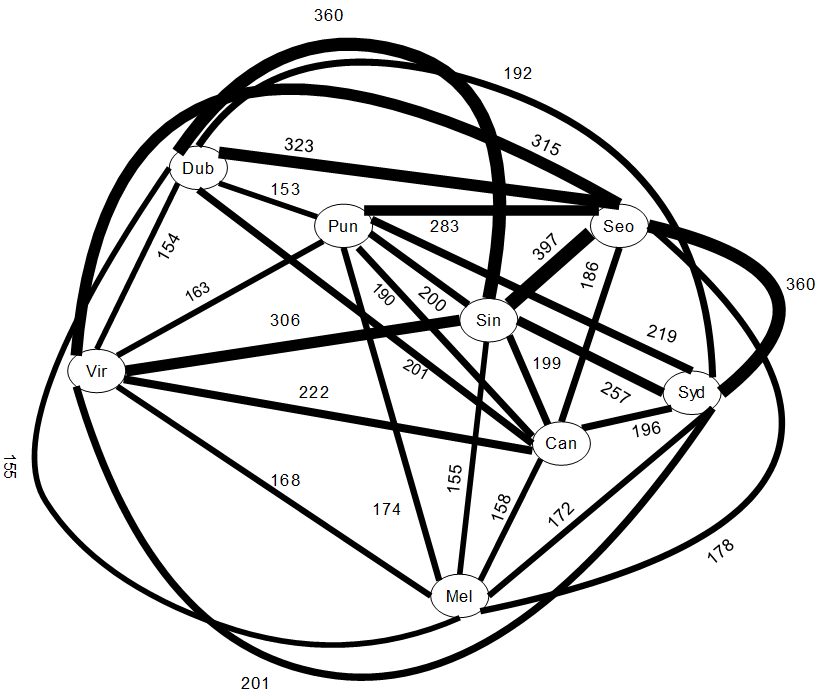}}
  \caption{Network traffic distribution across cluster nodes with latencies scaling factor of 5X for \textbf{workload E}.}
\label{fig:band-dis-e}
\vspace{-5mm}
\end{figure*}

\begin{figure*}[ht!]
  \centering
  \subfloat[Cassandra]{\label{fig:cassand-f-band-dis}\includegraphics[height=5cm,width=0.5\textwidth]{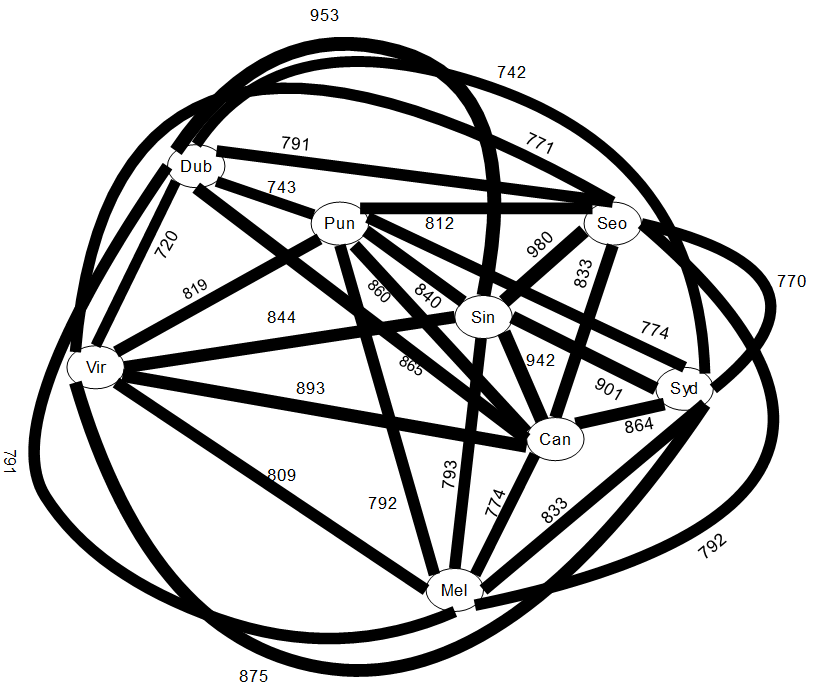}}
  \subfloat[Mongo]{\label{fig:mongo-f-band-dis}\includegraphics[height=5cm,width=0.5\textwidth]{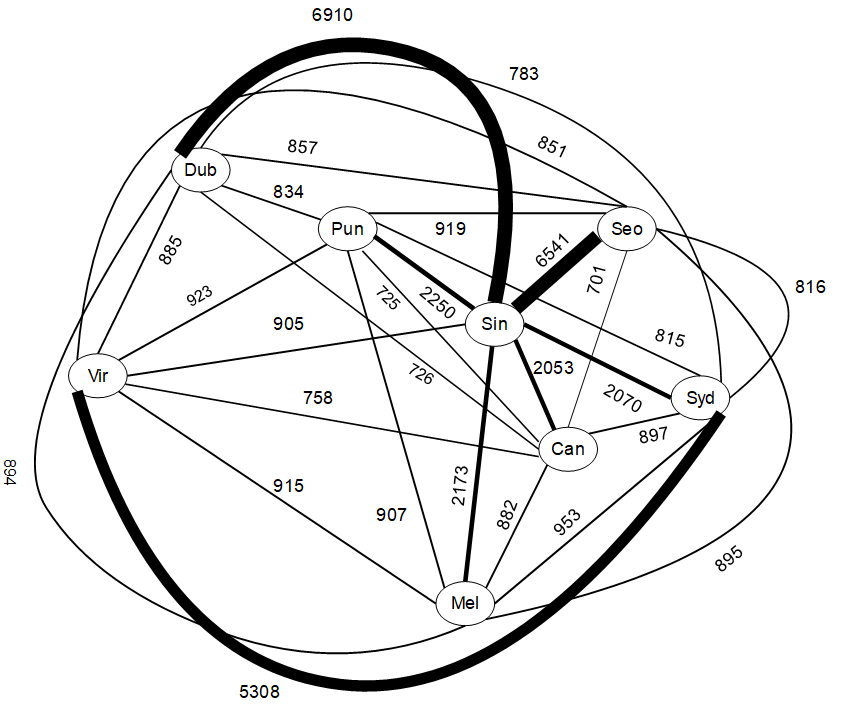}}\\
  \subfloat[Redis]{\label{fig:redis-f-band-dis}\includegraphics[height=5cm,width=0.5\textwidth]{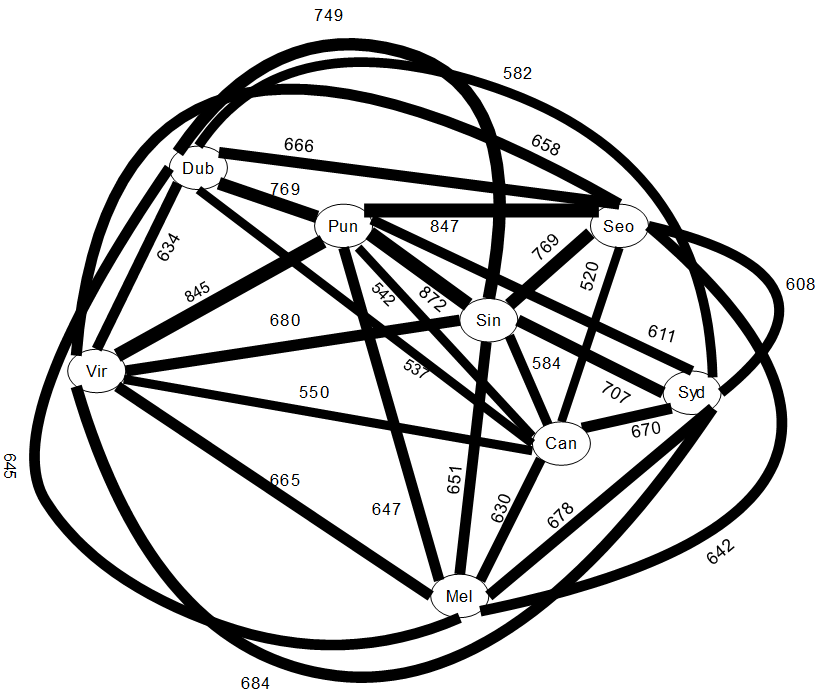}}
  \subfloat[MySQL]{\label{fig:mysql-f-band-dis}\includegraphics[height=5cm,width=0.5\textwidth]{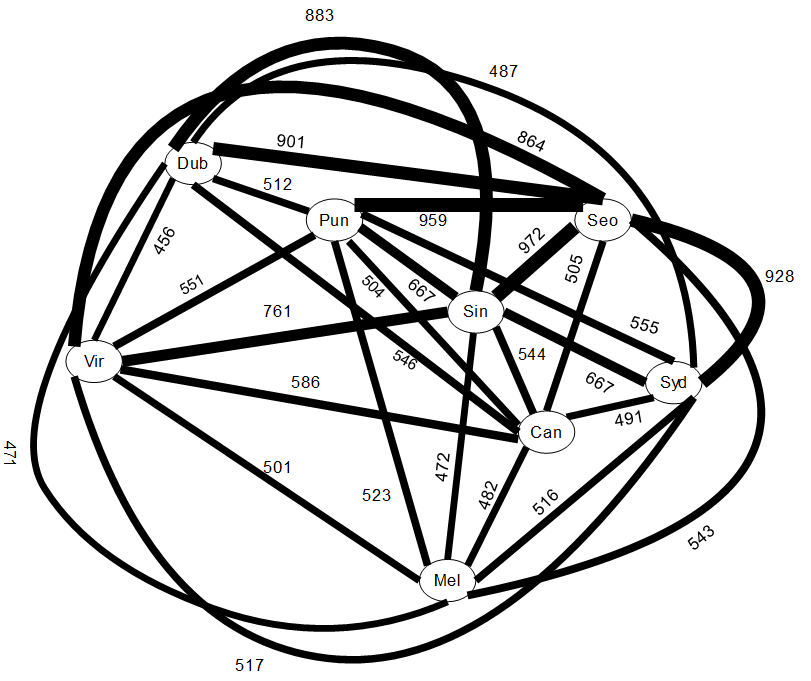}}
  \caption{Network traffic distribution across cluster nodes with latencies scaling factor of 5X for \textbf{workload F}.}
\label{fig:band-dis-f}
\vspace{-5mm}
\end{figure*}

Figs. \ref{fig:band-dis-a}-\ref{fig:band-dis-f} depicts network traffic distribution across cluster nodes with Latencies Scaling Factor (LSF) of 5X\footnote{Simliar to LSF=5X, LSF=X follows the same pattern in  distribution data transferred between nodes but is less in values.} for workloads A, B, E, and F\footnote{Please note that we skipped results for workloads C and D due to similarities in the results with workload B, where high latency dominates the workload types.}. In these grapgh,  each vertex represents a cluster node and each link illustrates a link between nodes. The thicker the link is in the graph, the greater the amount of data (in KB) is exchanged between nodes. Network traffic distribution exposes how much data is transferred between each pair of nodes individually. Moreover, it shows how many nodes actively participate in serving workloads.    

As can be seen in Fig. \ref{fig:band-dis-a}, for MongoDB, five nodes (Singapore, Seoul, Sydney, Virginia, and Dubai) actively participated to handle workload A, where the amount of transferred data reaches more than 4.4-6 MB. For MySQL, one more node, located in Pune, has participated in the data exchange (600-700 KB) in comparison to MongoDB. In contrast, Cassandra and Redis respectively exposed a balanced distribution of data among all nodes in the range of 600-700 KB and 500-600 KB. 

Fig. \ref{fig:band-dis-b} shows that nodes in Seoul, Sydney, Pune, Canberra, Melbourne, and Virginia actively serve workload B for MongoDB, where the amount of exchanged data is 1.2 - 1.6 MB. For MySQL, client uses nodes in  Pune, Seoul, Sydney, Dubai, and Virginia to perform workload B. Cassandra and Redis expose almost the same amount and distribution of data  across nodes as seen for workload A.

Fig. \ref{fig:band-dis-e} gives the details on data distribution across cluster nodes for workload E. MongoDB exploited all nodes to serve workload E as opposed to workloads A and B. However, for Cassandra and Redis, all nodes are involved to serve workload E though the amount of exchanged data is more than workloads A and B. For MySQL, the pattern of exchanged data for workload E is the same as for A and B though the amount of exchanged data is less. 

Fig. \ref{fig:band-dis-f} presents the data flow between cluster nodes for workload F. We observed the same pattern in data distribution between nodes but more in the amount for workload F as opposed to workload A. This is because workload F including data updating.

\subsubsection{The impact of down-sizing and up-sizing of cluster on the throughput}\label{sec:scalingupdown}
In this section, we discuss the impact of the down-/up-sizing of the cluster on the throughput of databases. Down-/up-sizing of the cluster has been conducted on-the-fly, unless indicated otherwise, during workloads processing. This evaluation answers how removing and adding node(s), as a consequence of nodes mobility, affect databases throughput. For this evaluation, we considered two cases as follows. 

\begin{figure*}[ht!]
  \centering
  \subfloat[Cassandra-removing]{\label{fig:cass-removing}\includegraphics[height=3cm,width=0.33\textwidth]{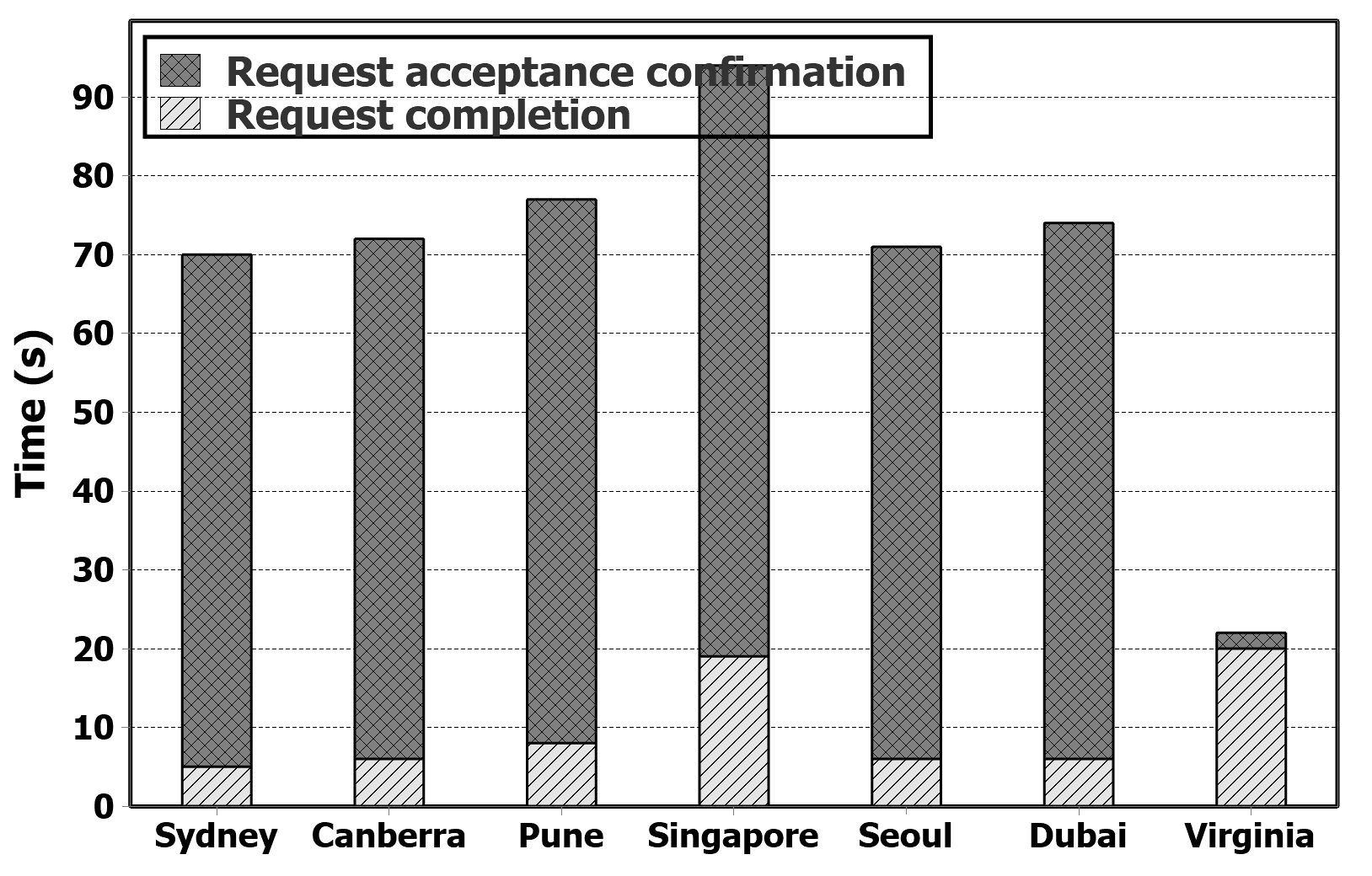}}
  \subfloat[Cassandra-adding]{\label{fig:cass-adding}\includegraphics[height=3cm,width=0.33\textwidth]{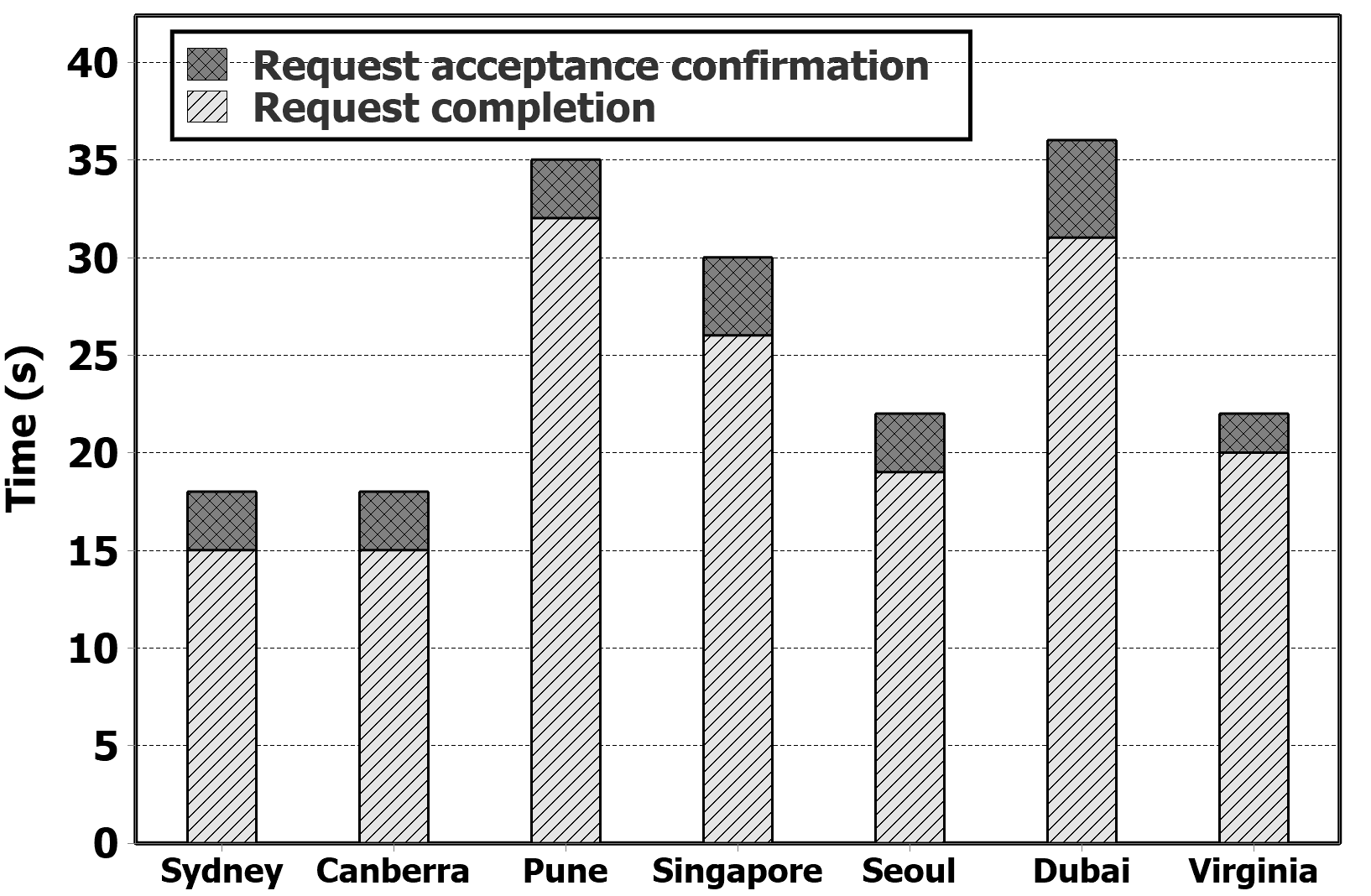}}
  \subfloat[MongoDB-removing]{\label{fig:mongo-removing}\includegraphics[height=3cm,width=0.33\textwidth]{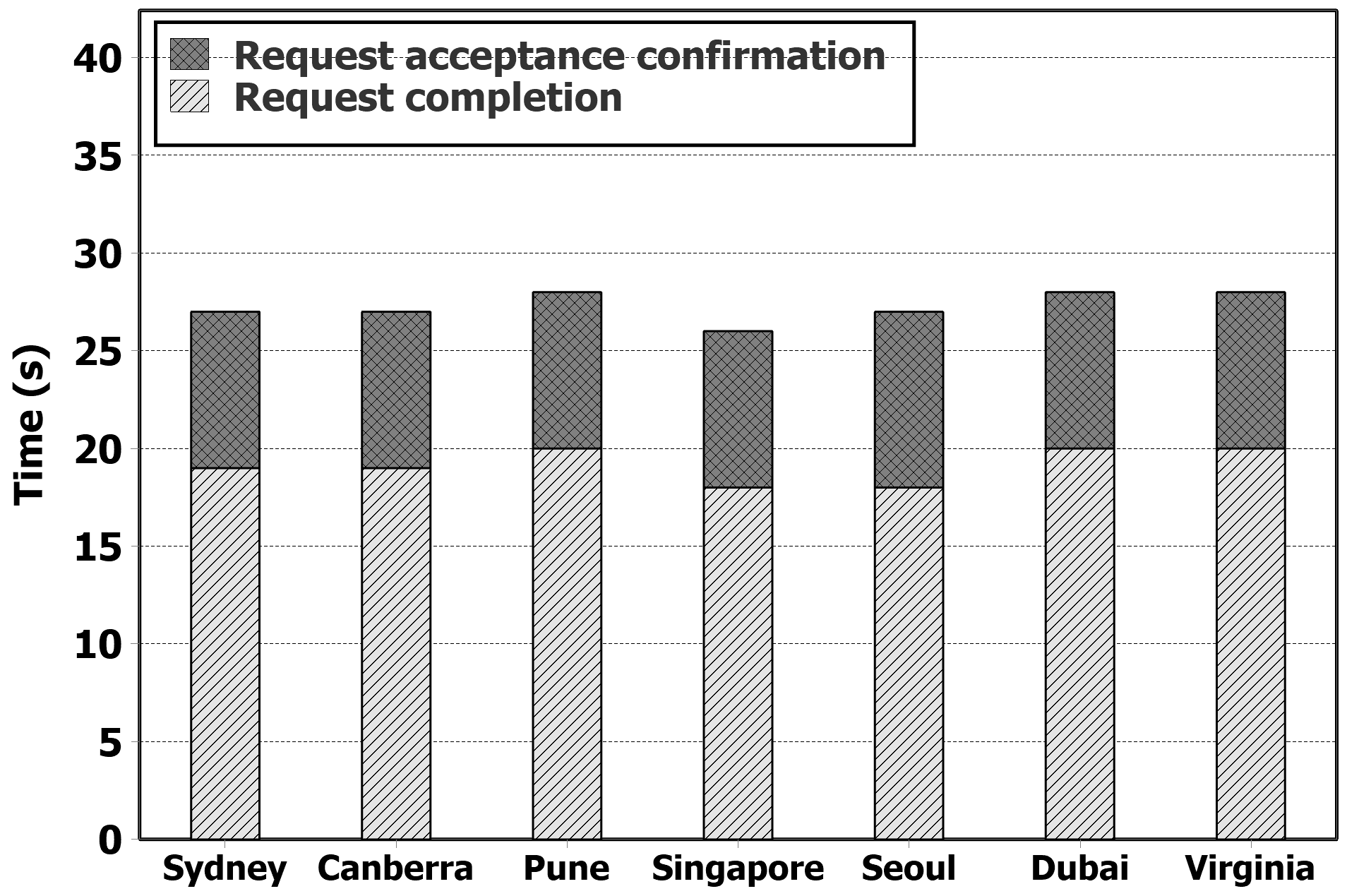}}\\
  \subfloat[MongoDB-adding]{\label{fig:mongo-adding}\includegraphics[height=3cm,width=0.33\textwidth]{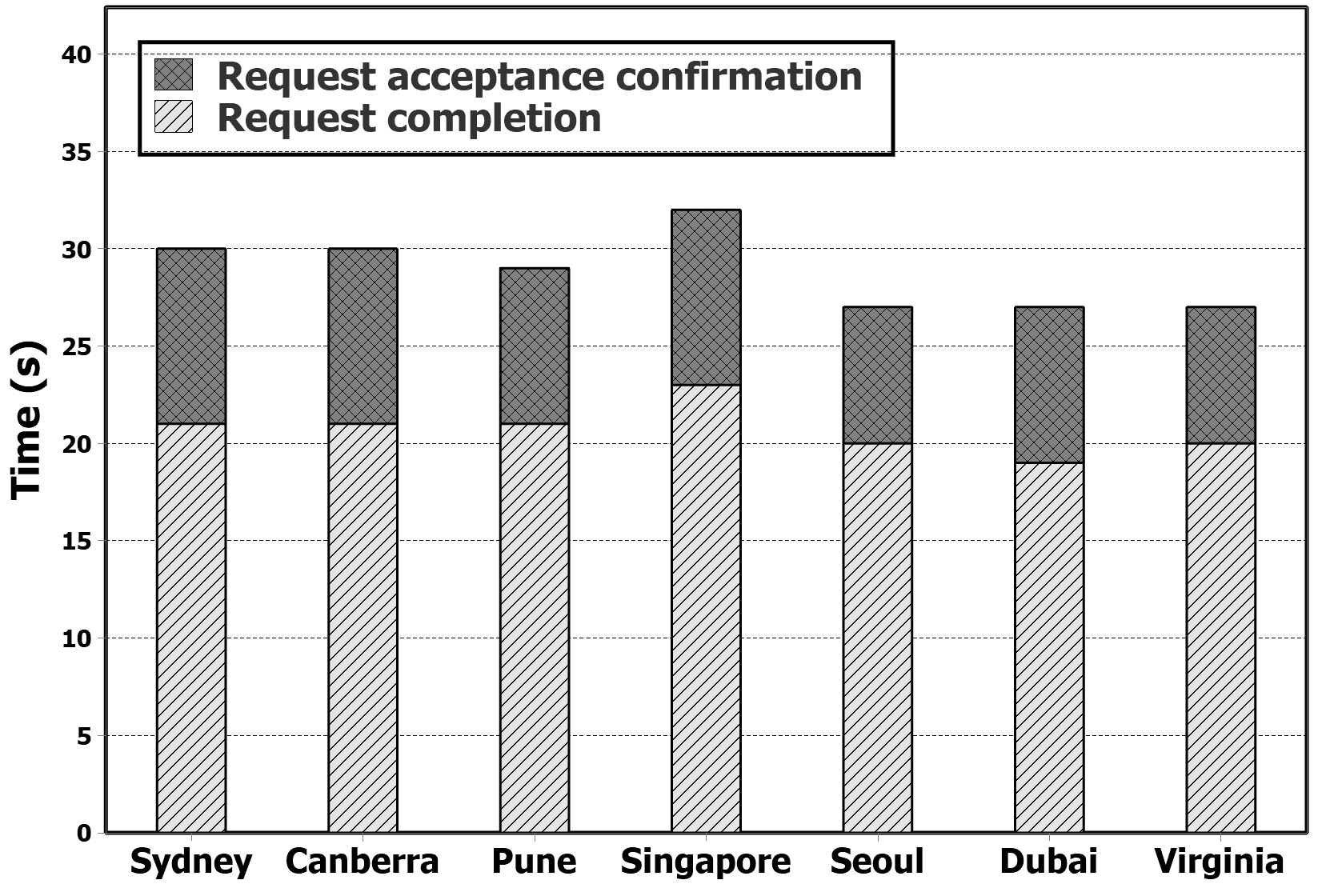}}
  \subfloat[Redis-removing]{\label{fig:redis-removing}\includegraphics[height=3cm,width=0.33\textwidth]{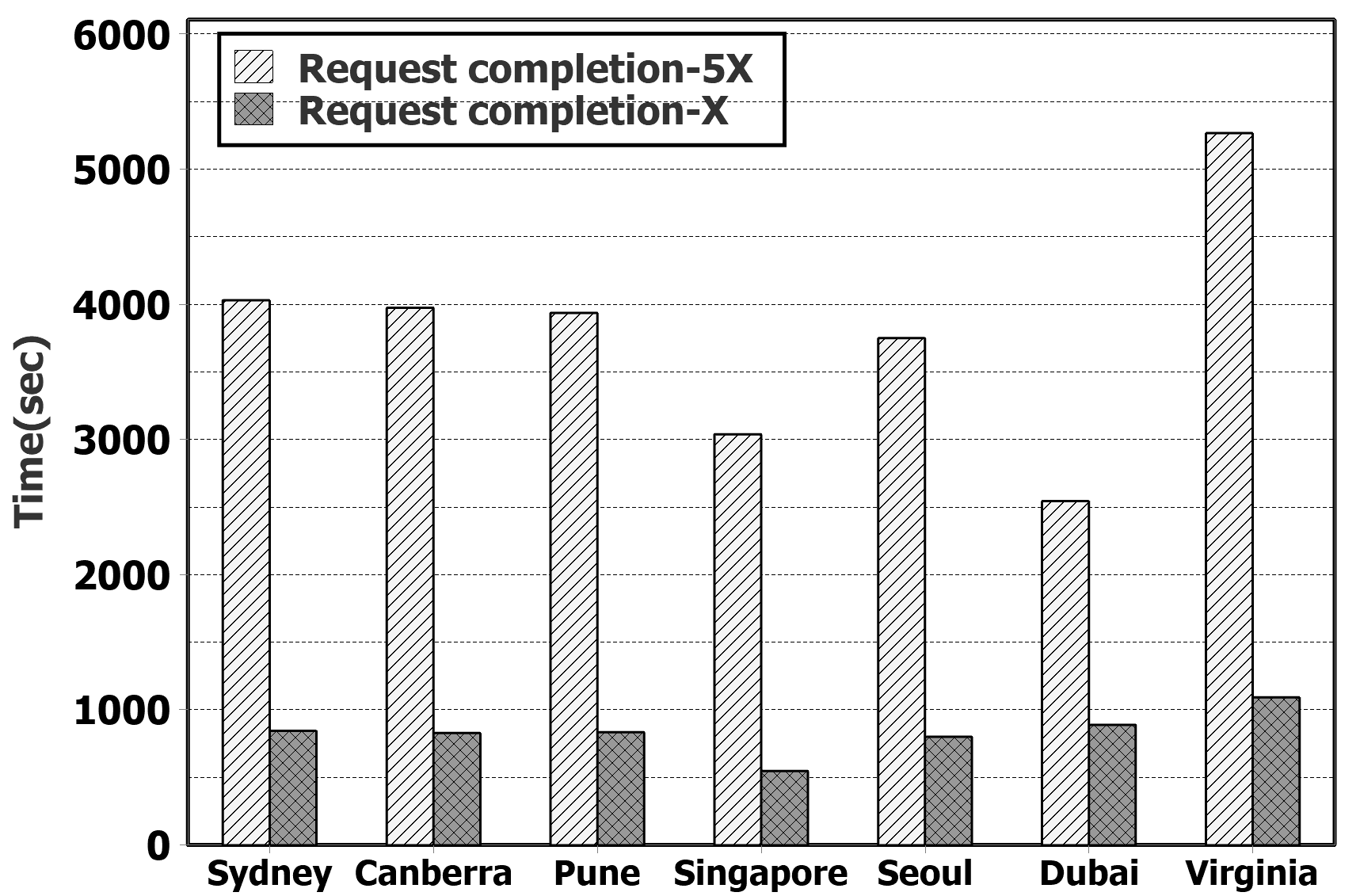}}
  \subfloat[Redis-adding]{\label{fig:redis-adding}\includegraphics[height=3cm,width=0.33\textwidth]{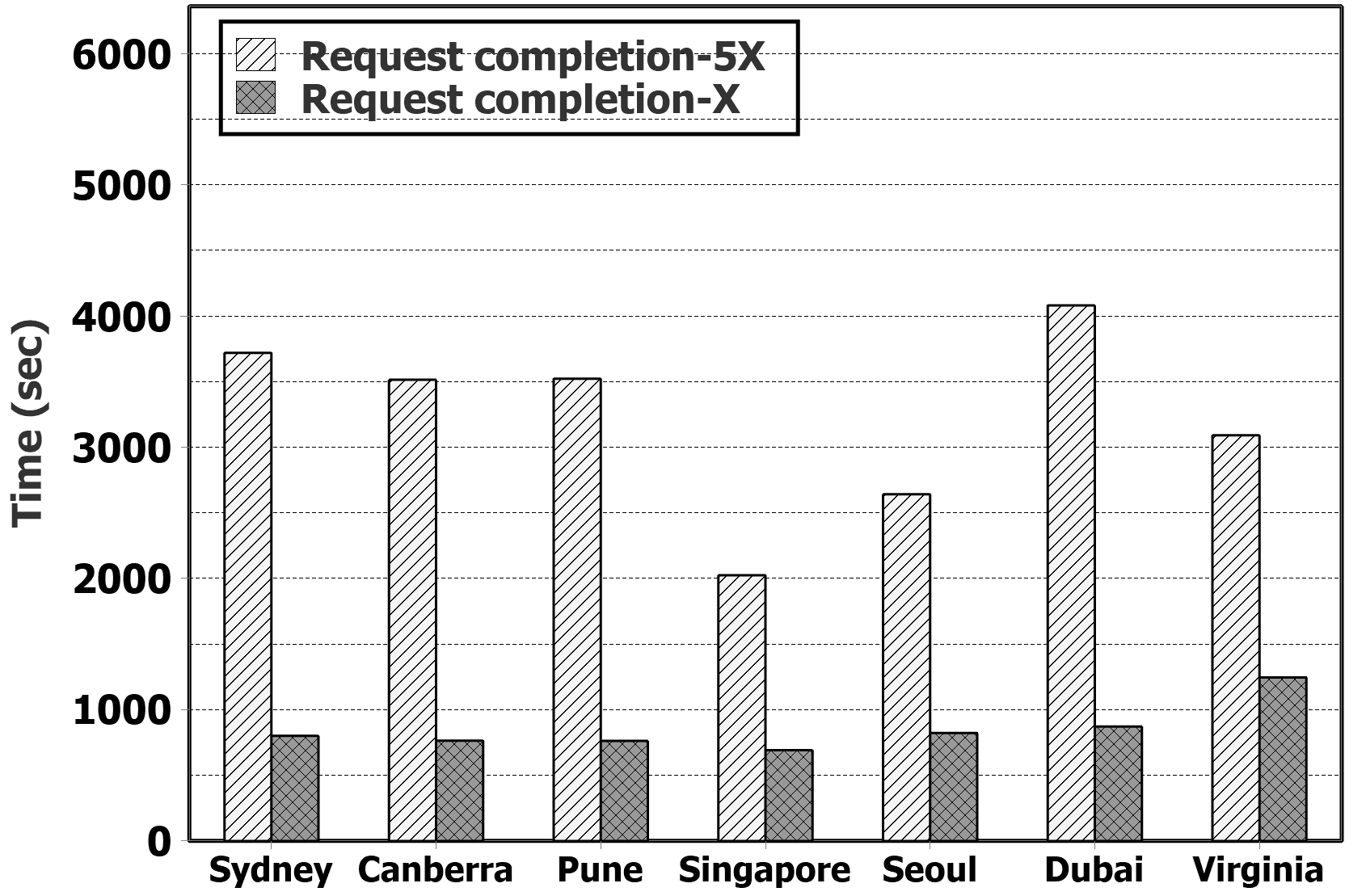}}
  \caption{Removing and adding a single node from and to a cluster with latencies scaling factor of 5X.}
\label{fig:removing-addine-single-node}
\vspace{-5mm}
\end{figure*}

\textit{Removing and adding a single node.}
We initially measured removing and adding a single node to better understand its possibility. Due to the asynchronous nature of communication with the cluster, removing and adding processes can be separated into two distinct time periods. \textit{Request acceptance confirmation:} This is the time that the database cluster needs to confirm the request for cluster resizing was accepted (in other words the client prints a success message). Usually, it takes a shorter time and is mostly the same for addition and removal. \textit{Request completion:} This is the actual time necessary to complete the resizing process, which depends on the amount of data in the database, network latencies/bandwidths between nodes, replication factor, replication strategy, and particularly on the amount of data stored on the node being removed. The primary difference between removal and addition process lies in the data rebalancing process. More specifically, if a node is removed, then its stored data should be moved to the remaining nodes. Conversely, if a new node is added (depending on the replication strategy) the stored  data in the cluster might be equally distributed across all nodes. In addition, the associated metadata-related overheads contribute to additional delay. In summary, the total time consists of starting the resizing process (short) and actually applying the desired changes (significantly longer).  

Fig. \ref{fig:removing-addine-single-node} plots the required time to remove and add a single node from and to databases cluster. Figs. \ref{fig:cass-removing} and \ref{fig:cass-adding} depict the process of removing and adding a single node in Cassandra cluster with LSF=5X require 20-90 and 15-35, respectively. Note that,  the seed node located in Melbourne is not removed.
 Figs. \ref{fig:mongo-removing} and \ref{fig:mongo-adding} depict that the request acceptance and request completion times respectively are about 5 and 20 seconds for removing and adding a single node in the MongoDB cluster with LSF=5X.
 Note that Melbourne and Virginia respectively host master and non-voting nodes, and we did not remove them from the cluster due to the limitation of MongoDB cluster. Thus, processing time for adding and removing a node in MongoDB cluster is less than in  Cassandra cluster because Cassandra requires to re-distribute data across new set of nodes while MongoDB does not.
Figs. \ref{fig:redis-removing} and \ref{fig:redis-adding} reveal that the required time for the process of removing and adding a single node in the Redis cluster is proportional to the latency between the node being removed/added with other nodes in the cluster; 2000-5000 seconds for 5X and 500-1200 seconds for X. 
This is because Redis requires re-sharding and re-balancing data across nodes as explained in Section \ref{sec:challanges}\footnote{Note that the request acceptance confirmation time for Redis is about 2 seconds, which is a very small value as compared to request completion.}.    

\textit{Removing and adding a set of nodes.}
We discuss the impact of removing and adding a set of nodes from and to the cluster as workloads are running. We considered three scenarios. \textbf{Downsizing:} this scenario is a reduction of 8 nodes to 2 nodes with a step of one for removal node. \textbf{Upsizing:} this scenario is an increment of 2 nodes to 8 nodes with a step of one node in reverse order as conducted for the downsizing scenario. \textbf{All-nodes:} we considered the all-nodes scenario as a benchmark in which the number of nodes is fixed to 8 when each workload is running. 
We removed/added nodes with the help of databases API (Section \ref{sec:challanges}) and bash scripts while architecture-based constraints for each database are preserved. We also measured \textit{settling down} and \textit{settling up} as the duration time in which a node respectively is removed from and added to a cluster for down-/up-sizing scenarios. The interval time between each removing/adding a node is 120 seconds and workloads run with enough operations so that all 6 nodes can be removed/added. For \textit{all-nodes} scenario, we ran experiments with the same number of operations as conducted for down-/up-sizing scenarios.

\begin{figure*}[ht!]
  \centering
  \subfloat[Workload A]{\label{fig:cass-scaling-a-5x}\includegraphics[height=3cm,width=0.33\textwidth]{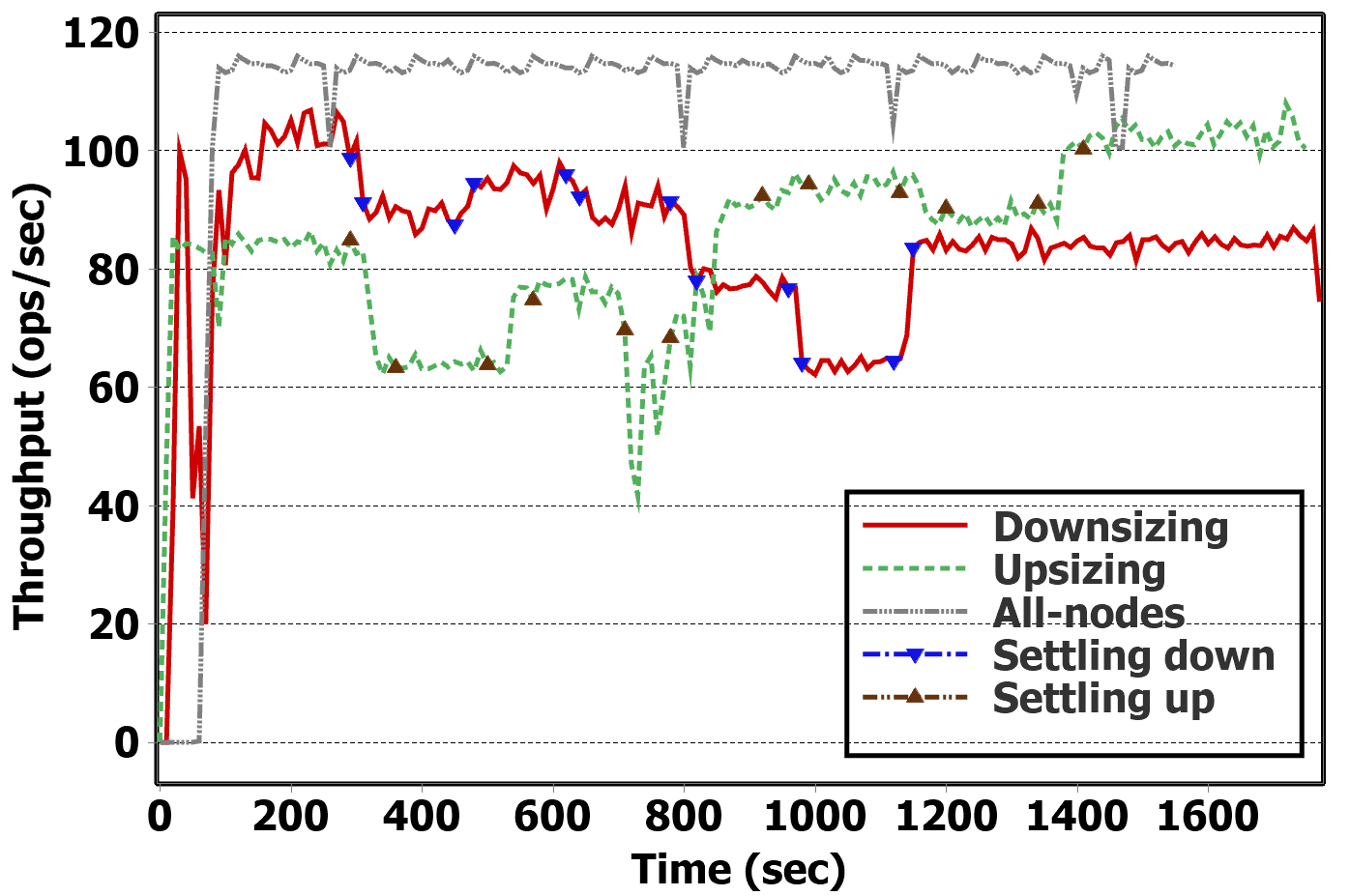}}
  \subfloat[Workload B]{\label{fig:cass-scaling-b-5x}\includegraphics[height=3cm,width=0.33\textwidth]{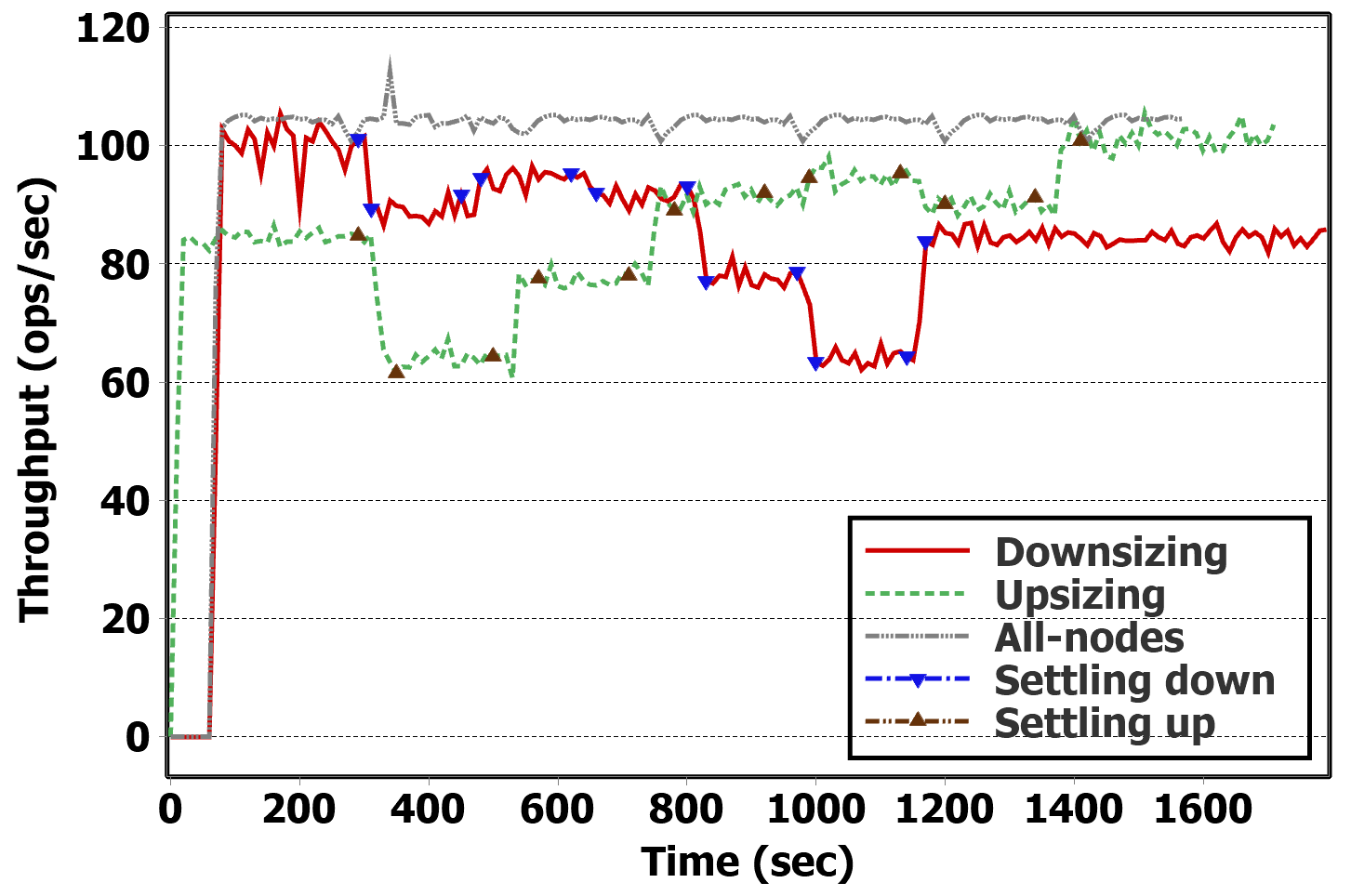}}
  \subfloat[Workload C]{\label{fig:cass-scaling-c-5x}\includegraphics[height=3cm,width=0.33\textwidth]{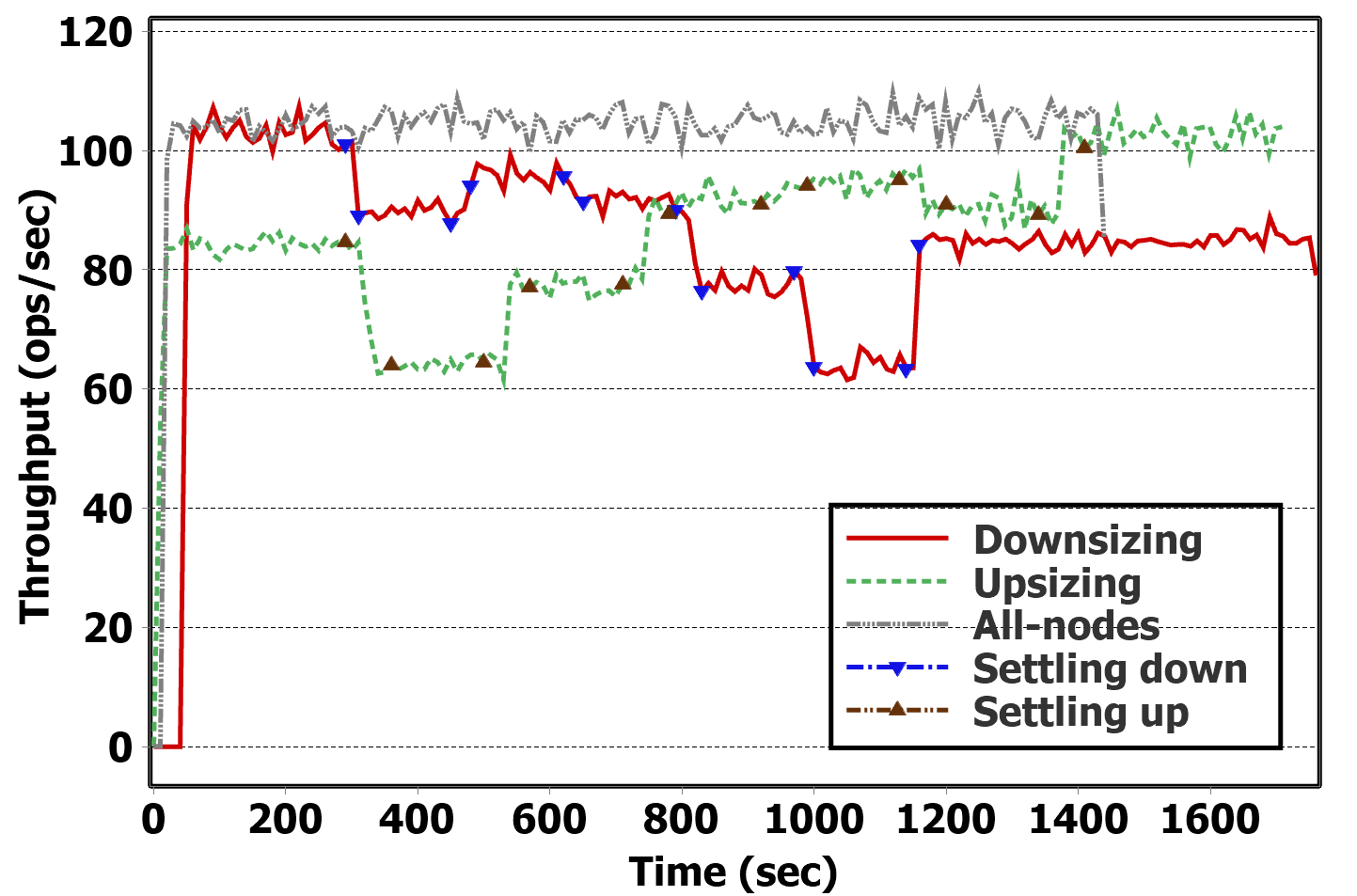}}\\
  \subfloat[Workload D]{\label{fig:cass-scaling-d-5x}\includegraphics[height=3cm,width=0.33\textwidth]{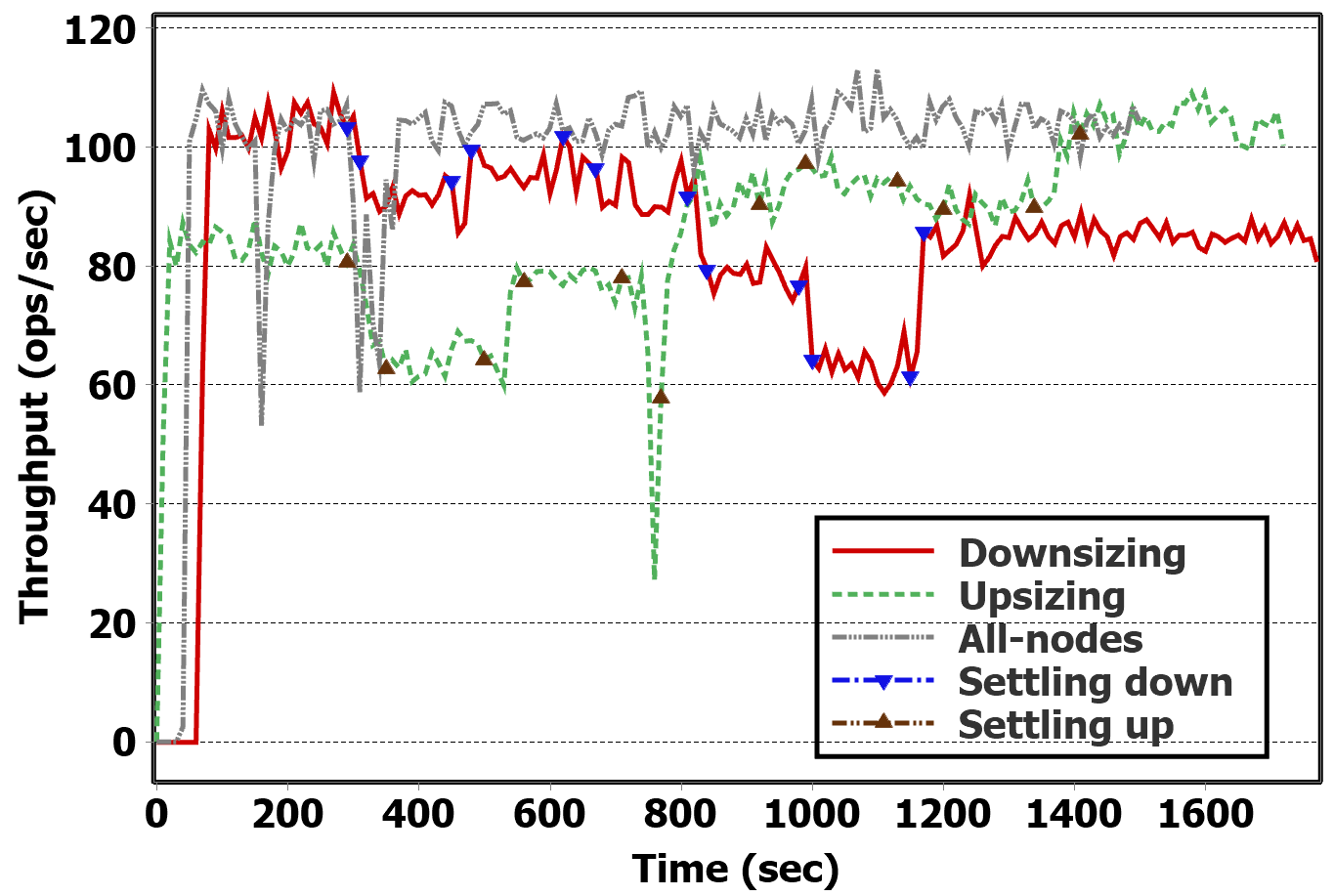}}
  \subfloat[Workload E]{\label{fig:cass-scaling-e-5x}\includegraphics[height=3cm,width=0.33\textwidth]{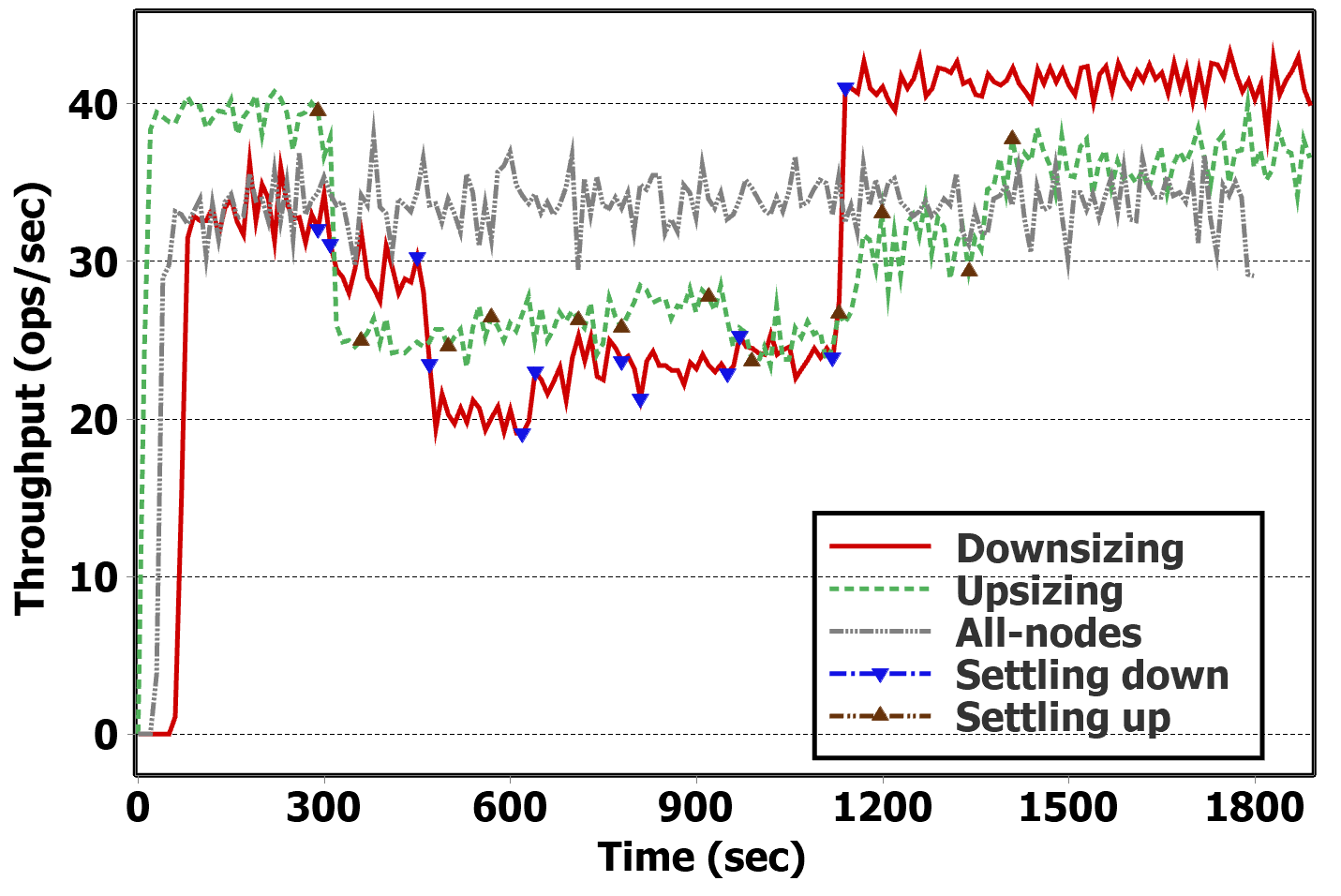}}
  \subfloat[Workload F]{\label{fig:cass-scaling-f-5x}\includegraphics[height=3cm,width=0.33\textwidth]{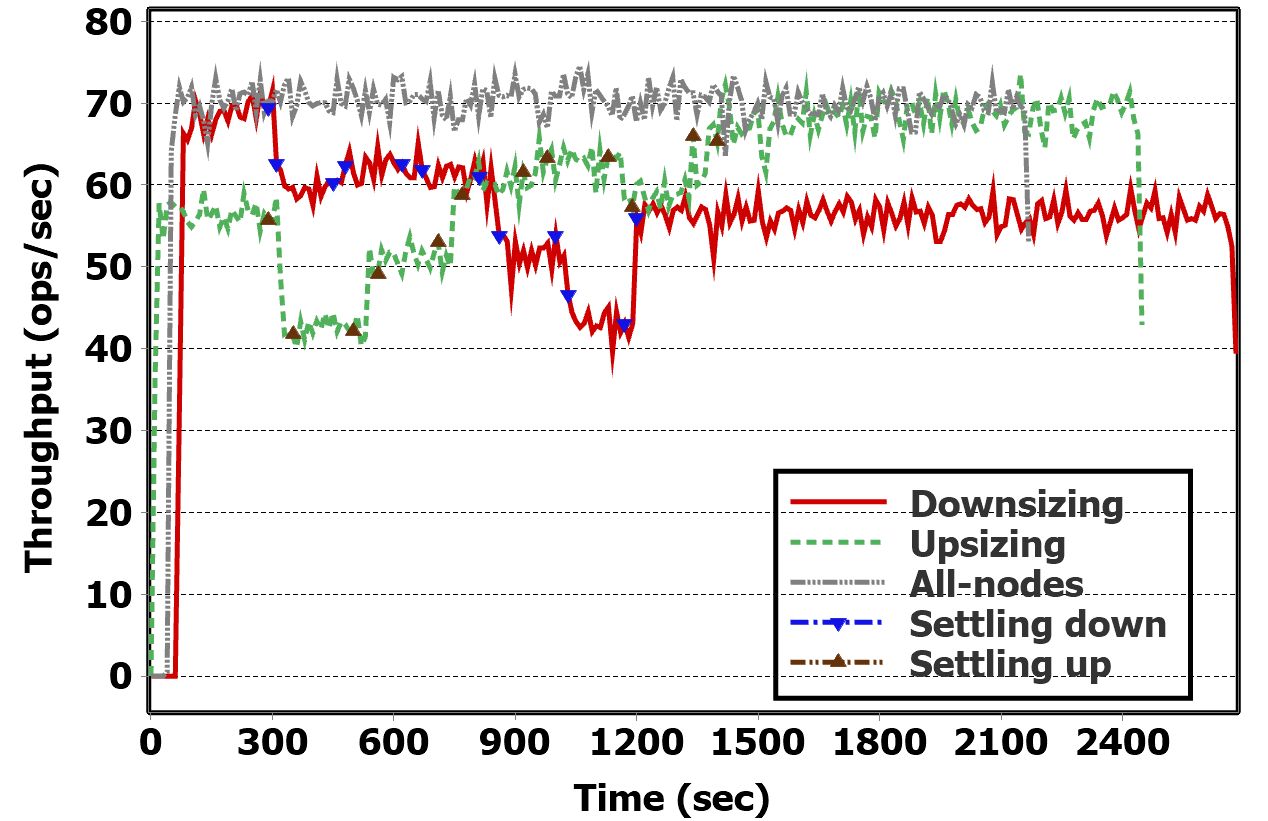}}
  \caption{The impact of on-the-fly Up-/Down-sizing scenarios on \textbf{Cassandra} throughput. Legends present \textit{Downsingin} scenario in which a node is removed from the cluster during \textit{Settling down} time, \textit{Upsizing} scenario in which a node is added to the cluster during \textit{Settling up} time, and \textit{All-nodes} scenario in which the size of the cluster is fixed to 8 nodes.}
\label{fig:cassandra-removing-adding-nodes-3r-rd-5x}
\vspace{-3mm}
\end{figure*}

Fig. \ref{fig:cassandra-removing-adding-nodes-3r-rd-5x} plots the throughput of Cassandra for down-/up-sizing and all-nodes scenarios over time. We removed nodes from clusters located in Singapore, Sydney, Canberra, Pune, Dubai and Virginia and added them to the cluster in reverse order as performed for the downsizing scenario. 
Fig . \ref{fig:cass-scaling-a-5x} shows that Cassandra throughput reduces from 100 to 60 Ops/sec for workload A as nodes are removed in the mentioned order except for the farthest node in Virginia; In fact, removing a node in Virginia raises throughput from 60 to more than 80 Ops/sec since the data are moved closer to a clients. In contrast, for upsizing, as the more nodes are added to a cluster, the more increments in throughput can be observed except for the farthest node in Virginia. As expected, the all-nodes scenario outperforms two other scenarios in throughput ($\approx$120 Ops/sec). 
For workloads B, C, and D (Fig. \ref{fig:cass-scaling-b-5x}-\ref{fig:cass-scaling-d-5x}), the values and pattern of  Cassandra throughput are the same as seen for workload A. This reveals that higher latency dominates the workload type, which results in achieving almost the same values for throughput. However, for workloads E and F (\ref{fig:cass-scaling-e-5x}-\ref{fig:cass-scaling-f-5x}), the same pattern can be seen for Cassandra throughput, though its value is less than the values for other workloads. This can be explained that workloads F and E are naturally heavier than other workloads in basic operations. 

\begin{figure*}[ht!]
  \centering
  \subfloat[Workload A]{\label{fig:mongo-scaling-a-5x}\includegraphics[height=3cm,width=0.33\textwidth]{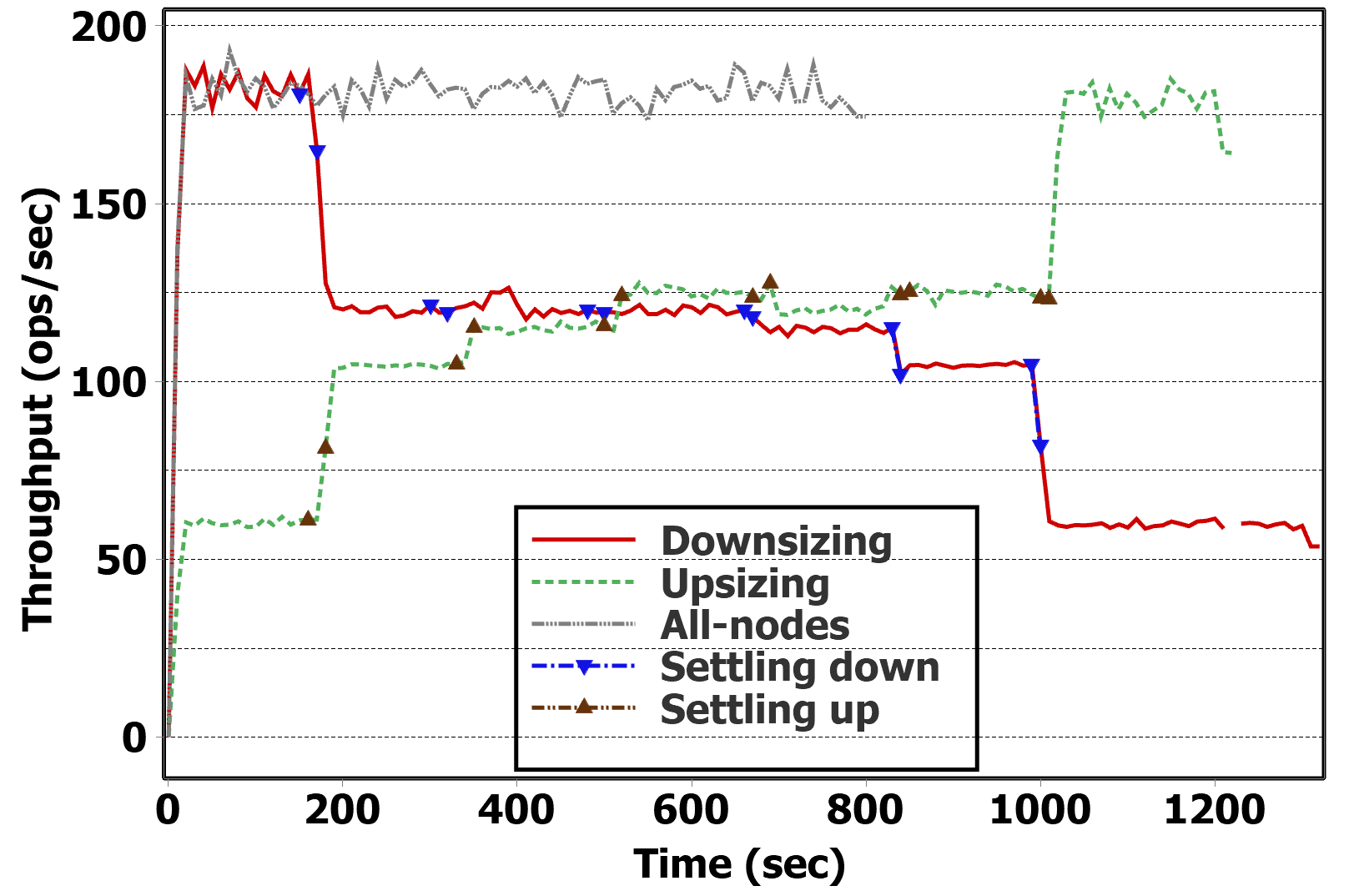}}
  \subfloat[Workload B]{\label{fig:mongo-scaling-b-5x}\includegraphics[height=3cm,width=0.33\textwidth]{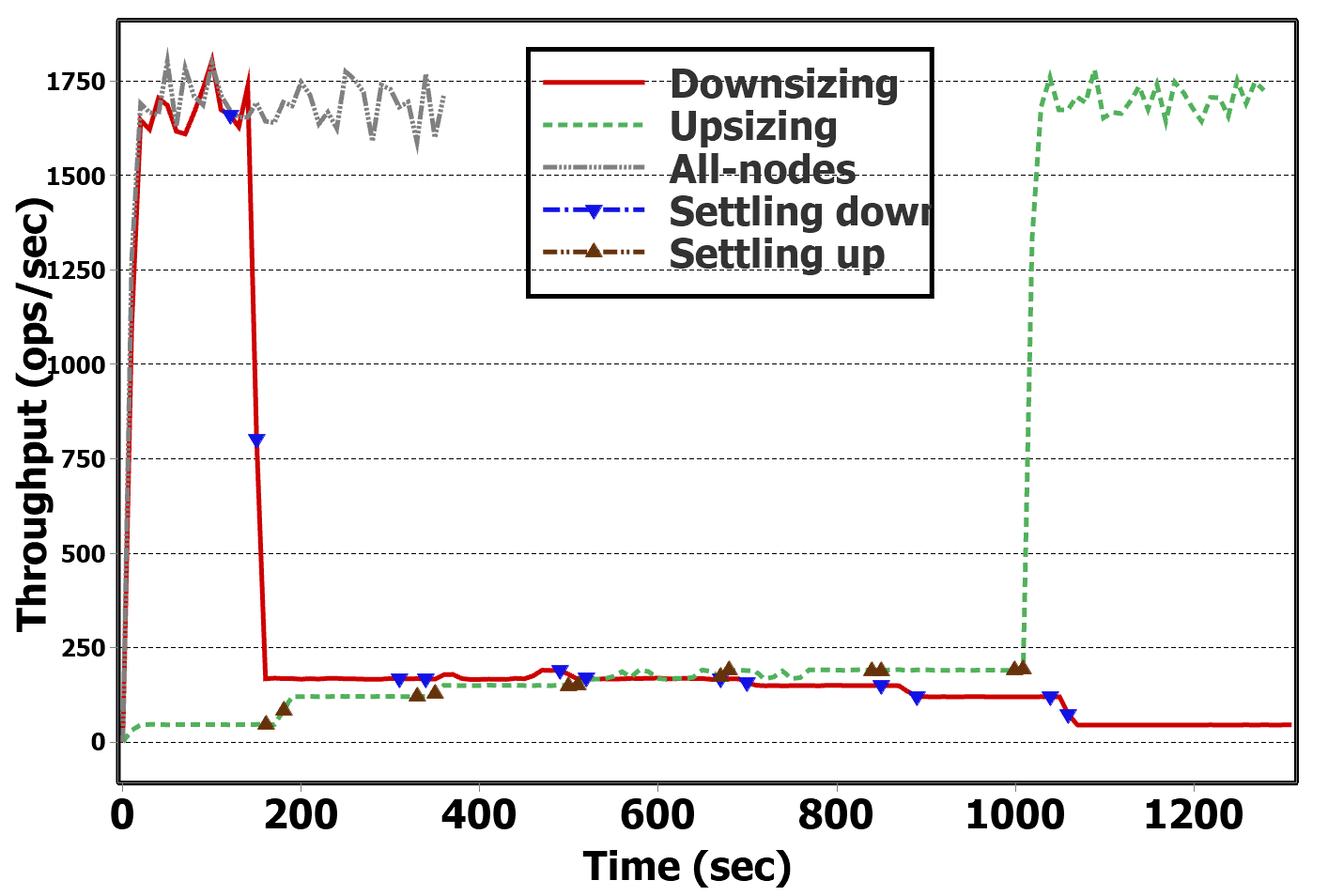}}
  \subfloat[Workload C]{\label{fig:mongo-scaling-c-5x}\includegraphics[height=3cm,width=0.33\textwidth]{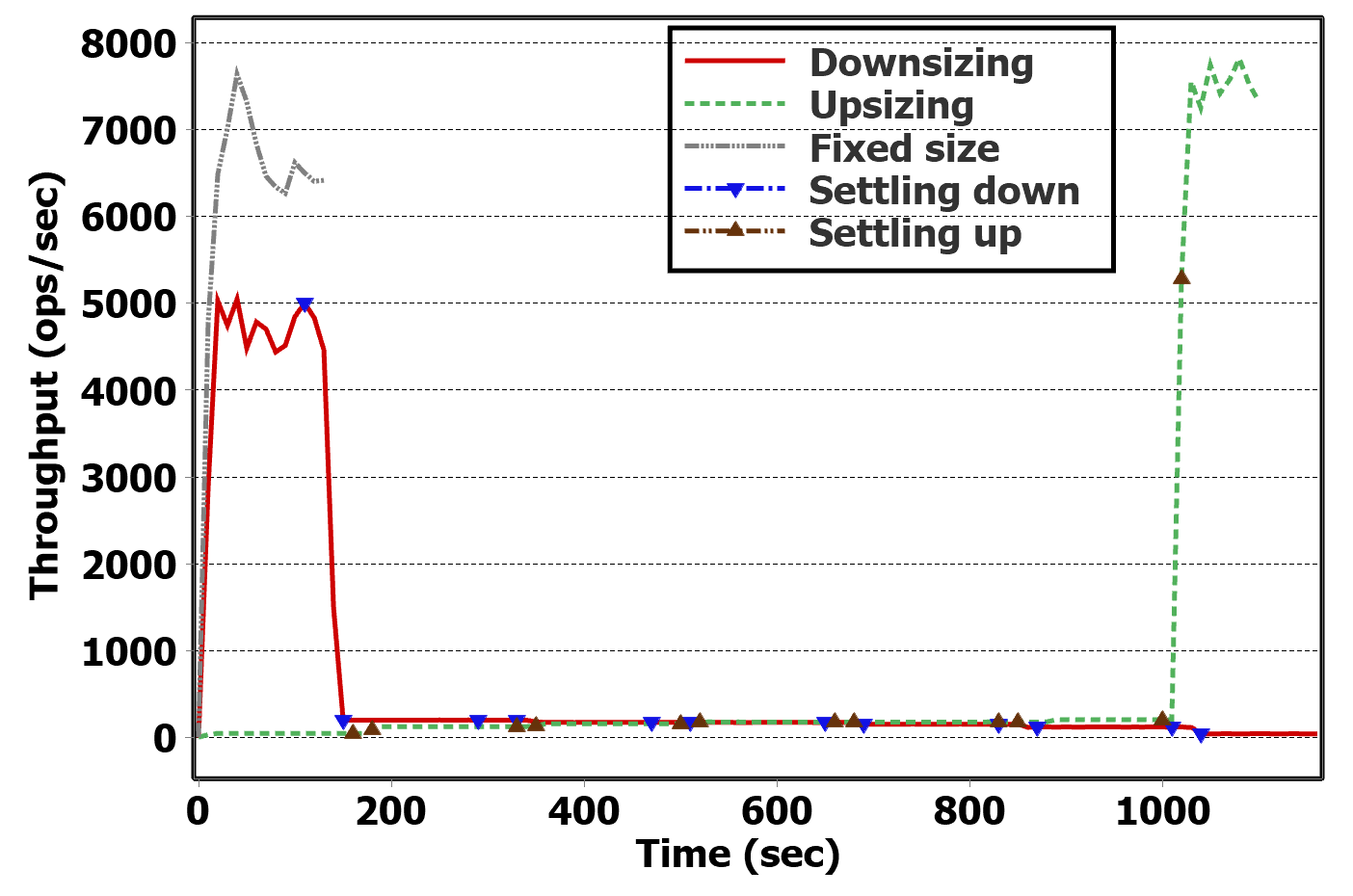}}\\
  \subfloat[Workload D]{\label{fig:mongo-scaling-d-5x}\includegraphics[height=3cm,width=0.33\textwidth]{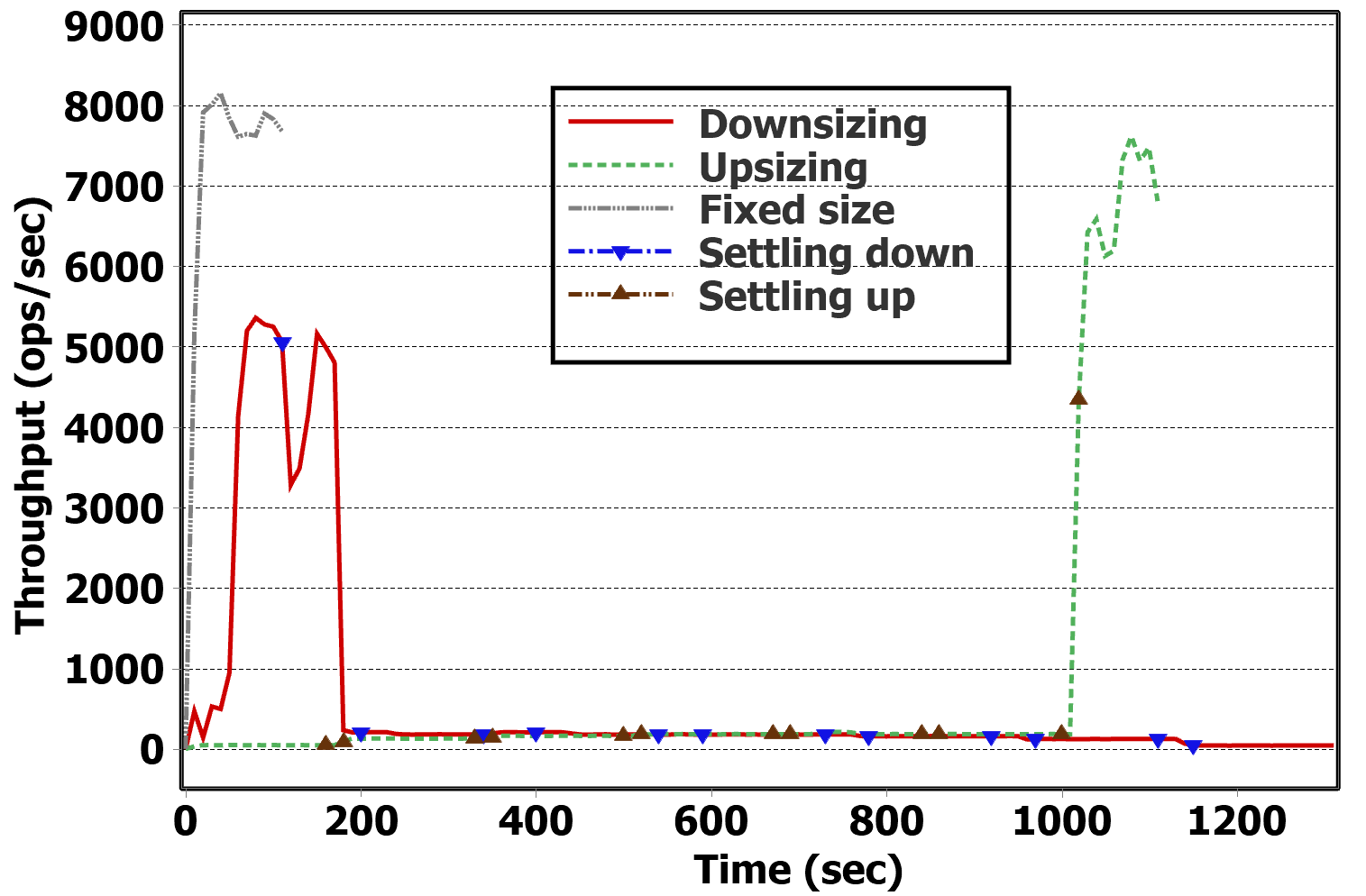}}
  \subfloat[Workload E]{\label{fig:mongo-scaling-e-5x}\includegraphics[height=3cm,width=0.33\textwidth]{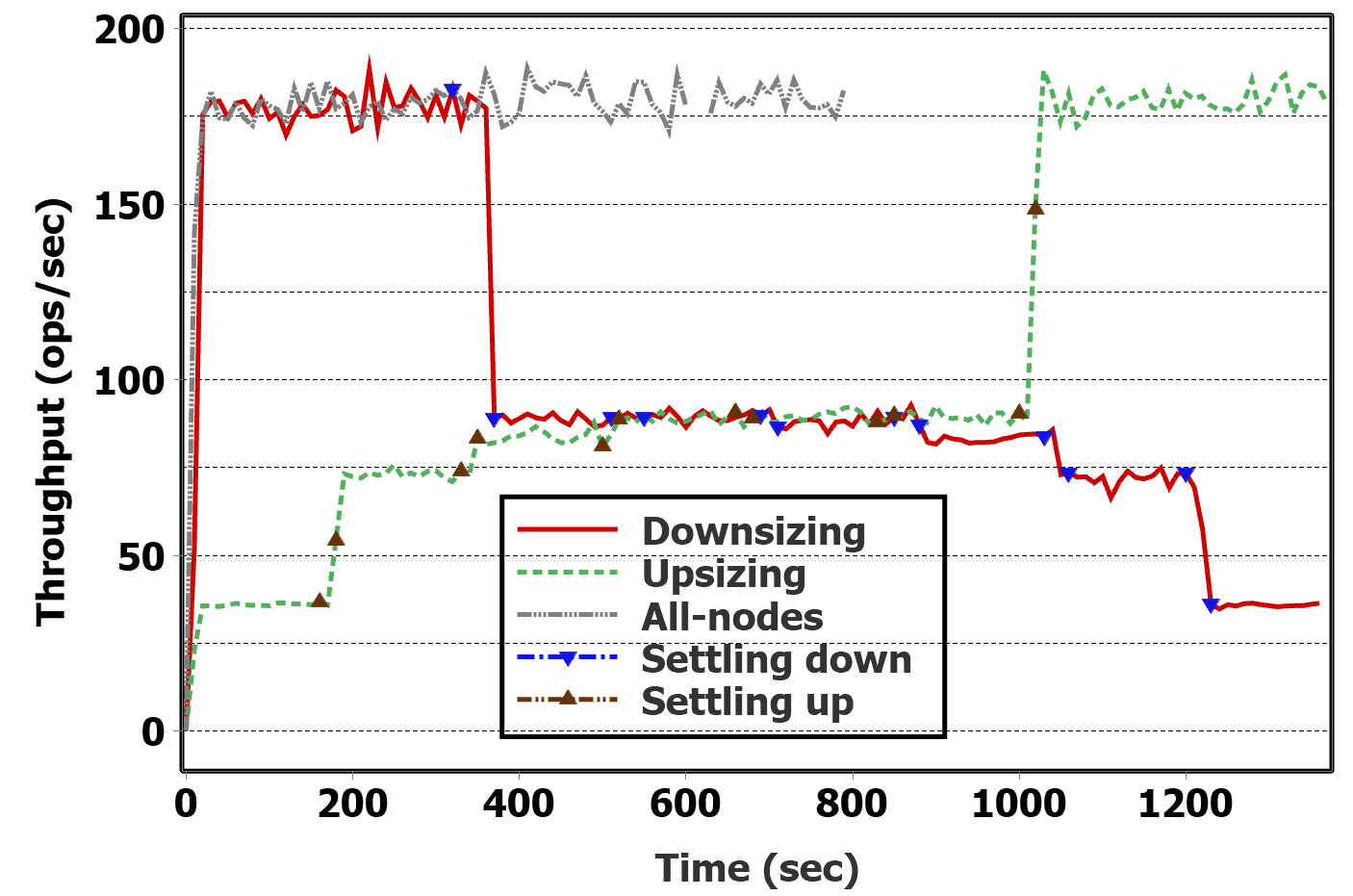}}
  \subfloat[Workload F]{\label{fig:mongo-scaling-f-5x}\includegraphics[height=3cm,width=0.33\textwidth]{Fig/mongo-f-remove-add-nodes-5x.png}}
  \caption{The impact of on-the-fly Up-sizing and Down-sizing of a \textbf{MongoDB} cluster on \textbf{MongoDB} throughput}
\label{fig:mongo-removing-adding-nodes-5x}
\vspace{-5mm}
\end{figure*}

Fig. \ref{fig:mongo-removing-adding-nodes-5x} illustrates MongoDB throughput for down-/up-sizing and all-nodes scenarios. Nodes are respectively removed in Singapore, Sydney, Canberra, Pune, Seoul, and Dubai; and in reverse, they are added in the upsizing scenario. We set the read policy to secondary nodes since this policy allows us to evaluate running workloads as nodes are removed/added. For workload A, as the node in Singapore is removed, throughput significantly reduces from 190 to 120 Ops/sec (Fig. \ref{fig:mongo-scaling-a-5x}) because read requests cannot be served by data in the client node anymore. Removing the next three nodes has slightly reduced in throughput whilst removing the last two nodes in Dubai and Seoul has more impact on the throughput. This can be justified as the client serves more requests based on data in Dubai. We observed a symmetric trend for upsizing scenario (Fig. \ref{fig:mongo-scaling-a-5x}). As expected, the all-nodes scenario requires 850 seconds to complete workload A, while down-/up-sizing scenarios need 1200 and 1290 seconds. 
For workloads B-D (Figs. \ref{fig:mongo-scaling-b-5x}-\ref{fig:mongo-scaling-d-5x}), removing and adding nodes close to the client has a significant impact on the throughput (at most 1700 Ops/sec for  B and 5000 Ops/sec for  C and D ). In addition, at the initial stage of adding nodes, MongoDB throughput for all-nodes scenario -- 8000 Ops/sec-- significantly is more than the throughput for other scenarios -- 5000 Ops/sec (Figs. \ref{fig:mongo-scaling-c-5x} and \ref{fig:mongo-scaling-d-5x}). This might be due to changing the master node from Melbourne to  Singapore during workload running. Workloads E and F follows similar trend for down-/up-sizing scenarios (Fig. \ref{fig:mongo-scaling-e-5x}-\ref{fig:mongo-scaling-f-5x}) as seen for workload A though the throughput values are different.   

\begin{figure*}[ht!]
  \centering
  \subfloat[Workload A]{\label{fig:redis-scaling-a-5x}\includegraphics[height=3cm,width=0.33\textwidth]{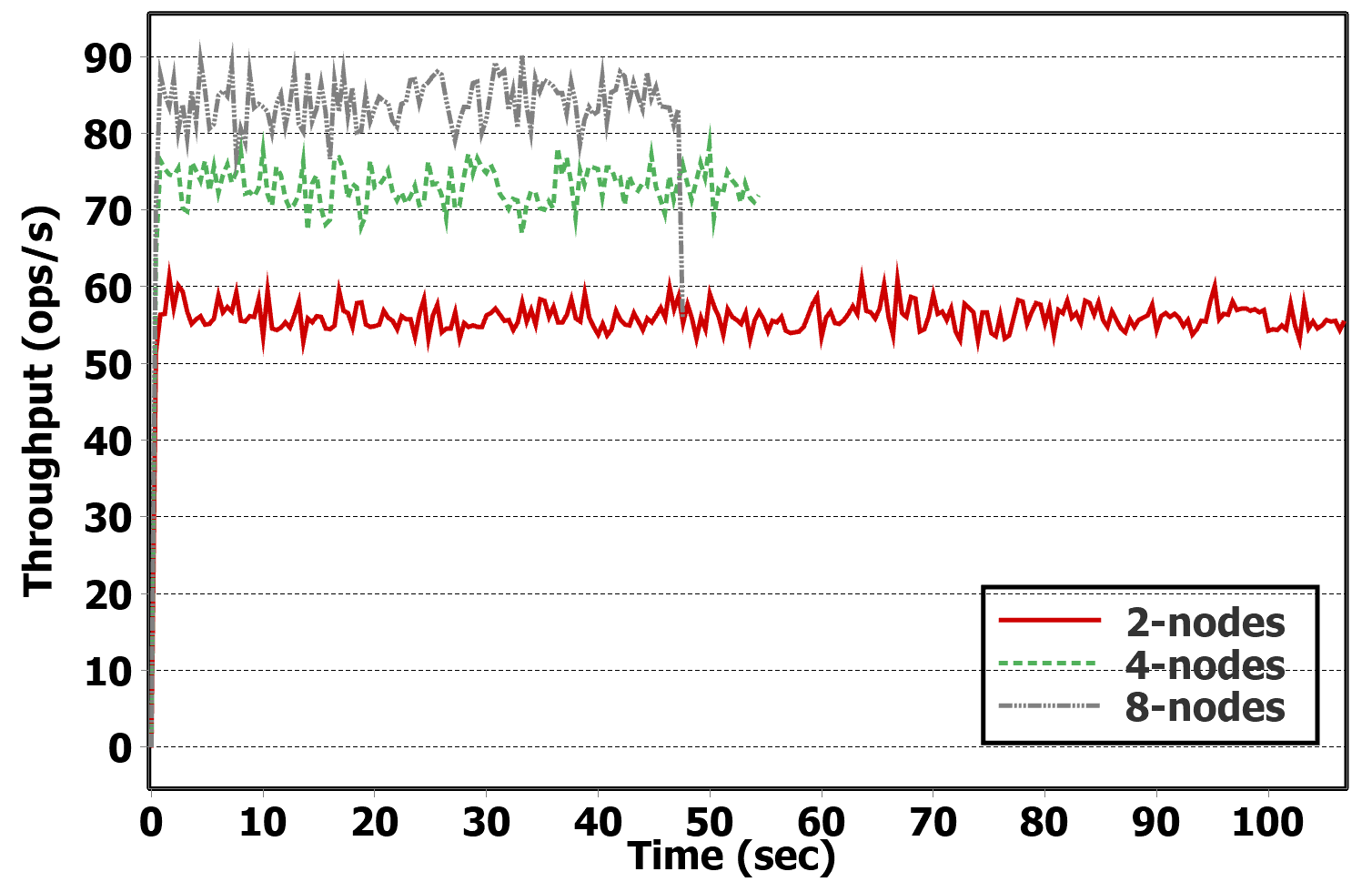}}
  \subfloat[Workload B]{\label{fig:redis-scaling-b-5x}\includegraphics[height=3cm,width=0.33\textwidth]{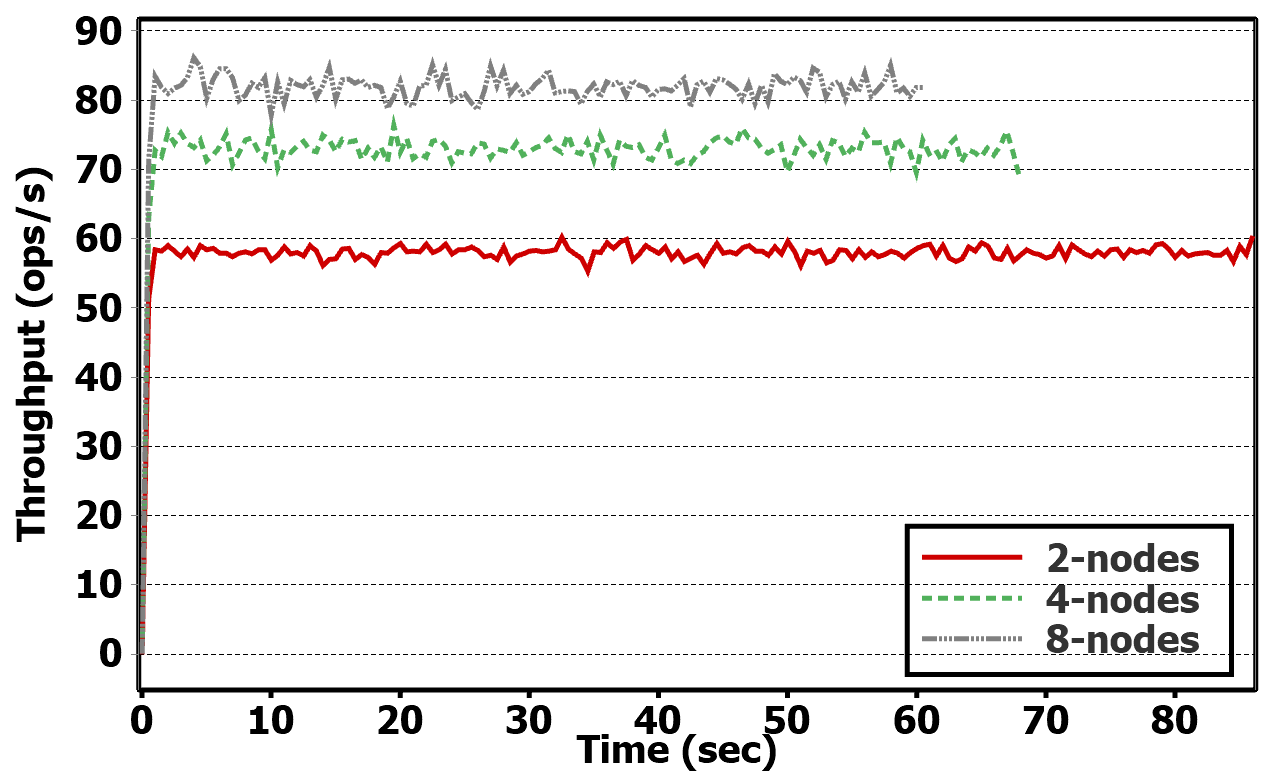}}
  \subfloat[Workload C]{\label{fig:redis-scaling-c-5x}\includegraphics[height=3cm,width=0.33\textwidth]{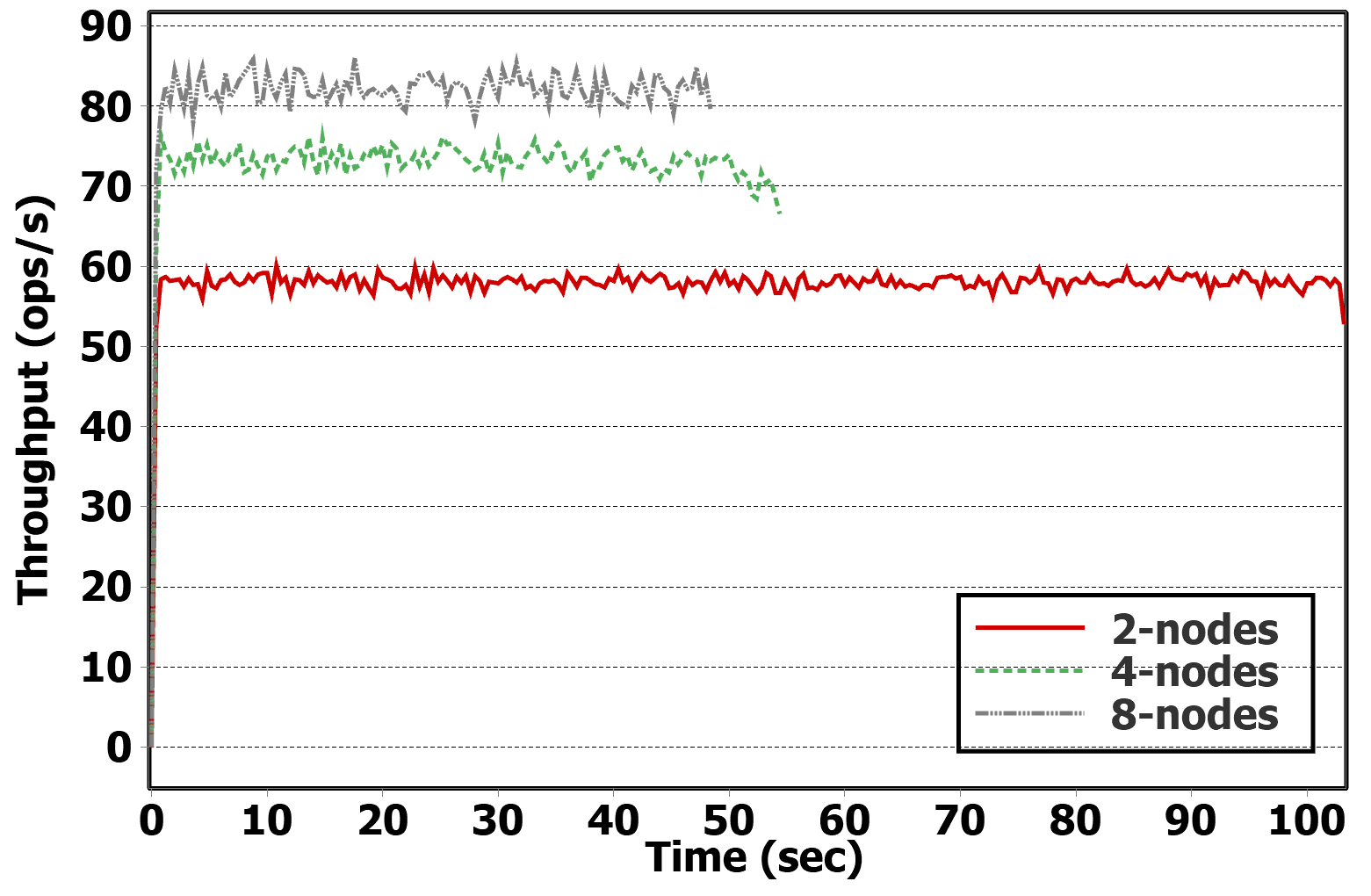}}\\
  \subfloat[Workload D]{\label{fig:redis-scaling-d-5x}\includegraphics[height=3cm,width=0.33\textwidth]{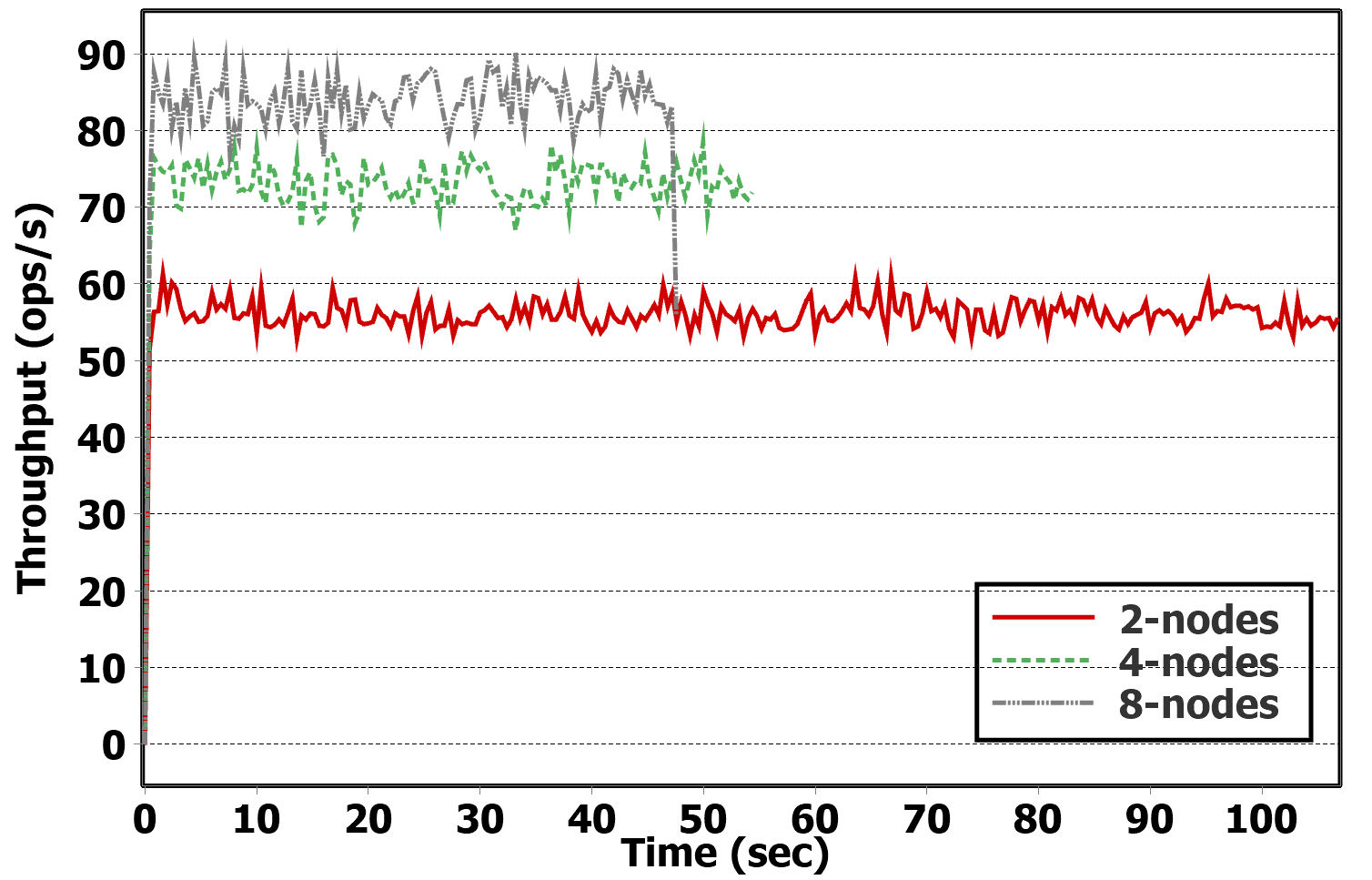}}
  \subfloat[Workload E]{\label{fig:redis-scaling-e-5x}\includegraphics[height=3cm,width=0.33\textwidth]{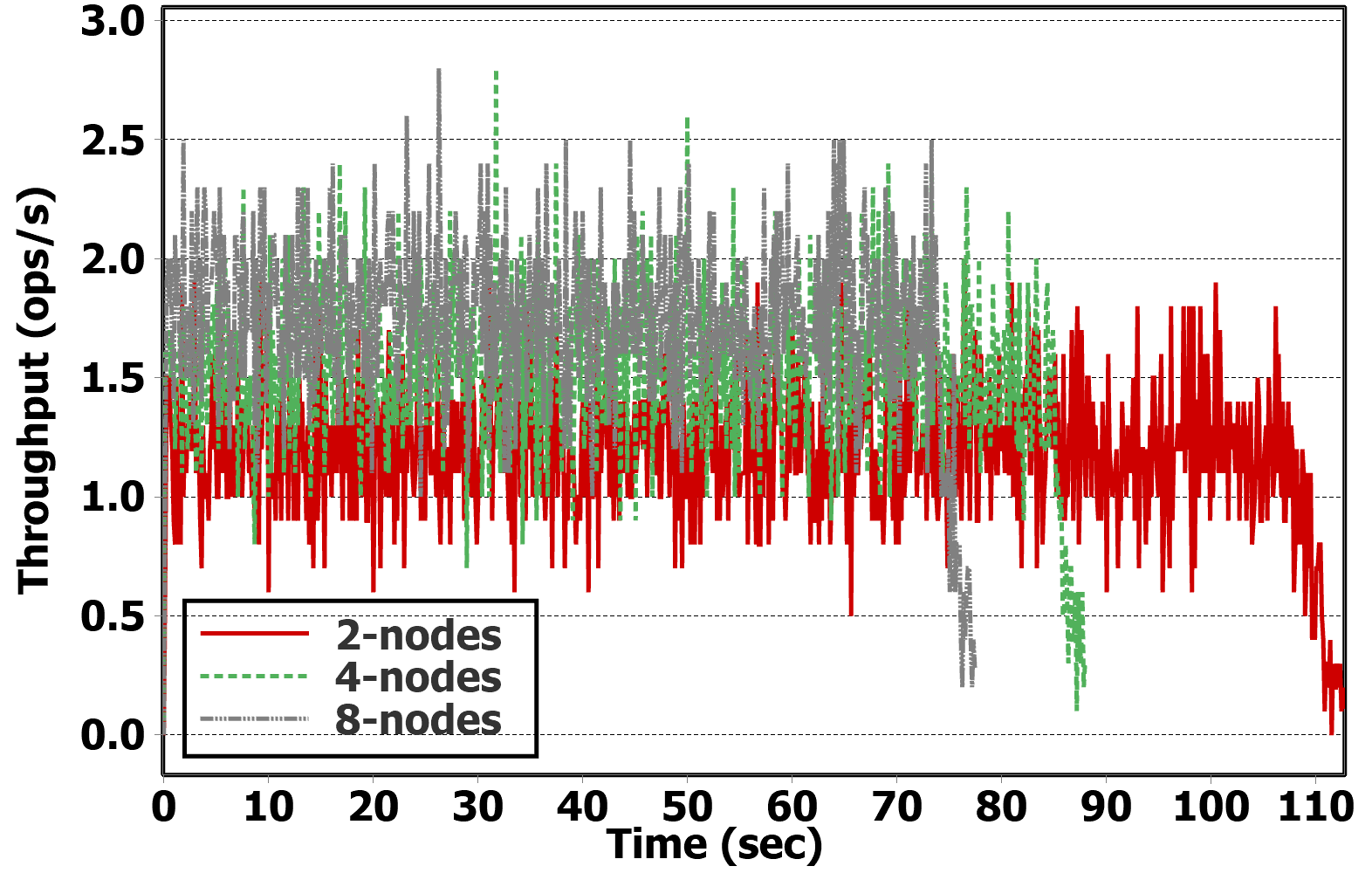}}
  \subfloat[Workload F]{\label{fig:redis-scaling-f-5x}\includegraphics[height=3cm,width=0.33\textwidth]{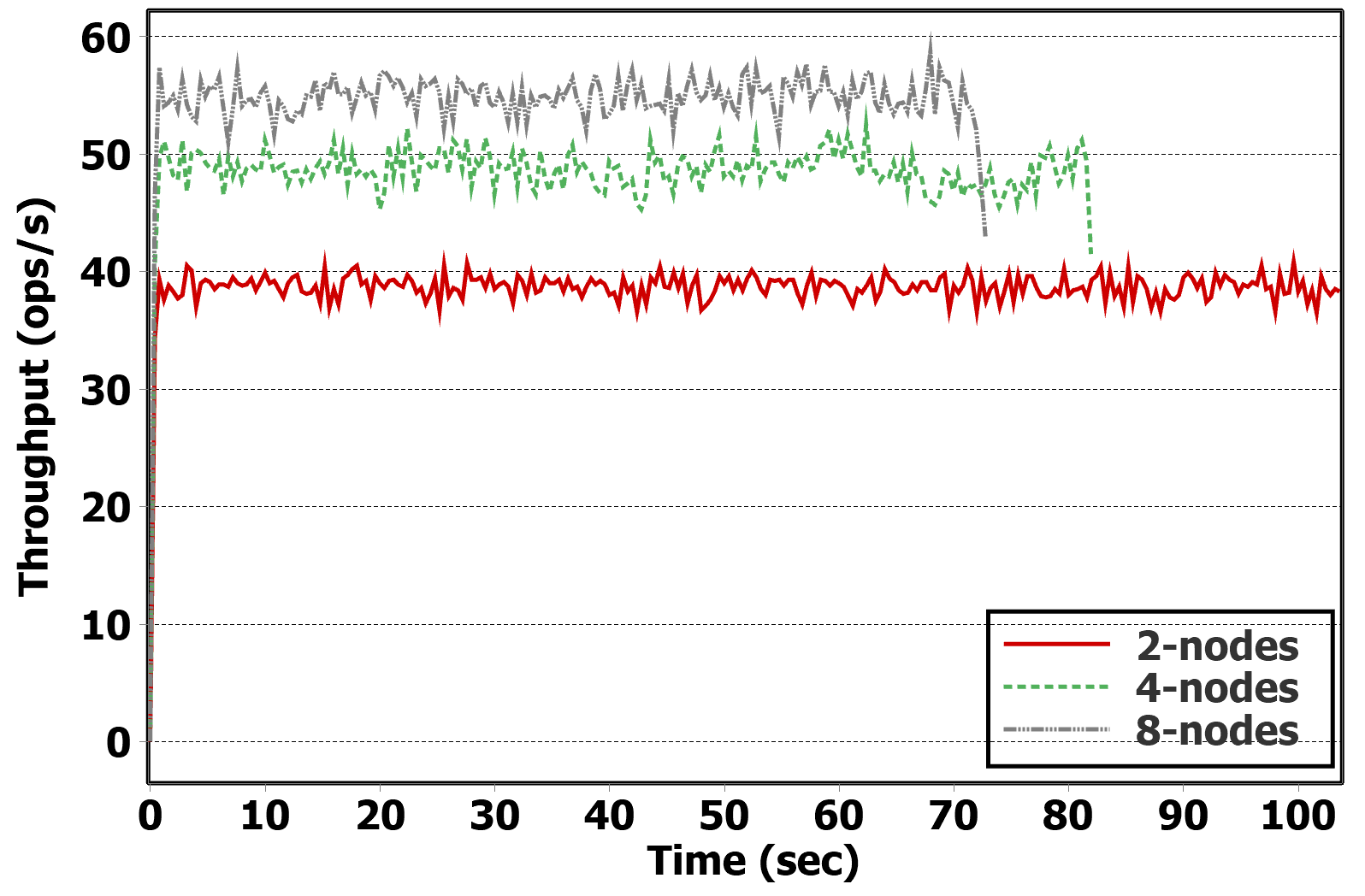}}
  \caption{The impact of offline Up-sizing and Down-sizing of a \textbf{Redis} cluster on \textbf{Redis} throughput }
\label{fig:redis-offline-scaling}
\vspace{-5mm}
\end{figure*}
We investigated  impact of \textbf{offline} down-/up-sizing scenarios on Redis throughput since on-the-fly removing and adding a node requires long time (see Figs. \ref{fig:redis-removing} and \ref{fig:redis-adding}).
We started with two nodes and 6 preserved nodes are removed and added in the same order as conducted for MongoDB. The results show that the more nodes are in a cluster, the higher throughput we achieve for all workloads (Fig. \ref{fig:redis-offline-scaling}). The throughput of the 8-nodes Redis cluster is at most two times of the throughput for a 2-nodes Redis cluster (Figs. \ref{fig:redis-scaling-a-5x}, \ref{fig:redis-scaling-c-5x}, and \ref{fig:redis-scaling-d-5x}). The workloads A-D generate almost the same throughput because of the latency domination to other factors while this value for the workload F is 2/3 of throughput for the workloads A-D. In contrast, Redis cannot operate effectively for the workload E across the nodes with high latency and thus we skipped here.

\textbf{\begin{table}[t]
	\caption{Scaling up of Redis for workloads A, B, E, and F with LSF=X. }\label{tab:redis-scalingup-x}
	\centering
	\small
	\begin{tabular}{p{2cm}p{1cm}p{1cm}p{1cm} p{1cm}}
		\hline
		        &A   &B &E &F  \\\hline\hline
	2-nodes	    &5x      &5x          &5.5x    &5x\\
	4-nodes	    &5.3x    &5.3x        &6x      &5.2x\\
	8-nodes	    &5.8x    &5.8x        &6x      &6x\\\hline
	\end{tabular}
	\vspace{-6mm}
\end{table}}
So far, we conducted the removing and adding nodes from and to a cluster with LSF=5X.  Due to the space constraints, we summarize the results for LSF=X, where we observed the same pattern in throughput for all databases and all workloas as seen for LSF=5X. From a values perspective, it can be different for databases and workloads. For MongoDB and Cassandra, as LSF reduces from 5X to X, the throughput increases by (3.6-3.8) and  (4.18-5) times, respectively. This value linearly increases by latency (5-6 times) for Redis (Table \ref{tab:redis-scalingup-x}). These results imply that MongoDB is less affected by latency in comparison to Cassandra due to full replication, which in turns, less than Redis. In other words, less latency has more improvements in Redis throughput.

\subsubsection{The impact of links removal on the performance}\label{sec:linkremoval}

In this section, we discuss how a link removal has an impact on network accessibility, which in turn, impacts the throughput of databases. We provided accessibility through redundant links (mesh network) on the virtualized network layer by using WireGuard. On top of the mesh network layer, we used {\myfont{tinc}} to handle self-routing data if links are removed in its underlying layer (i.e., mesh network). To evaluate such an arrangement of network layers, we deployed 8 VMs and created a random mesh network layer across them (\textit{densely network}). Then, we assigned actual latency and bandwidth across 8 VMs (Table \ref{tab:latency-bandwidth-measure}) to the links through \textit{tcconfig}. We also generated random numbers in the range of actual latency and bandwidth values for redundant links. According to this setup, we created 64 links across 8 VMs and then remove 85\% of them -- 50 out of 64-- (\textit{sparsely network}). For each link removal, we evaluate databases 60 seconds before and after removing the link. Thus, removing 50 links for particular workload evaluation requires $50 \times 120 = 6000$ seconds. Thus, we ran the experiment for a particular workload at least for 6000 seconds.

\begin{figure*}[ht!]
  \centering
  \subfloat[Workload A]{\label{fig:cass-removing-link-a}\includegraphics[height=3cm,width=0.25\textwidth]{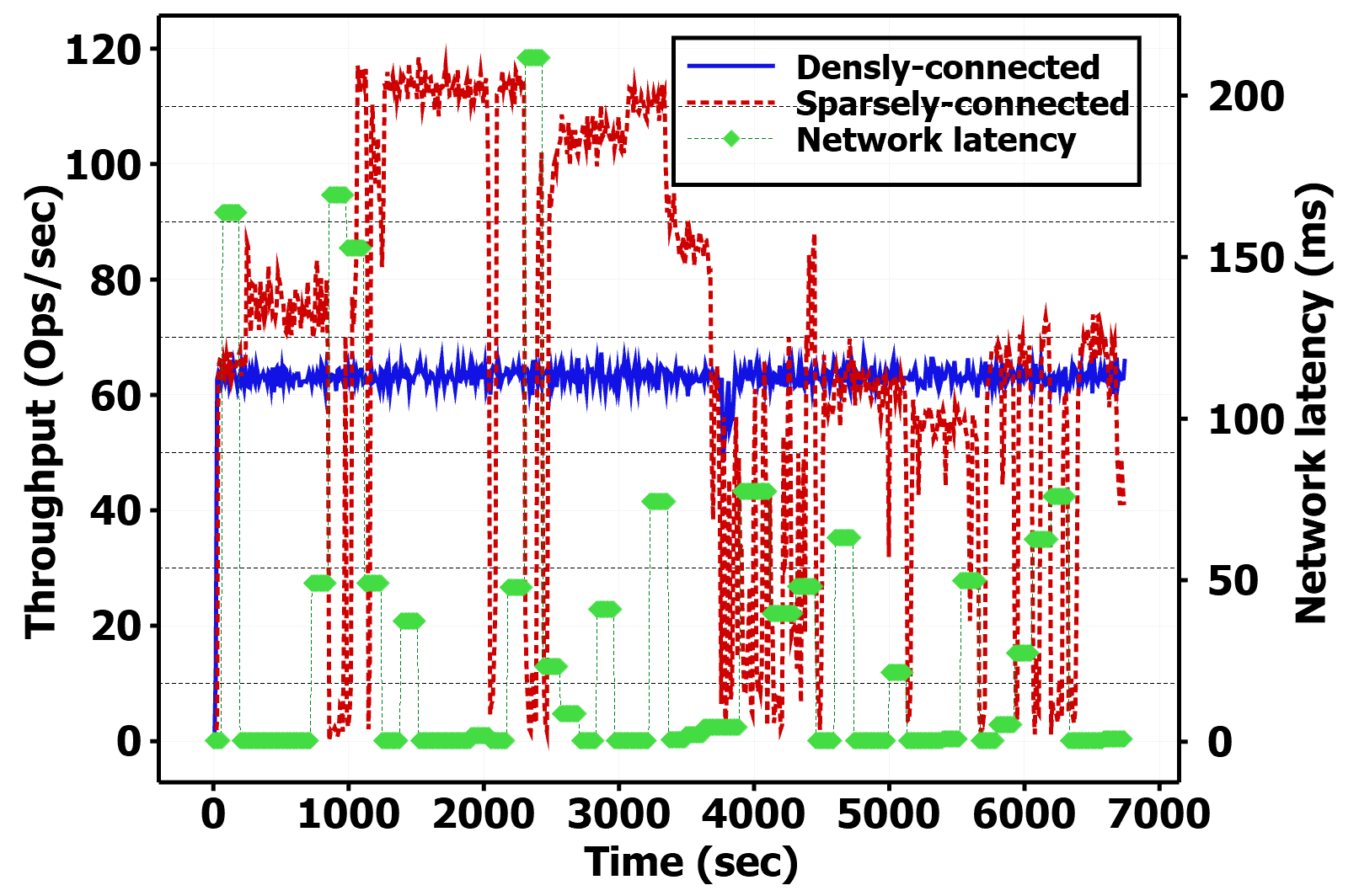}}
  \subfloat[Workload B]{\label{fig:cass-removing-link-b}\includegraphics[height=3cm,width=0.25\textwidth]{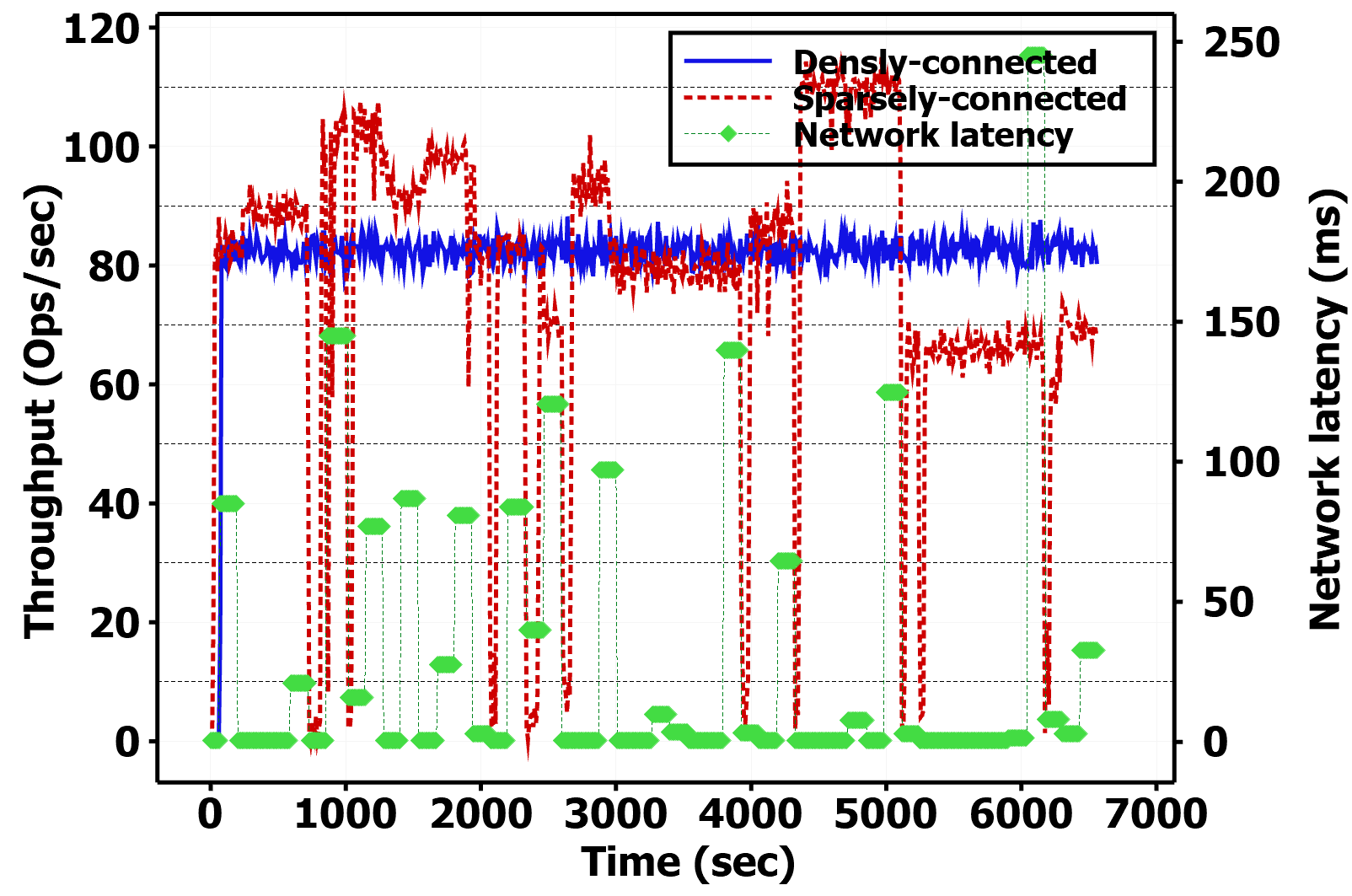}}
  \subfloat[Workload E]{\label{fig:cass-removing-link-e}\includegraphics[height=3cm,width=0.25\textwidth]{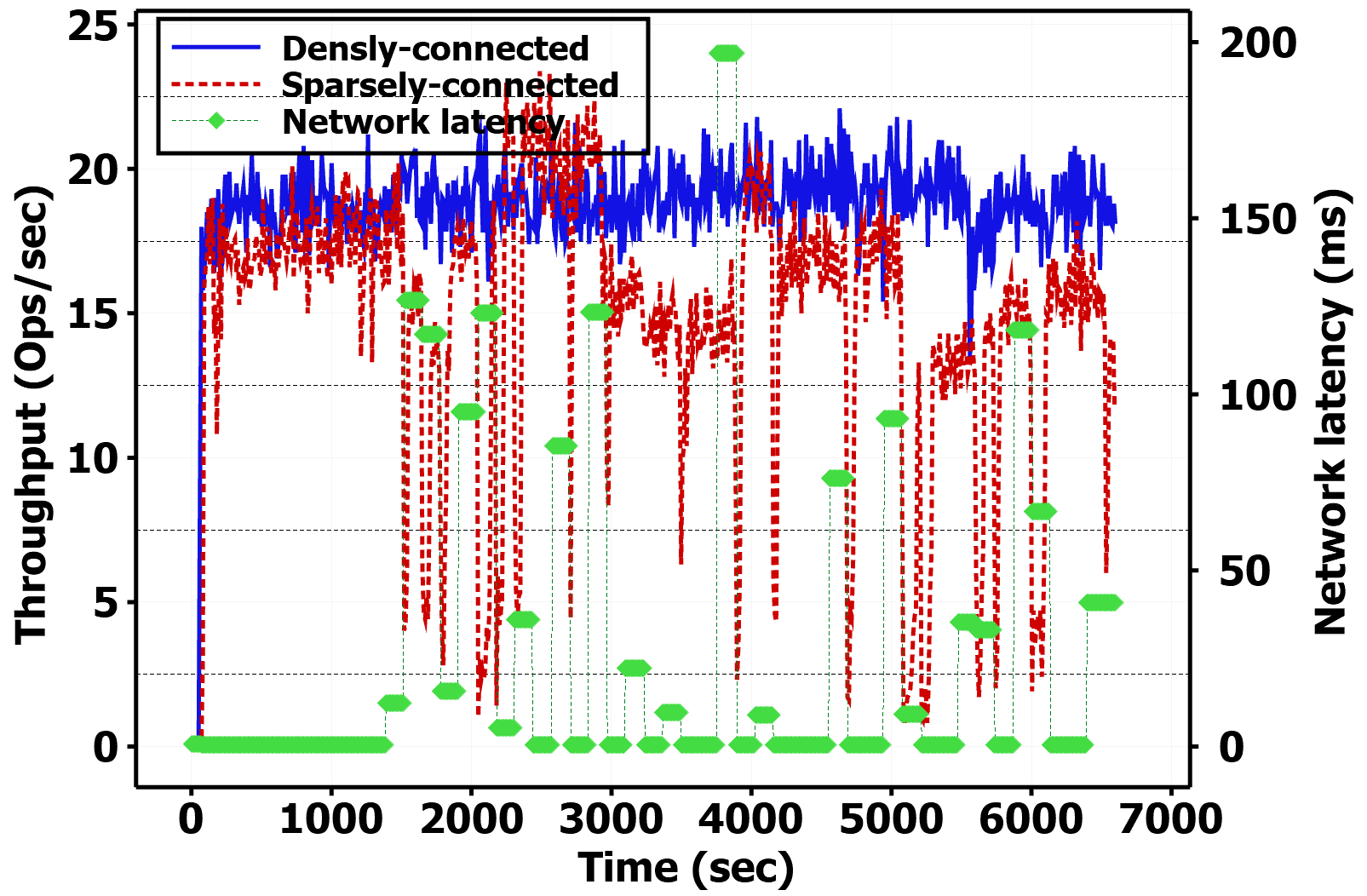}}
  \subfloat[Workload F]{\label{fig:cass-removing-link-f}\includegraphics[height=3cm,width=0.25\textwidth]{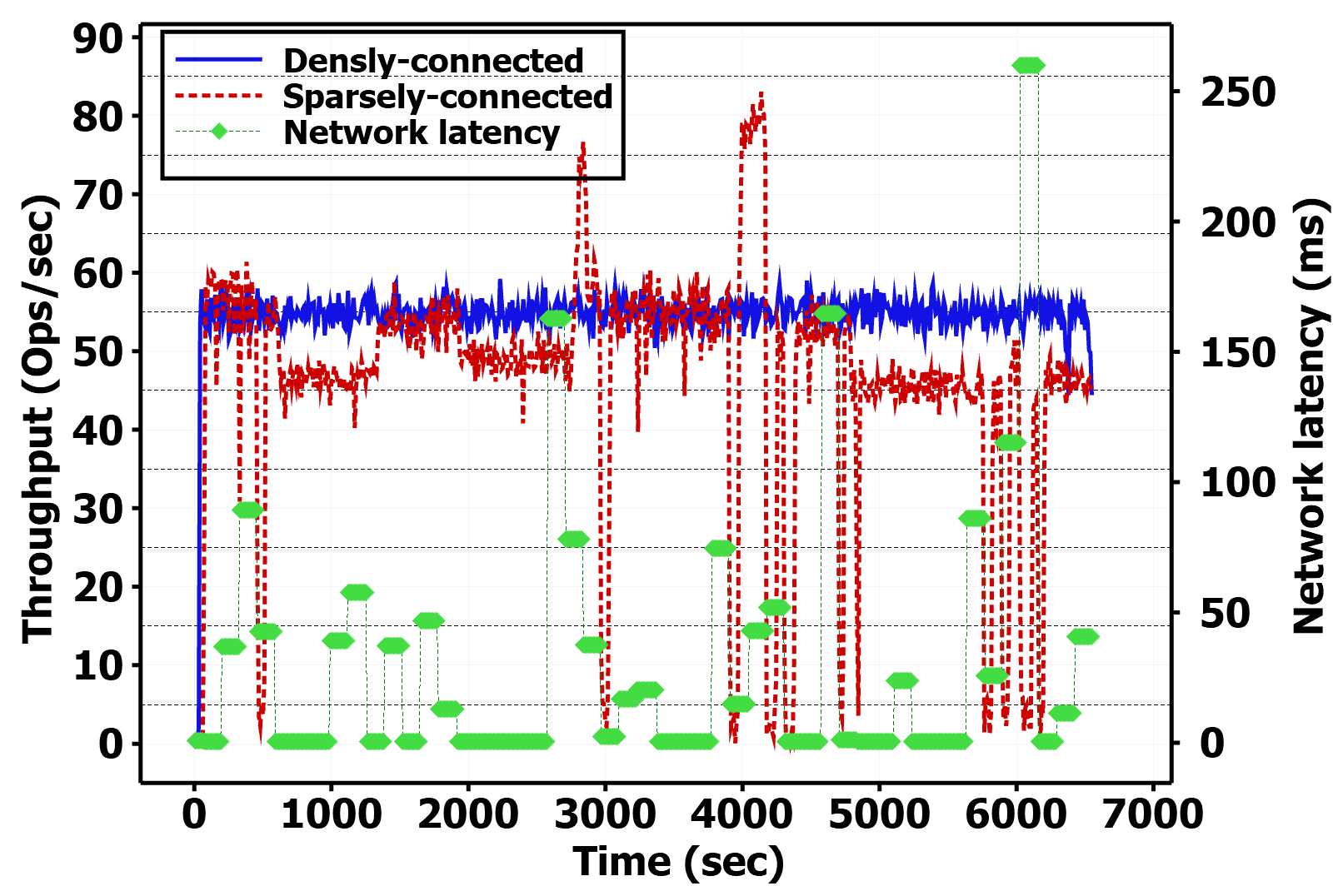}}\\
  \caption{Performance impact of links removal on \textbf{Cassandra} cluster}
\label{fig:cass-removing-link}
\vspace{-5mm}
\end{figure*}

Fig. \ref{fig:cass-removing-link} plots the throughput of Cassandra for \textit{densely} and \textit{sparsely} networks against \textit{network latency} across mesh network.
Network latency relates to the changes in average network latency between all VMs before removing a link and after removing a link for 10 times measuring through the Ping command. From the results, we observed that in the densely network, Cassandra provides a throughput of  60 Ops/sec for workload A, 80 Ops/sec for workload B, 20 Ops/sec for workload E, and 55 Ops/sec for workload F. As links are removed from the mesh network, the following observations can be made from the results. First, mesh network significantly incurs changes in network latency in some cases because {\myfont{tinc}} re-routed the data to send through other redundant links, which might be shorter or longer paths from a latency perspective. This improves the throughput, as an example, for workload A between 1000-2000 seconds and 3000-4000 seconds (Fig. \ref{fig:cass-removing-link-a}). In contrast, Re-routing of data can degrade the throughput of Cassandra as can be observed in all workloads.
Second, as expected, during a new link selection to re-route data, Cassandra might become unresponsive for a very short time and its throughput gets close to zero/exactly zero for all workloads. However, this phenomenon can be seen less for workload B due to the lesser engagement of links for reading requests.   

\begin{figure*}[ht!]
  \centering
  \subfloat[Workload A]{\label{fig:mongo-removing-link-a}\includegraphics[height=3cm,width=0.25\textwidth]{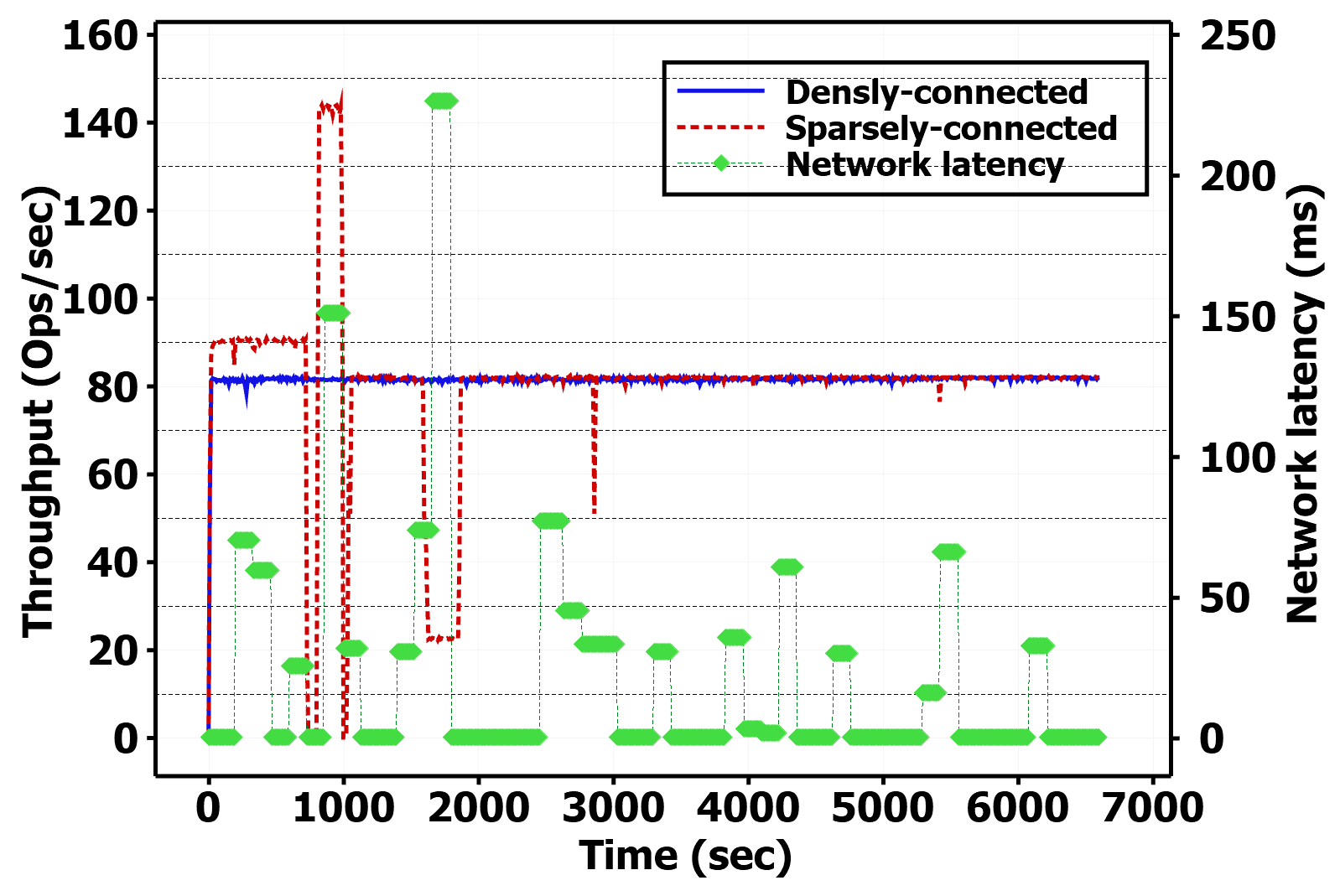}}
  \subfloat[Workload B]{\label{fig:mongo-removing-link-b}\includegraphics[height=3cm,width=0.25\textwidth]{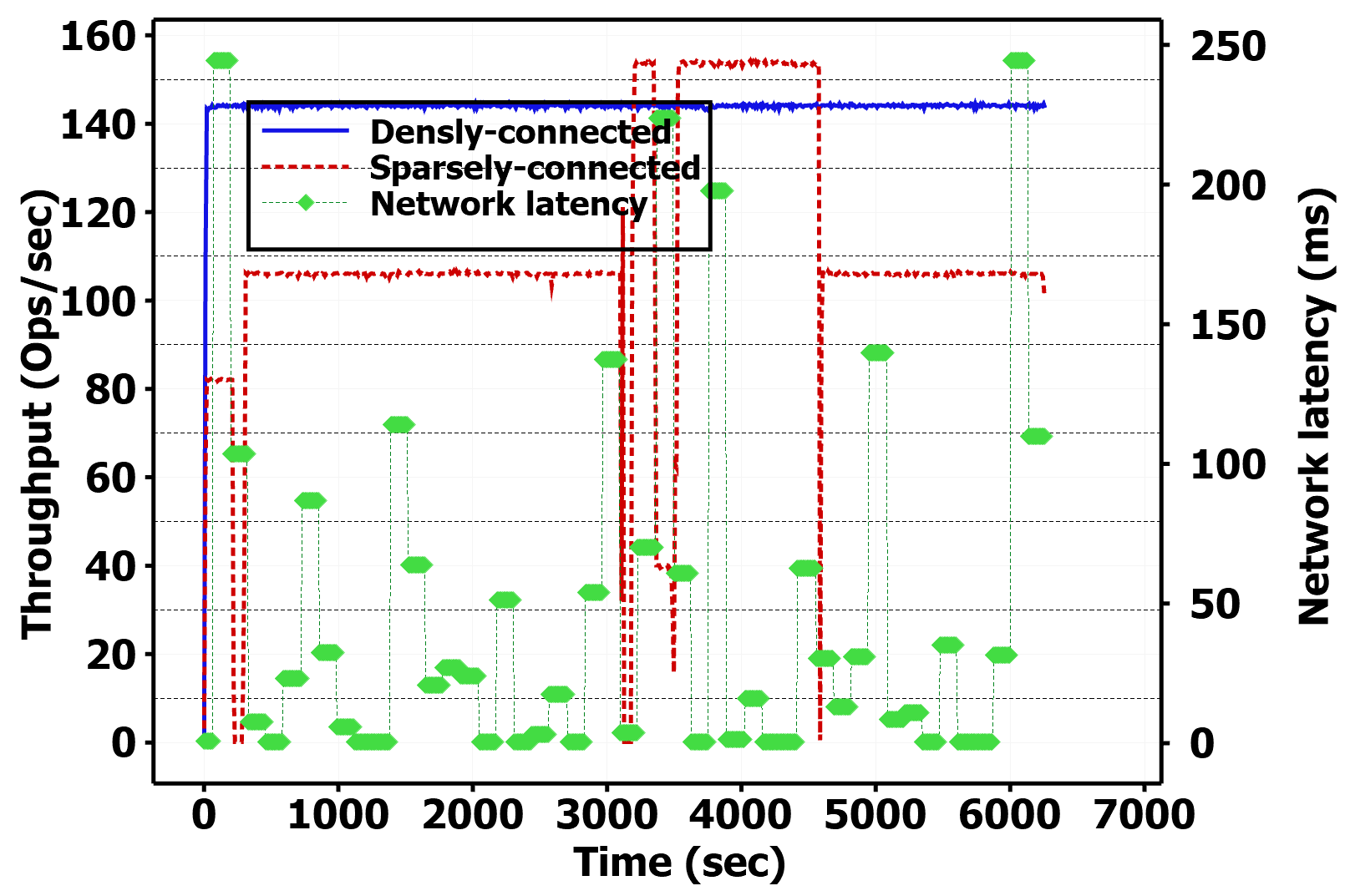}}
  \subfloat[Workload E]{\label{fig:mongo-removing-link-e}\includegraphics[height=3cm,width=0.25\textwidth]{Fig/mongo-a-cut-link-5x.png}}
  \subfloat[Workload F]{\label{fig:mongo-removing-link-f}\includegraphics[height=3cm,width=0.25\textwidth]{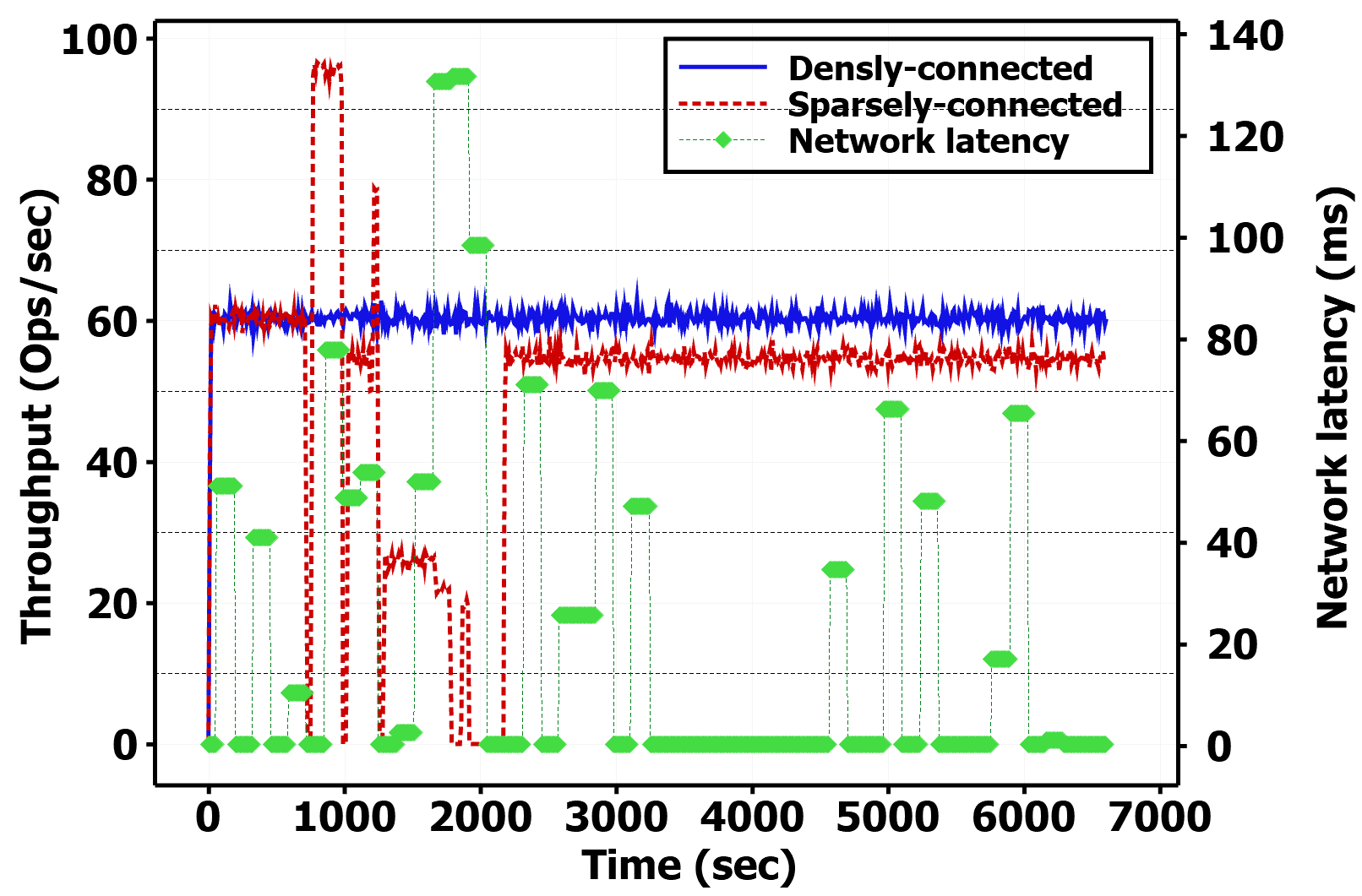}}\\
  \caption{Performance impact of links removal in \textbf{MongoDB} cluster}
\label{fig:mongo-removing-link}
\vspace{-5mm}
\end{figure*}

Fig. \ref{fig:mongo-removing-link} shows the performance impact of links removal in the MongoDB cluster. Compared to Cassandra, MongoDB shows a better resilience during the removal of the links since it was unresponsive to requests for 3-4 times (i.e., zero in throughput for 5-10 seconds). This is because the requests can be served by any secondary or primary node, not by several nodes as required for Cassandra. As a result, the probability of all nodes being unresponsive is much less than a subset of nodes (e.g., at least three nodes should be available in the Cassandra cluster) despite links removal. As expected, after every unresponsiveness, {\myfont{tinc}} handles re-routing data and in some cases it finds a better path to send requests and receive data. As an example, we can see a throughput of 140 Ops/sec for workload A.   

\begin{figure*}[ht!]
  \centering
  \subfloat[Workload A]{\label{fig:redis-removing-link-a}\includegraphics[height=3cm,width=0.33\textwidth]{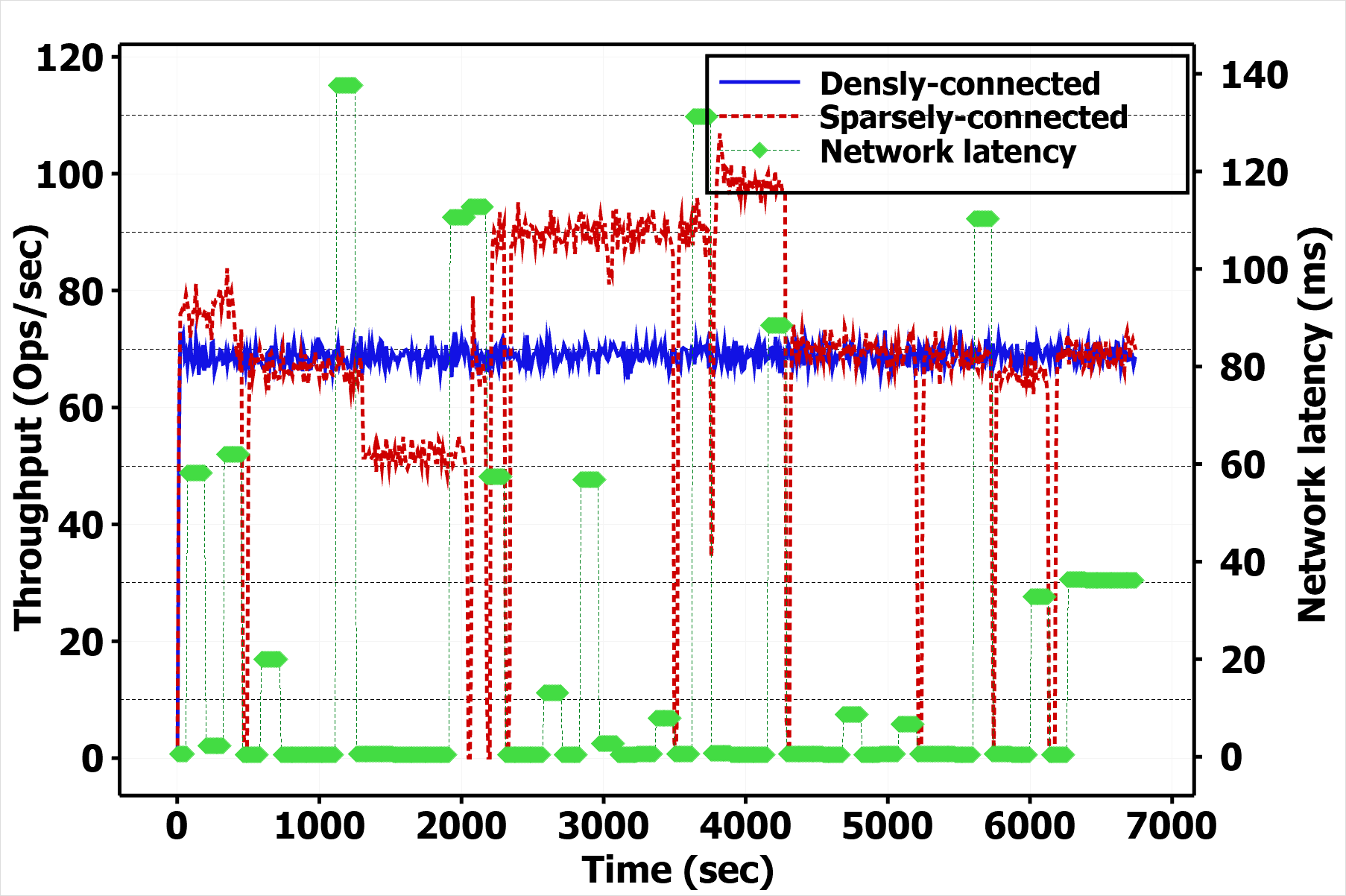}}
  \subfloat[Workload B]{\label{fig:redis-removing-link-b}\includegraphics[height=3cm,width=0.33\textwidth]{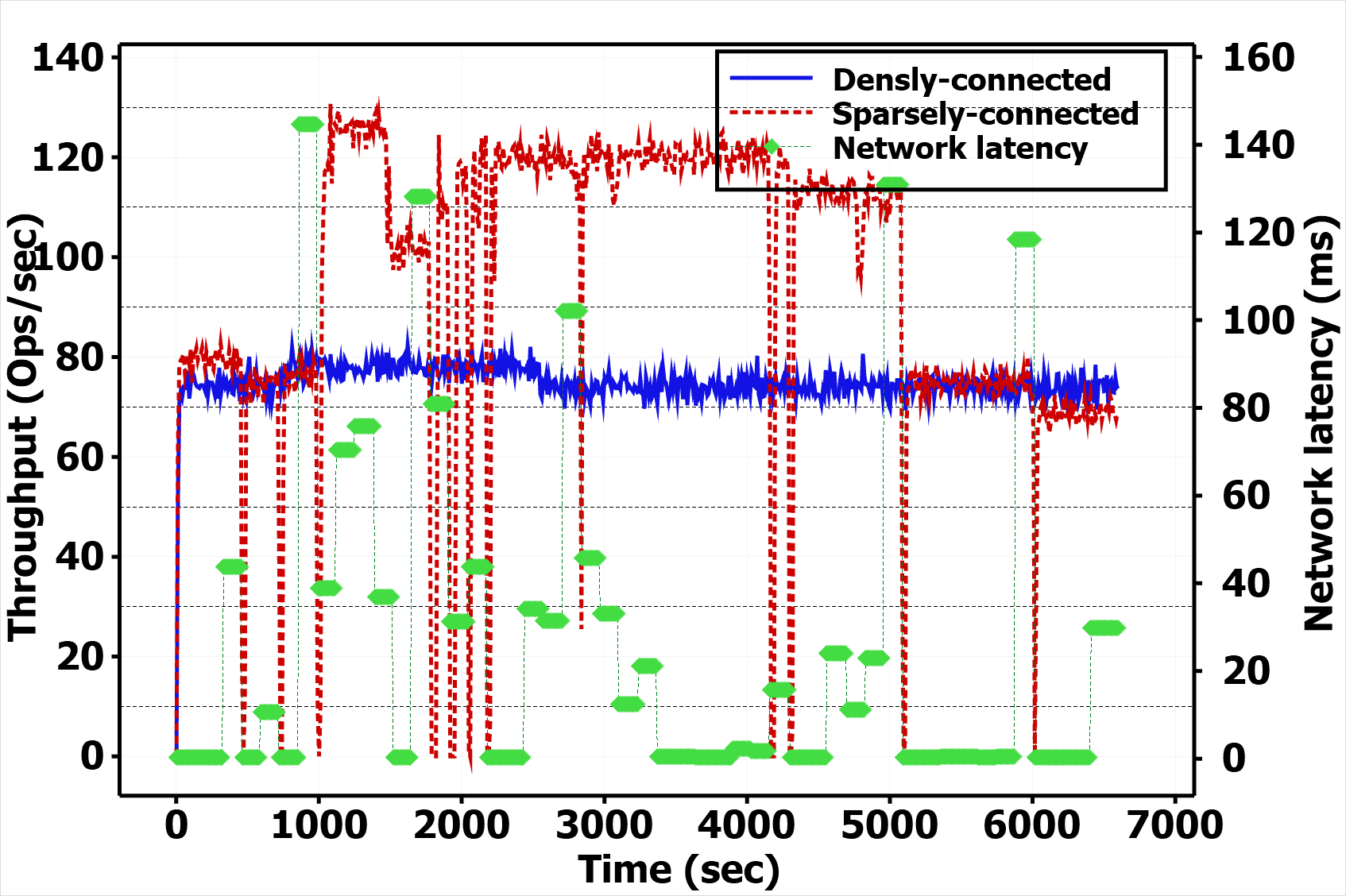}}
  \subfloat[Workload F]{\label{fig:redis-removing-link-f}\includegraphics[height=3cm,width=0.33\textwidth]{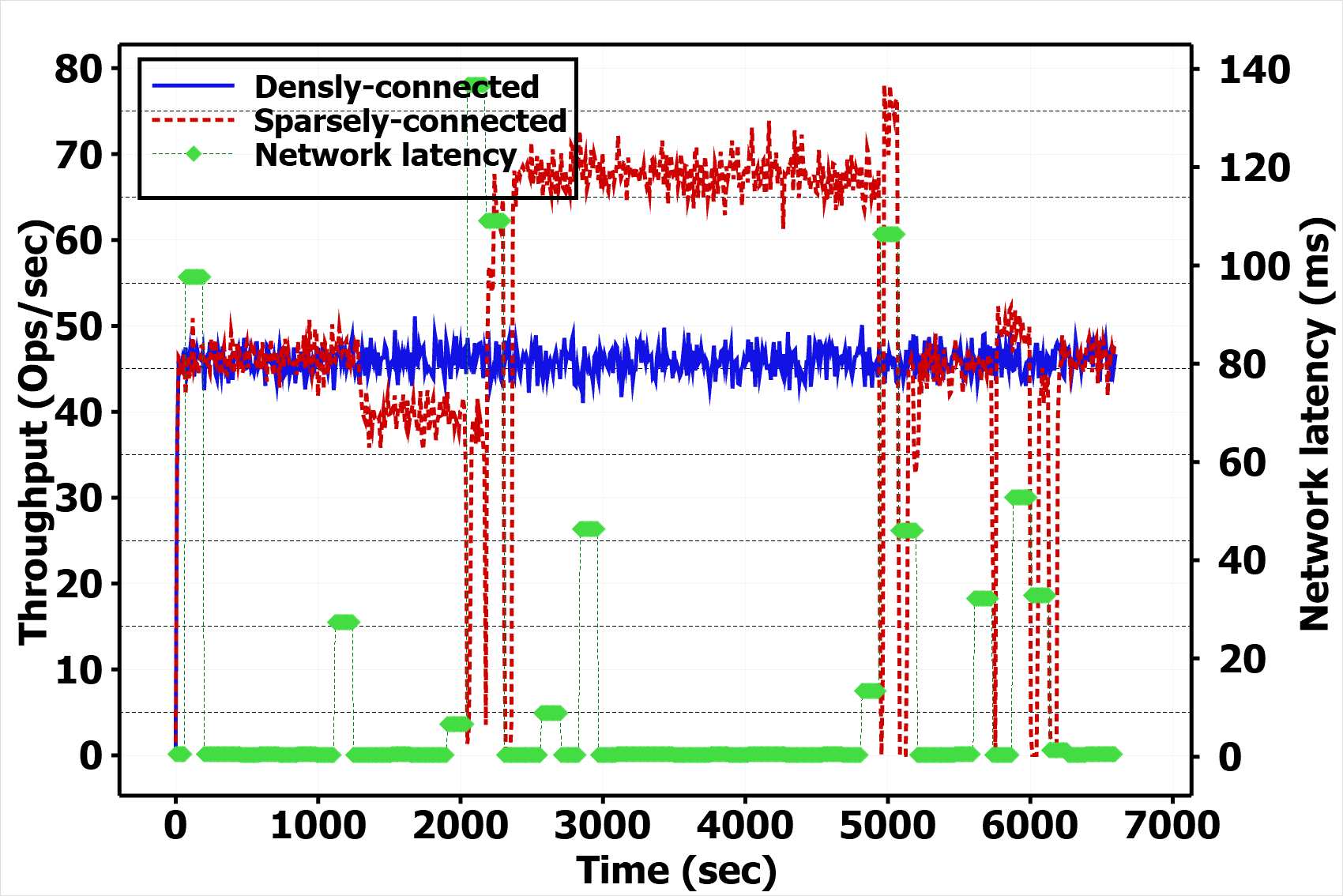}}\\
  \caption{Performance impact of links removal in \textbf{Redis} cluster}
\label{fig:redis-removing-link}
\vspace{-5mm}
\end{figure*}

Fig. \ref{fig:redis-removing-link} shows the impact of links removal on Redis throughput. We used Jedis ({\myfont{v2.10.2}})\footnote{Jedis:\url{https://github.com/redis/jedis}} as a client library in Java to support re-sending requests to Redis as data server. Moreover, we modified YCSB ({\myfont{v0.17.0}}) to re-send requests during re-routing data by {\myfont{tinc}} due to links removal. The results show that Redis become unresponsive during  re-routing data by {\myfont{tinc}} although the number of such situations for Redis is less than Cassandra (about 10 times) and unresponsive duration time is between 20 and 40 seconds. Furthermore, as thlinks are removed from the mesh network, the throughput of Redis can be improved or degraded depending on the path of re-routing data.

\begin{figure*}[ht!]
  \centering
  \subfloat[Workload A]{\label{fig:mysql-removing-link-a}\includegraphics[height=3cm,width=0.25\textwidth]{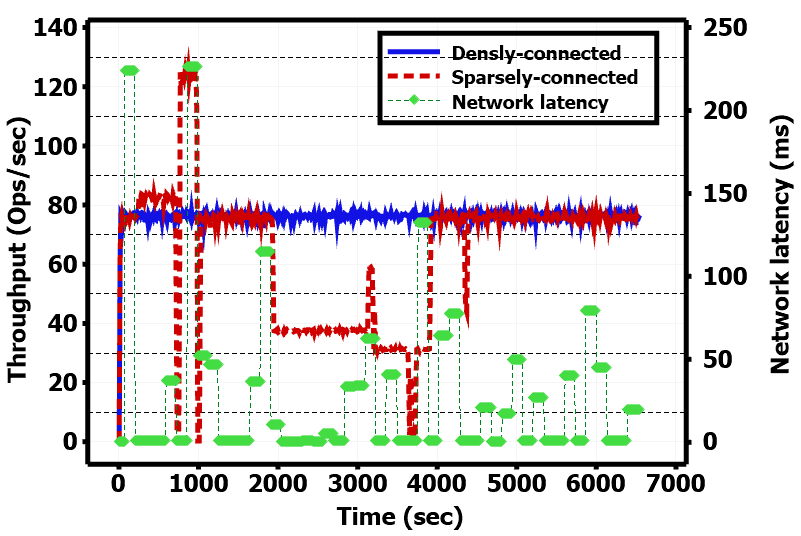}}
  \subfloat[Workload B]{\label{fig:mysql-removing-link-b}\includegraphics[height=3cm,width=0.25\textwidth]{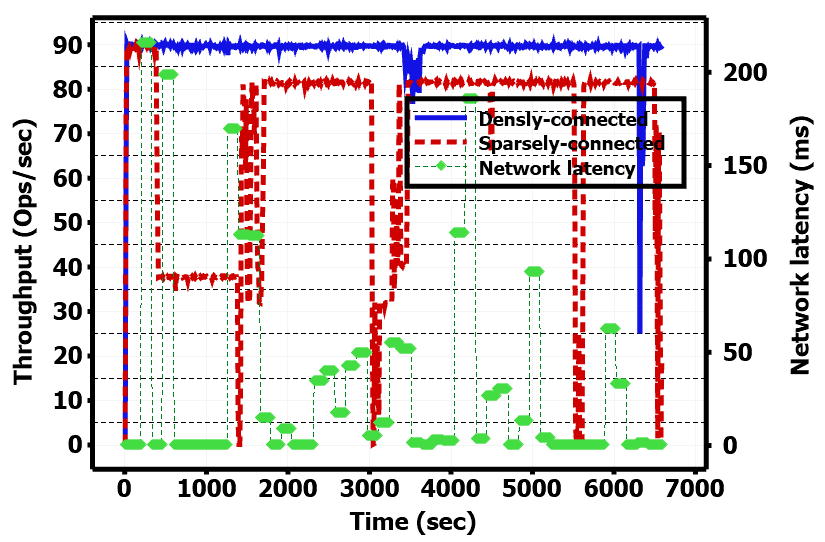}}
  \subfloat[Workload E]{\label{fig:mysql-removing-link-e}\includegraphics[height=3cm,width=0.25\textwidth]{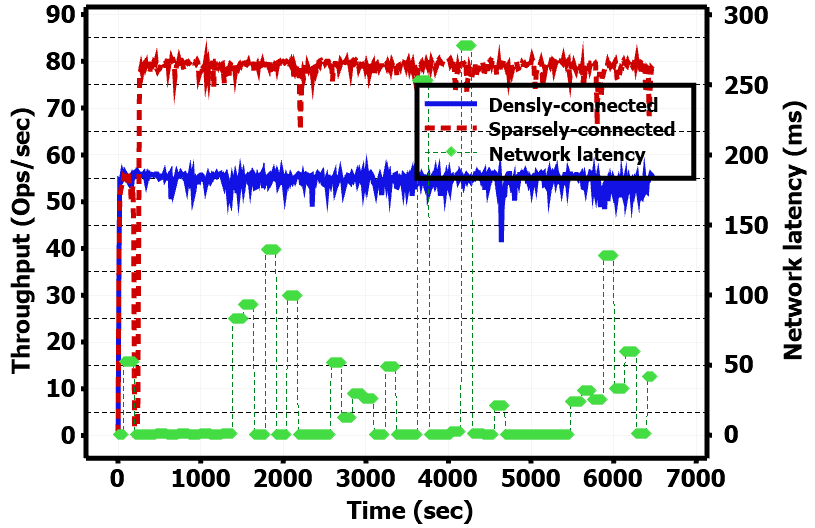}}
  \subfloat[Workload F]{\label{fig:mysql-removing-link-f}\includegraphics[height=3cm,width=0.25\textwidth]{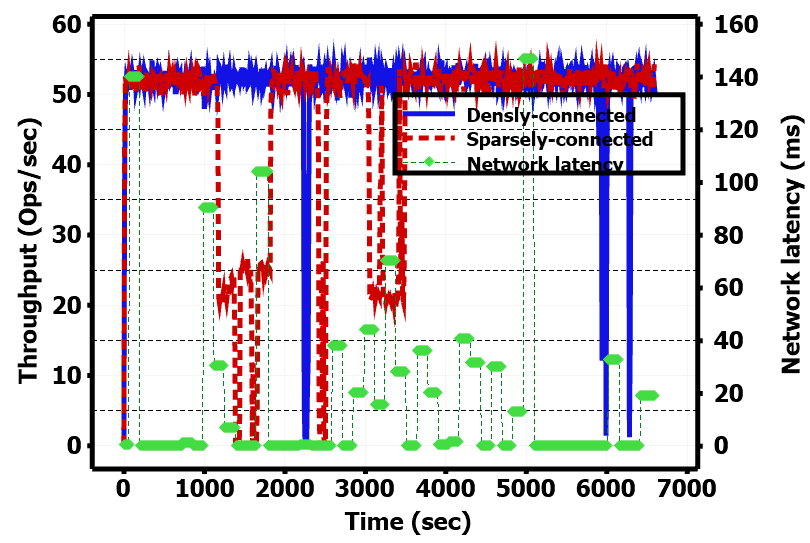}}
  \\
  \caption{Performance impact of links removal in \textbf{MySQL} cluster}
\label{fig:mysql-removing-link}
\vspace{-7mm}
\end{figure*}

Fig. \ref{fig:mysql-removing-link} shows the effect of links removal on  MySQL throughput. Unresponsiveness of MySQL to the requests is four times for workload A, three times for workload B, one time for workload E, and three times for workload F. Thus, MySQL exposed more or less the same values in terms of unresponsiveness in comparison to MongoDB. This can be explained that both databases keep a full replication of data in the data groups (for MySQL) or nodes (for MongoDB).

\section{Lessons learnt}
\textbf{Fragmented Hybrid Cloud Implementation:}
We implemented FHC by means of Terraform for infrastructure deployment, Linux-based utilities to create an arbitrary mesh network, and applying traffic shaping rules in terms of latency and bandwidth. We have also written a bunch of bash scripts to leverage database APIs for conducting down-/up-sizing databases at the application level and links removal. Whilst each task is relatively simple to conduct, coherent orchestration of all tasks together poses extra challenges. {\myfont {tcconfig}} as a tc command wrapper makes it easy to setup network rules for a single link. As we come up with applying multiple rules on a single link between two nodes, {\myfont {tcconfig}} cannot handle it because {\myfont {tcconfig}} generates repetitive random id numbers for these rules. This leads to requiring troubleshooting in terms of the unique id numbers for each rule. To figure out this issue, we generate an executable script that includes tc commands and assign a unique id number to each traffic shaping rule. Moreover,  {\myfont {tcconfig}} forces links to have less bandwidth size as the latency increases for values more than 250 ms. This most likely did not affect our environment because the highest latency between two data centers is 223 ms (Table \ref{tab:latency-bandwidth-measure}). {\myfont {tinc}} as a peer-to-peer VPN requires 30-40 seconds to re-route data as a link appears/disappears. However, {\myfont {tinc}} cannot find the shortest path in terms of latency; instead, it finds the shortest path based on the number of hops between two nodes. Thus, to optimize network latency, it is required to change tinc source code at the network level to select links-based paths between two nodes based on latency. 

\textbf{Mobility:} This feature causes variation in distance between cluster nodes in FHC. We translated distance to latency, where the more distance is the more latency is observed \cite{Wu2013}. Similarly, the same correlation between distance and bandwidth can be considered. {\myfont{tcconfig}} was used to change variation in latency and bandwidth to emulate the impact of latency on the performance of databases deployed across mobile cluster nodes. The results show that Cassandra has the most tolerance against the latency variation for all workloads, where data is largely distributed across cluster nodes. In contrast, Redis was the most impacted in all workloads except in read-related workloads (B, C, D) in which MySQL flipped Redis. MongoDB has an intensively competitive rank with Redis (in workloads A and F) and MySQL (in the read-related workloads). The results also show that there is a correlation between data node distance and the amount of transferred data for all databases with all workloads. For the workloads A and F, MongoDB incurs the lowest increment in data transfer rate, followed by Cassandra. For the read-related workloads, distance dominates the workload type and there is no clear win in terms of the lowest increment in the data transfer rate. For the workload E, the highest rate of data transferred belongs to Redis, implying that Redis has the lowest throughput in it. Furthermore, for MongoDB, the  lowest number of nodes engaged in serving all the workloads except the workload E. MySQL comes after MongoDB, where more nodes are involved to serve the workloads. In Cassandra and Redis, however, all the nodes have actively engaged to handle the workloads, whilst the transferred data in the Redis cluster is more than in the Cassandra cluster. 

\textbf{Down-/Up-sizing Cluster:}
Databases provide a rich set of APIs to manage data nodes. The arrangement of these APIs usage is highly-dependent on database architecture. Cassandra, MongoDB, and Redis facilitate more flexible APIs than MySQL to conduct down-/up-sizing clusters. This can be justified that MySQL supports a pre-defined  data model and dependency across data tables. Furthermore, in MongoDB, all the requests are directed to the master node to shrink/expand the MongoDB cluster, whilst in Cassandra and Redis, the requests are submitted to the node that is being removed/added. There is no need to redistribute data in MongoDB as a node is removed/added; whilst, Redis reshards and re-balances data across cluster nodes and Cassandra goes through re-balancing data in the background. MySQL compiles a time-consuming process to remove/add node group rather than a node. These differences in architectural perspective are reflected in databases performance as the cluster is down-/up-sized. Cassandra's performance is subject to fluctuation during the down-/up-sizing cluster since it struggles to re-balance and re-establish data across cluster nodes. MongoDB, in contrast, takes smooth and prompt action to remove/add a node, and changes in its performance depend on latency between the node being removed/added and the client node. Redis demands a long-time process for down-/up-sizing clusters and its performance correlates to cluster size.

\textbf{Network Accessibility:}
We implemented a resilient mesh network for distributed database deployment through redundant links by means of WireGuard. We leveraged {\myfont{tinc}} to re-route data in network in response to the removal of links.  {\myfont{tinc}} guarantees database services running against changes in network connectivity although with 30-40 network disconnectivity. This leads to fluctuation and unresponsiveness in database performance depending on the data replication factor. The result shows that MongoDB and MySQL are more resilient during link removal, while Cassandra and Redis are the worst. 

\section{Conclusion and Future work}
We implemented a layered Fragmented Hybrid cloud (FHC) in which computing nodes are hosted by mobile nodes. Mobility of computing nodes leads to the features for FHC in terms of time-varying latency and bandwidth, on-the-fly joining/leaving nodes, and network inaccessibility. We leveraged {\myfont{tccofig}} to create variable latency and bandwidth over time, API at databases level to remove and add nodes, and Linux utilities ({\myfont{WireGuard}} and {\myfont{tinc}}) to guarantee network accessibility. Our evaluation of a selected set of distributed databases (i.e., Cassandra, MongoDB, Redis, and MySQL) deployed on FHC provides  criteria for selecton of a suitable database. Cassandra's performance was impacted the least with latency increment for write-related and scan workloads, whilst  Redis was affected the most. For the read-related workloads, MySQL received the most impact from latency increment followed by Redis.  MongoDB and MySQL rely on fewer nodes to handle the workloads (except workload E), whilst Cassandra and Redis depend on all the nodes to serve workloads. Removing and adding nodes from and to the cluster nodes causes significant fluctuation in Cassandra's performance. In contrast, MongoDB smoothly transits to a new cluster configuration, although changes in MongoDB performance depend on the latency between the client node and the node being removed/added. Redis, however,  incurs a relatively time-consuming procedure to remove/add nodes and its performance depends on cluster size. Lastly, MongoDB and MySQL are the most tolerable databases against network failure due to full replication usage at the node and node group level, respectively.  Differently, Cassandra and Redis are the least because of distributing data across cluster nodes with a limited replicas number. 

For  the future work, we come out from the default-setting zone of database parameters to evaluate the performance of databases on mobile and static cluster nodes deployed within and across clouds. In particular, we are interested in the impact of replicas number, consistency modes, and read/write policies on database throughput. Furthermore, we also zoomed into client parameters configuration to modify the generated workloads on databases in terms of data size and request number.

\ifCLASSOPTIONcompsoc
\else
\fi


\ifCLASSOPTIONcaptionsoff
  \newpage
\fi

\bibliographystyle{IEEEtran}
\vspace{3 mm}
\bibliography{references}
\vspace{-11 mm}
\begin{IEEEbiography}[{\includegraphics[width=1in,height=1.25in,clip, keepaspectratio]{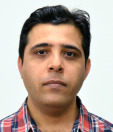}}]{Yaser Mansouri} received the B.S. and M.S.  and PhD degrees in computer science from National University of Iran, Tehran, Ferdowsi University of Mashhad, Mashhad, and The University of Melbourne, Melbourne, respectively. He is currently a researcher with the Centre for Research on Engineering Software Technologies (CREST) at the University of Adelaide, Adelaide. His research interests cover the broad area of Cloud Computing and  Distributed systems with special emphasis on algorithm designs and empirical performance evaluation of large-scale systems.
\end{IEEEbiography}

\vspace{-11 mm}
\begin{IEEEbiography}[{\includegraphics[width=1in,height=1.25in,clip, keepaspectratio]{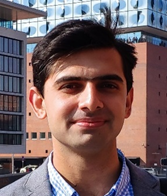}}]{Faheem Ullah} 
 is a lecturer and cyber security program coordinator at the University of Adelaide, Australia. Faheem Ullah is also a member of CREST - Centre for Research on Engineering Software Technologies. He previously completed his PhD from the University of Adelaide. Faheem's research and teaching interests include big data analytics, cyber security, software engineering, and cloud computing. Faheem is a two times gold medalist, one-time silver medalist, and receiver of 6 academic distinctions. He has supervised more than 40 undergrad/master/PhD students.
\end{IEEEbiography}

\vspace{-11 mm}
\begin{IEEEbiography}[{\includegraphics[width=1in,height=1.25in,clip, keepaspectratio]{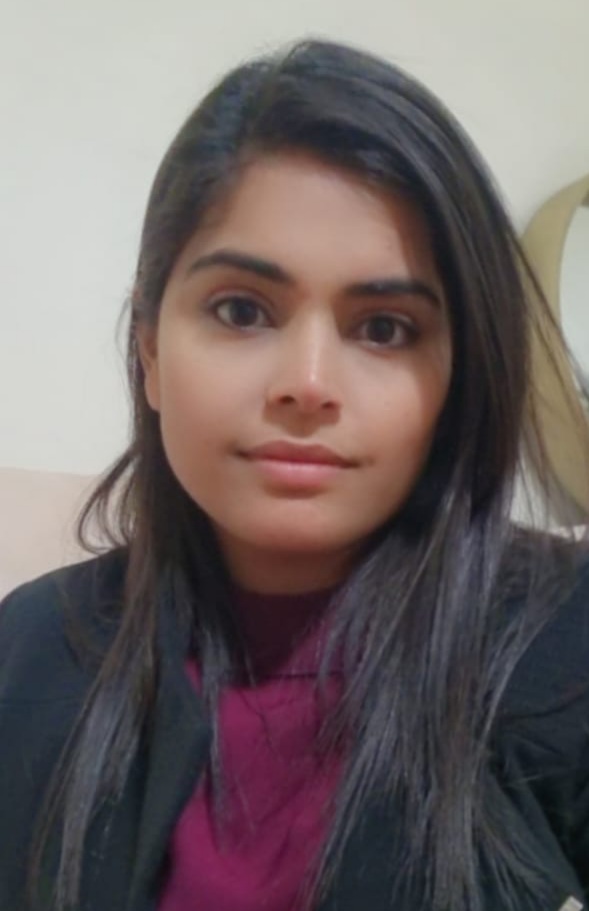}}]{Shagun Dhingra} is a Software Engineer with the Centre for Research on Engineering Software Technologies (CREST) at the University of Adelaide. As a part of the research team, Shagun’s work is completely focused on research, design, implementation, and maintaining software programs to fulfill the requirements of research projects. Shagun also focuses on implementing, evaluating, and tuning big data analytical frameworks in private cloud and public Clouds (Azure and AWS). Shagun has completed Master of Data Science from The University of Adelaide. 
\end{IEEEbiography}

\vspace{-11 mm}
\begin{IEEEbiography}[{\includegraphics[width=1in,height=1.25in,clip, keepaspectratio]{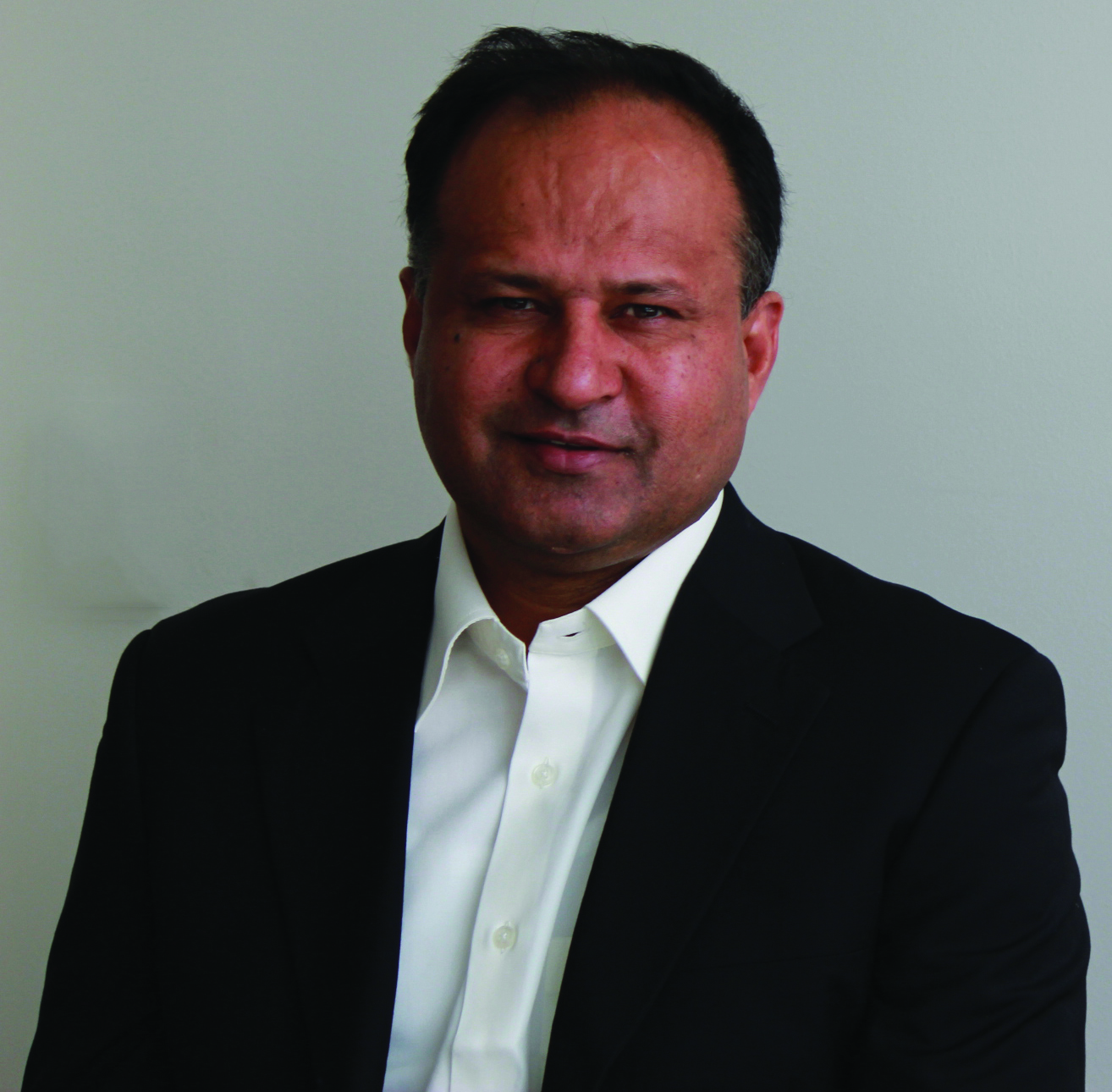}}]{M. Ali Babar}
is a Professor in the School of Computer Science, University of Adelaide. He is a honorary visiting professor at the Software Institute, Nanjing University, China. Prof Babar has established an interdisciplinary research centre, CREST - Centre for Research on Engineering Software Technologies, where he leads the research and research training of more than 30 (10 PhD students) members. He leads a theme, Platforms and Architectures for Cybersecurity as Service, of the Cyber Security Cooperative Research Centre (CSCRC). Prof Babar has authored/co-authored more than 220 peer-reviewed publications through premier Software Technology journals and conferences. In the area of Software Engineering education, Prof Babar led the University's effort to redevelop a Bachelor of Engineering (Software) degree that has been accredited by the Australian Computer Society and the Engineers Australia (ACS/EA). He coordinates both undergraduate and postgraduate programs of Software Engineering at the University of Adelaide. Prior to joining the University of Adelaide, he spent almost 7 years in Europe (Ireland, Denmark, and UK) working as a senior researcher and an academic. Before returning to Australia, he was a Reader in Software Engineering with the Lancaster University.
\end{IEEEbiography}

\end{document}